\documentclass[12pt,preprint]{aastex}

\newcommand{\teff}{$T_{\rm eff}$}

\newcommand{\kms}{km\,s$^{-1}$}

\slugcomment{\today}

\shorttitle{Fundamental Properties of O-Type Stars}
\shortauthors{Heap et al.}

\begin{document}

\title{Fundamental Properties of O-Type Stars\altaffilmark{1}}
\author{Sara R. Heap, Thierry Lanz\altaffilmark{2}, and Ivan Hubeny\altaffilmark{3}}

\begin{abstract}
We present a comprehensive analysis of high-resolution, far-ultraviolet
HST/STIS, FUSE, and optical spectra of 17 O stars in the Small Magellanic
Cloud. Our analysis is based on the OSTAR2002 grid of NLTE metal
line-blanketed model atmospheres calculated with our NLTE model atmosphere
code TLUSTY. We systematically explore and present the sensitivity of
various UV and optical lines to different stellar parameters. We have
obtained consistent fits of the UV and the optical spectrum to derive the
effective temperature, surface gravity, surface composition, and
microturbulent velocity of each star. Stellar radii, masses, luminosities
and ages then follow directly. Similarly to recent but more limited studies,
we derive cooler temperatures than the standard effective temperature
calibration of O stars. We propose a new calibration between the spectral
type and effective temperature based on our results from UV metal lines as
well as optical hydrogen and helium lines. For stars of the same spectral
subtype, we find a general good agreement between $T_{\mathrm{eff}}$\
determinations obtained with TLUSTY, CMFGEN, and FASTWIND models. The lower
effective temperatures translate into lower bolometric corrections and lower
ionizing fluxes. We derive ionizing luminosities that are smaller by
a factor of 3  compared to luminosities inferred from previous standard
calibrations. The
chemical composition analysis reveals that the surface of about 3/4 of the
program stars is moderately to strongly enriched in nitrogen, while showing
the original helium, carbon, and oxygen abundances. Our results support the
new stellar evolution models that predict that the surface of fast rotating
stars becomes nitrogen-rich during the main sequence phase because of
rotationally-induced mixing. We find, however, enrichment factors (up to a
factor of 40) that are larger than predicted by stellar evolution models.
Most stars exhibit the ``mass discrepancy'' problem, that is,
the spectroscopic masses are lower than the evolutionary masses. We
interpret this discrepancy as a result of fast rotation which lowers the
measured effective gravity. Nitrogen enrichment and low spectroscopic masses
are therefore two manifestations of fast rotation. Our study thus emphasizes
the importance of rotation in our understanding of the properties of massive
stars and provides a framework for investigating populations of low
metallicity massive stars at low and high redshifts.
\end{abstract}


\altaffiltext{1}{Based on observations with the NASA/ESA \textsl{Hubble
Space Telescope\/}, obtained at the Space Telescope Science Institute, which
is operated by the Association of Universities for Research in Astronomy,
Inc., under NASA contract NAS5-2655. Also based on observations made with
the NASA-CNES-CSA \textsl{Far Ultraviolet Spectroscopic Explorer\/}, which
is operated for NASA by the Johns Hopkins University under NASA contract
NAS5-32985.}

\affil{Laboratory of Astronomy and Solar Physics, NASA Goddard Space Flight
Center, Greenbelt, MD 20771}

\altaffiltext{2}{Department of Astronomy, University of Maryland, College
Park, MD 20742} 

\altaffiltext{3}{Steward Observatory, University of Arizona, Tucson, AZ 85721}

\email{Sara.R.Heap@nasa.gov, tlanz@umd.edu, hubeny@as.arizona.edu}

\setcounter{footnote}{3}


\section{INTRODUCTION}

We present a determination of the the fundamental properties of O-type stars 
in order to provide a solid basis for understanding the evolution of 
massive stars forming today or long ago, and  for determining the properties 
of unresolved starbursts both near and far. A knowledge of a star's
mass, luminosity, and effective temperature is essential for
understanding the evolution of massive stars, the initial mass function, and
ages and properties of clusters and \ion{H}{2} regions.
The fundamental properties of O-type stars are also essential for interpreting the spectra
of unresolved star-forming regions and of starburst galaxies. 
Most efforts to synthesize the spectra of starburst 
galaxies use observed spectra of relatively nearby stars as templates. 
However, even these so-called empirical
methods involve a calibration of spectral type in terms of  
fundamental stellar parameters (e.g. mass, effective temperature, bolometric
luminosity) and then a subsequent calibration of these parameters in
terms of initial mass and age via a comparison with evolutionary tracks. 

There are already several spectral calibrations in general use today. One is
by Vacca et al. (1996), which is based on spectral analyses of galactic O stars 
carried out in
the 1980's. Another, which has been incorporated in the \textsc{starburst99}
code of Leitherer et al. (1999), uses the spectral calibration of
Schmidt-Kaler (1982). In the last twenty years, great strides have been
taken in increasing the quality of the observed spectra and in increasing
the physical realism of model atmospheres and spectra. Our work takes
advantage of both advances. We make use of the high-quality, high-resolution
spectra of O-type stars in the Small Magellanic Cloud (SMC) ($\S 2$). In this respect,
we are extending the recent studies of selected types of O
stars (e.g. O~supergiants: \citep{bianchi02, crowther02, HL02a, martins02, massey04},
O~dwarfs: \citep{herrero02, NGC346, massey04}. Our study encompasses 17 stars in the
SMC that are well distributed in spectral type and luminosity class while
homogeneous in metallicity. The
breadth of the sample enables us to evaluate the sensitivity of optical and ultraviolet
spectral features to temperature, gravity, and microturbulence 
($\S 4-5$)

To interpret these spectra, we use the new, fully line-blanketed NLTE model atmospheres 
and spectra calculated by Lanz \& Hubeny (2003) ($\S 3)$.
We paid particular attention to blanketing by metal lines on the structure
of the atmosphere and emergent spectrum. We find that back-warming of the atmosphere
produced by the millions of metal lines -- all treated in NLTE -- not only affects
the metal-line spectrum but also the helium lines, leading to a lower
lower (i.e. normal) helium abundance in O-type stars and a lower 
temperature scale ($\S 6$). The lower temperature scale translates to lower 
the bolometric corrections, bolometric luminosities, and ionizing fluxes.
A new calibration of ionizing luminosities with spectral type is presented in \S6.4.

Including the myriad of metal lines in our models brings other benefits as well.
The most obvious is the expanded set of spectral diagnostics in the ultraviolet,
which is replete with metal lines. A less obvious benefit is the high level
of internal consistency of the model spectrum: ionization balance of helium, iron,
and carbon now yield the same effective temperature; and helium singlets now give
the same results as the helium triplets. 

We find that the temperature scale depends on
metallicity. Based on TLUSTY models, O-type stars of solar metallicity are about
1,000 to 2,000 K cooler than SMC O stars of the same spectral type. We also find that  
luminosity class is more closely correlated with the ratio of luminosity to 
spectroscopic mass than
it is to bolometric luminosity.  However, spectroscopic mass is
often lower than the mass derived from a star's position on the HR diagram
-- a circumstance known as the ``mass discrepancy''. We find that the mass
discrepancy is generally greatest for stars whose atmospheres are highly
enriched in nitrogen. Both anomalies -- the low spectroscopic masses and
high nitrogen abundances -- can be explained if the stars are rapid rotators %
\citep{HeapRot02}. Conclusions are presented in $\S 7$.


\section{OBSERVATIONS}

The program stars in this study are the same as those selected for an ultraviolet
spectroscopic survey of O-type stars with the \textsl{Space Telescope
Imaging Spectrograph\/} (STIS) on the \textsl{Hubble Space Telescope\/}
(HST) (HST GO Program 7437, Lennon=PI). This survey targeted 17 O-type
stars in the Small Magellanic Cloud for ultraviolet spectroscopy. The 
resultant ultraviolet spectra along with optical spectra obtained at ground-based
telescopes are displayed in an atlas of O-type stars in the Small Magellanic Cloud 
constructed by \citet{walborn00}.

We selected O-type stars in the Small Magellanic Cloud 
because of the known distance, low metallicity,
and generally low reddening of this nearby satellite galaxy. In
this study, we adopt a distance, 
$d=63\pm 3$ kpc, or equivalently, a distance modulus, $(m-M)=19.0\pm 0.1$,
in agreement with recent determinations based on Cepheids [$(m-M)=19.04\pm
0.17$, Groenewegen 2000; $(m-M)=19.05\pm 0.13$, Kov\'{a}cs 2000] and 
the tip of the Red Giant Branch [$(m-M)=18.99\pm 0.08$, Cioni et~al.
2000].) Because the program stars are at the same distance, we can
examine the correspondence between luminosity class and bolometric
luminosity. The low metallicity of the SMC ($Z=0.2Z_{\odot }$; Venn 1999; 
Bouret et~al. 2003) makes it an ideal surrogate 
for star-forming galaxies at high redshift, which generally have a
low metal content. The low metallicity also gives us a practical benefit:
the winds of O-type stars in the SMC, which are driven by radiation pressure in metal lines,
are particularly weak, so that our approach of photospheric analysis is justified. 

The program stars form  an unbiased sample
of SMC O-type stars except that they were chosen for their narrow-lined spectra in order
to avoid problems with line-blending in the analysis. Since O stars tend to
be rapid rotators, this one bias means that we are viewing the stars pole-on.

The observational data include not only the STIS and ground-based
optical spectra of 17 program stars (Walborn et~al. 2000), but
also far-UV spectra (905-1187\,\AA) from the 
\textsl{Far Ultraviolet Spectroscopic Explorer\/} (FUSE) for 9 of the
program stars. The FUSE data broaden the UV spectral coverage
to 905-1700\,\AA\ and enable better temperature determinations
through ionization balance of elements such as C, N, Si, and S.
Table~\ref{StarTab} lists the program stars, their spectral
types \citep{walborn00, walborn02}, the observatory at which the optical
spectrograms were obtained, and the availability of STIS or FUSE spectra.
Our sample contains two more stars, NGC~346 MPG~487 and MPG~682, than
Walborn et~al.'s because STIS spectroscopy of these stars was secured at a
later date.

We examined each of the STIS target-acquisition images, which consists of a $%
5\arcsec\times 5\arcsec$ picture obtained with a wide-band, red filter ($%
\lambda _{c} = 7230$\,\AA). None of the targets appears to be multiple. Of
course, at a 2-pixel resolution of $0.1\arcsec$, a target-acquisition image
cannot resolve systems whose separations are less than 6300 AU, so this was
not a stringent test. We also checked that the target-acquisition sequence
found the right star. We found that STIS failed to acquire NGC 346 MPG 682,
in that the initial field was blank. When this happens, the acquisition
software still tries to center the brightest spot in the image (undoubtedly
a cosmic-ray hit) and repeats the sequence of taking an image and centering.
We found that the (unknown) star that was actually observed was hotter than
would be expected from its optical spectrum, so we concluded that the wrong
star was observed. Acquisition of NGC 346 MPG 12 may have also failed, but
since the STIS and ESO spectra of NGC 346 MPG 12 lead to the same spectral
classification, we retained the STIS spectrum of this star for further
analysis.

Both STIS and optical data and data-processing are described in detail by
Walborn et al. (2000). Here, we only describe the FUSE observations. See
\citet{FUSE2000}, \citet{FUSE2000b}, for a description of the FUSE
instrument and performance. All publicly available FUSE spectra of the
program stars were obtained with the star placed in the LWRS ($30\arcsec%
\times 30\arcsec$) aperture. For the majority of targets, the starfield is
sufficiently sparse that the target was the dominant star in the field of
view. However, LWRS spectra of stars in NGC 346 are clearly composite: they
disagree with STIS spectra in the region of overlap (1150-1187\,\AA). We
therefore discarded them. In our Cycle~2 FUSE program B134, we observed
MPG~12 using the MDRS ($4\arcsec\times 20\arcsec$) aperture to avoid the
crowding problem. However, the resulting spectral coverage was limited
because of the imperfect alignment of FUSE optical channels. Accordingly, we
only used data obtained in the LiF channels, covering the range 990-1187\,\AA.

A FUSE spectrum is recorded in eight segments, which together cover the
spectral region between 905 and 1187 \AA\ at a resolving power, $R=20,000$.
We acquired the raw data segments from the Multi-Mission Archive at Space
Telescope (MAST) and reprocessed the data using the \textsc{calfuse}
(v2.0.5) reduction software. We next co-added multiple exposures, if any. We
then registered, adjusted the flux levels, and combined the 8 spectral
segments. Finally, we registered and scaled the spectral flux to match the
STIS spectrum in the region of overlap (1150--1187\,\AA ), which includes
the \ion{C}{3}\,$\lambda$1176 multiplet, which was used as a wavelength
fiducial. The combined spectrum was binned to a 2-pixel resolving power, $%
R=20,000$. The typical signal-to-noise ratio in the continuum is 15-20 per
resolution element.


\section{MODEL ATMOSPHERES AND SPECTRA}

\label{Models}

We used the new, fully line-blanketed NLTE model atmospheres and spectra of 
\citet[OSTAR2002]{OS02}.\footnote{%
Available at http://tlusty.gsfc.nasa.gov/} These model atmospheres cover
the parameter range: 27,500\,K $\leq$$T_{\mathrm{eff}%
}$$\leq 55,000$\,K, with a sampling step of 2500\,K; $\log g\leq 4.75$, down
to the Eddington limit with intervals of 0.25\,dex; 10 metallicities, from
metal-rich models (twice the solar metallicity) to extremely metal-poor
models. All models in the grid have a normal helium abundance, He/H~=~0.1 by 
number density, and a microturbulent velocity, $\xi_{\mathrm{t}} = 
10$\,km\,s$^{-1}$. OSTAR2002 models relevant to the Small Magellanic Cloud are 
the ``S'' models ($Z = 0.2 Z_\odot$) and the ``T'' models ($Z = 0.1 Z_\odot$). 
If needed to achieve an overall match to observations, we interpolated logarithmically 
between model atmospheres for the chosen stellar parameters
and then computed the spectrum of the interpolated model
with SYNSPEC. See \citet{OS02} for a discussion comparing different
interpolation strategies in the \mbox{OSTAR2002} grid.

We calculated the NLTE model atmospheres with our 
computer code, TLUSTY, using the Complete Linearization/Accelerated Lambda
Iteration method \citep{NLTE1}. The program solves the basic structural
equations in a plane-parallel geometry: radiative transfer, hydrostatic
equilibrium, radiative equilibrium, statistical equilibrium, and charge and
particle conservation. The critical ingredients of such models are the completeness 
of opacity sources and energy levels treated explicitly in the statistical equilibrium
equations. The model atmospheres include  the opacities of about 2 million
lines dynamically selected from a list of 12 million lines, and they account for 
about 100,000 individual atomic levels of 45 ions. The spectrum is
typically sampled at 180,000 to 200,000 frequencies. Further information about the 
models, Opacity Distribution Functions, and Opacity Sampling options to treat the line
opacity of iron-peak elements is given in \citet{OS02}. Finally,
we calculated detailed spectra with our spectrum synthesis code, SYNSPEC, based on the 
atmospheric structure and NLTE populations provided by TLUSTY. 

The reliability of the metal-line spectrum is highly dependent on the
quality of the model atoms. Since UV absorption lines tend to arise from
ground-state or low-lying energy levels, the model atoms for those ions are
thought to be quite reliable. They should also be adequate for
higher energy levels involved in the formation of optical lines. The only
exception is the \ion{C}{3} model atom, which oversimplifies
the very high energy levels and may therefore  lead to an underestimate of
the total recombination rates and, consequently, the optical recombination lines
are predicted too weak \citep{NGC346}. However, the net global error on
the ionization level of carbon remains
insignificant and, in any case, it can be checked through a 
comparison of the effective temperature derived from carbon ionization
balance with temperatures derived from other elements. 

\subsection{Photospheric vs. Unified Models}
\label{ModCompar}

Since O-type stars are known to have have strong, supersonic winds, the validity of
static, photospheric models must be addressed. However, in low-metallicity stars 
like those in the SMC, only a few strong resonance lines show the signature of
a wind; most lines are symmetric, indicating that they are formed in a quasi-static 
photosphere. Even so, the hydrostatic density structure of the photosphere is 
complex because it is determined by the {\em effective} gravity, i.e. the Newtonian gravity 
corrected for radiative acceleration due to the outward-directed force of radiation 
pressure. Realistic models must account for all the lines, especially those in spectral
regions of high flux. In fact, it is preferable to use sophisticated photospheric 
models such as the OSTAR2002 models for stars with moderate or weak winds rather
than rely on ``unified'' (photosphere+wind) models with a simplified
description of the deep, quasi-static layers. 

To verify the validity of photospheric models as applied to SMC
O-type stars, we compared  TLUSTY/SYNSPEC spectra to spectra
computed with CMFGEN, Hillier and Miller's (1998) code for ``unified'' models.
Like TLUSTY, CMFGEN treats NLTE line blanketing in detail 
and thus describes the photospheric structure with similar
sophistication. We performed extensive comparisons in collaboration with
John Hillier, and both codes were upgraded in the process. We found that
spectral lines formed in the inner region of the atmosphere show identical
profiles in the CMFGEN and TLUSTY/SYNSPEC spectra even for SMC
supergiants \citep{AV83, HL2001, NGC346}.
[However,  static models are definitely inappropriate for 
Wolf-Rayet stars and for extreme 
supergiants, especially at high metallicities.]

\subsection{Line-Blanketed vs. H-He Models}
\label{HHeModels}

\begin{figure}[tbp]
\epsscale{0.88} \plotone{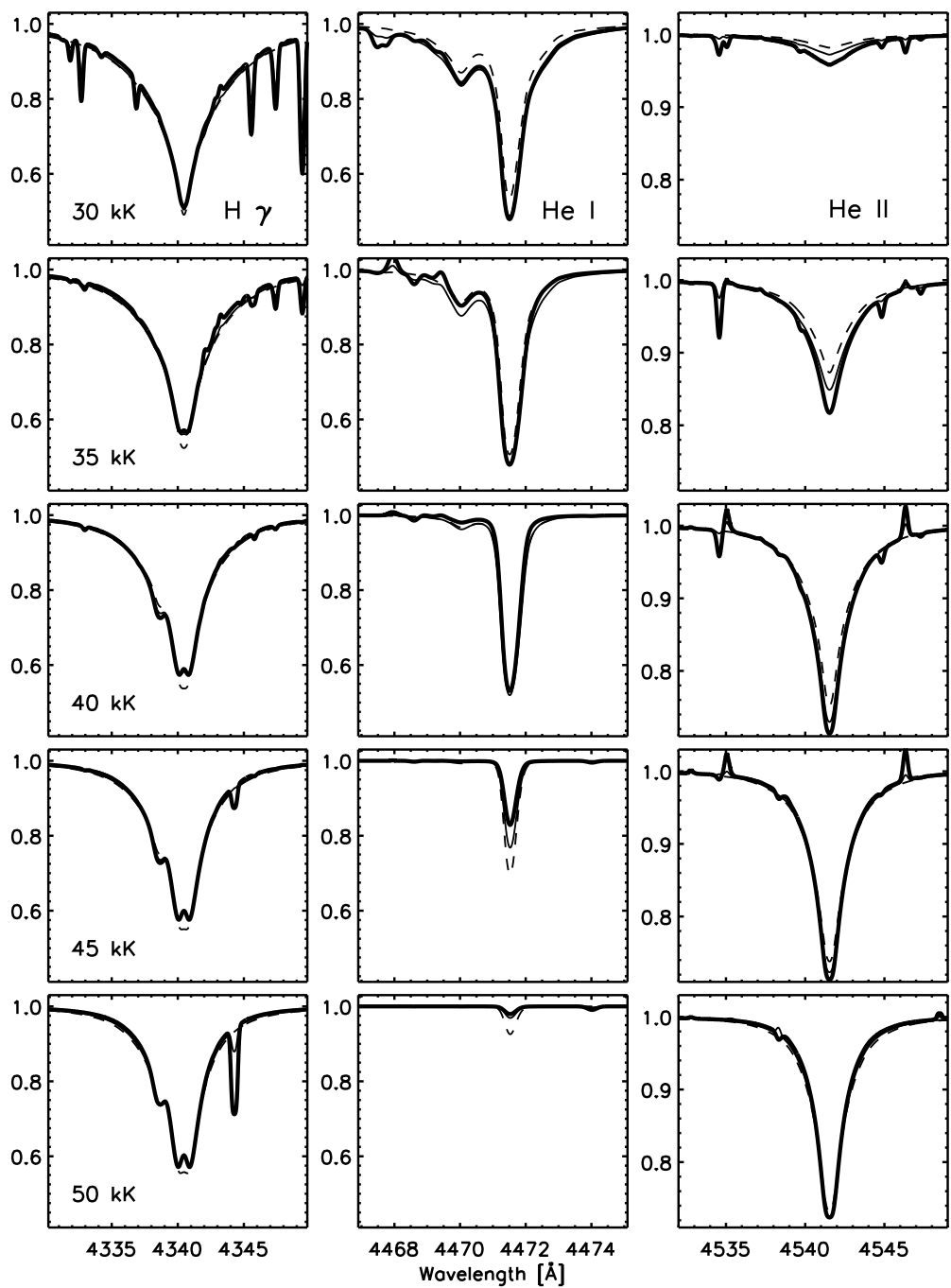} \figurenum{1}
\caption{Effect of metallicity on the main spectral-classification lines of
 O-type stars. The figure shows the  predicted line profiles of H$\gamma$ $\lambda 4340$ 
 (left column), HeI $\lambda 4471$ (middle), and HeII $\lambda 4542$ (right) for three
 metallicities: a solar metallicity (bold), SMC metallicity (solid line) and zero metalllicity, 
i.e. H and He line blanketing only (dashes). The models span the effective temperature 
range from 30,000\,K (top) to 50,000\,K (bottom) at a fixed surface gravity ($\log g = 4.0$).  
Although the gravity indicator, H$\gamma$ is insensitive to metallicity,
the temperature indicator, HeII $\lambda 4542$/HeI $\lambda 4471$ is quite clearly
sensitive to metallicity. }
\label{BLANKfig}
\end{figure}

\begin{figure}[tbp]
\epsscale{.9} \plotone{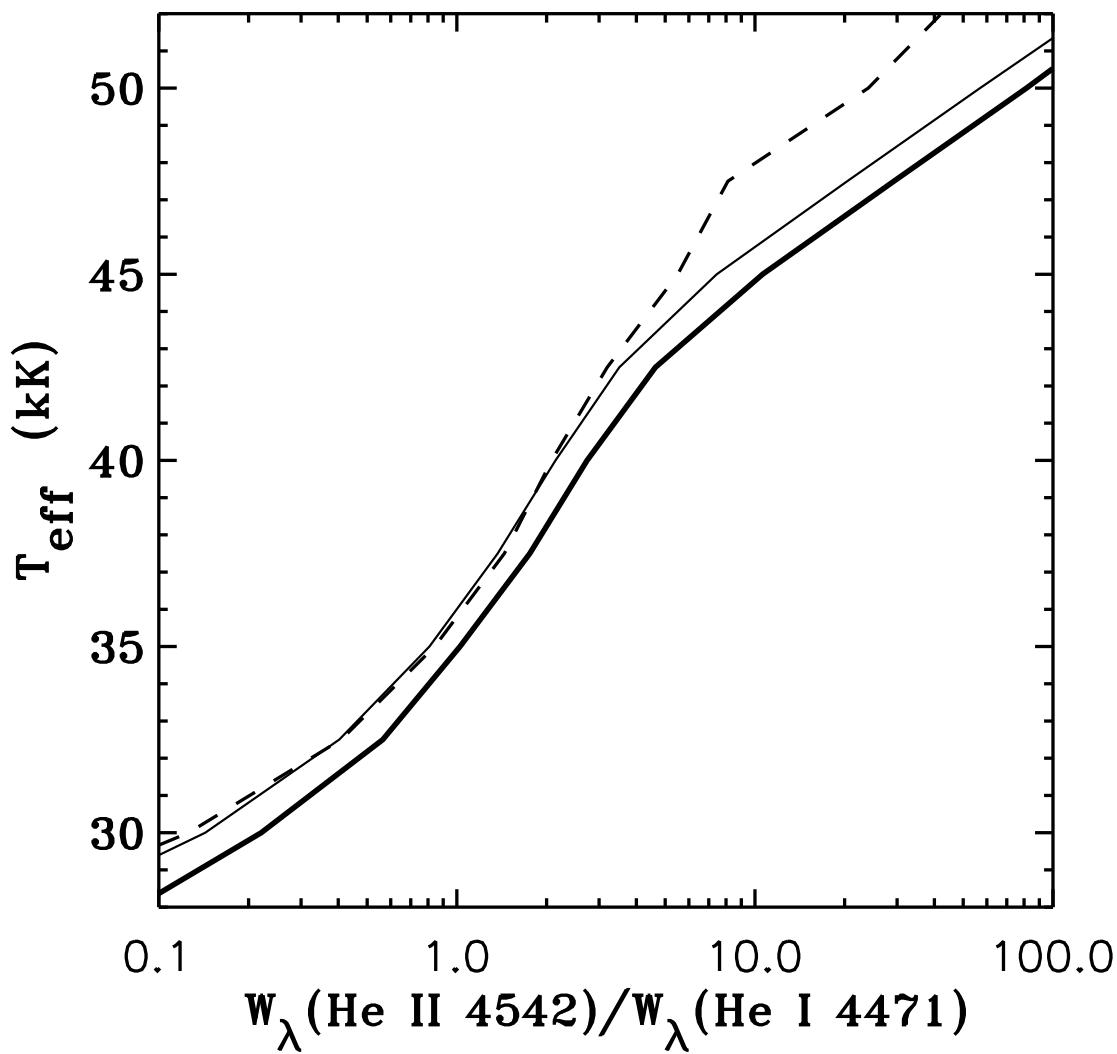} \figurenum{2}
\caption{Dependence of the temperature diagnostic -- the ratio of equivalent widths of  
\ion{He}{2}\,$\lambda$4542 to \ion{He}{1}\,$\lambda$4471 --  to metallicity in $\log g = 4.0$  
models. For this figure only, we omitted the metal-line opacity in synthesizing the spectrum 
in order to isolate the effect of metal-line blanketing on the structure of the atmosphere.
This ratio of equivalent widths yields a lower effective temperature for line-blanketed  models 
of solar composition (bold) than for H-He (dashes) or SMC-metallicity (solid line) models.}
\label{WHEfig}
\end{figure}

Previous studies show that absorption by numerous metal lines causes cooling
at the surface and heating of the inner regions of the atmosphere \citep{SA78}.
This back-warming effect raises the ionization level of
many diagnostic spectral lines. Figure~\ref{BLANKfig} shows the effects of metal 
line-blanketing on the hydrogen and helium lines used for spectral
classification and calibration. NLTE line-blanketing  produces only
slightly narrower profiles of H\,$\gamma$\  \citep{Lanz96} thereby 
yielding marginally higher (0.05 to 0.1~dex) surface gravities. 
Its main effect is on effective
temperature. \ion{He}{1}\,$\lambda$4471 line is strengthened at temperatures 
below 40,000\,K but weakened above 40,000\,K. \ion{He}{2}\,$\lambda$4542 is also
strengthened by line blanketing at all temperatures, although less so at
higher temperatures. The net result, shown in Fig.~\ref{WHEfig}, is that
line blanketing by metals produces a decrease of about 1000 -- 3000\,K in 
$T_{\mathrm{eff}}$\ derived from solar composition models compared to H-He
models. At lower metallicities, this systematic decrease in $T_{\mathrm{eff}}$ 
is smaller and is mostly found in hotter models. These results are in good
agreement with previous studies \citep{HHL98, martins02}.

\begin{figure}[tbp]
\epsscale{.9} \plotone{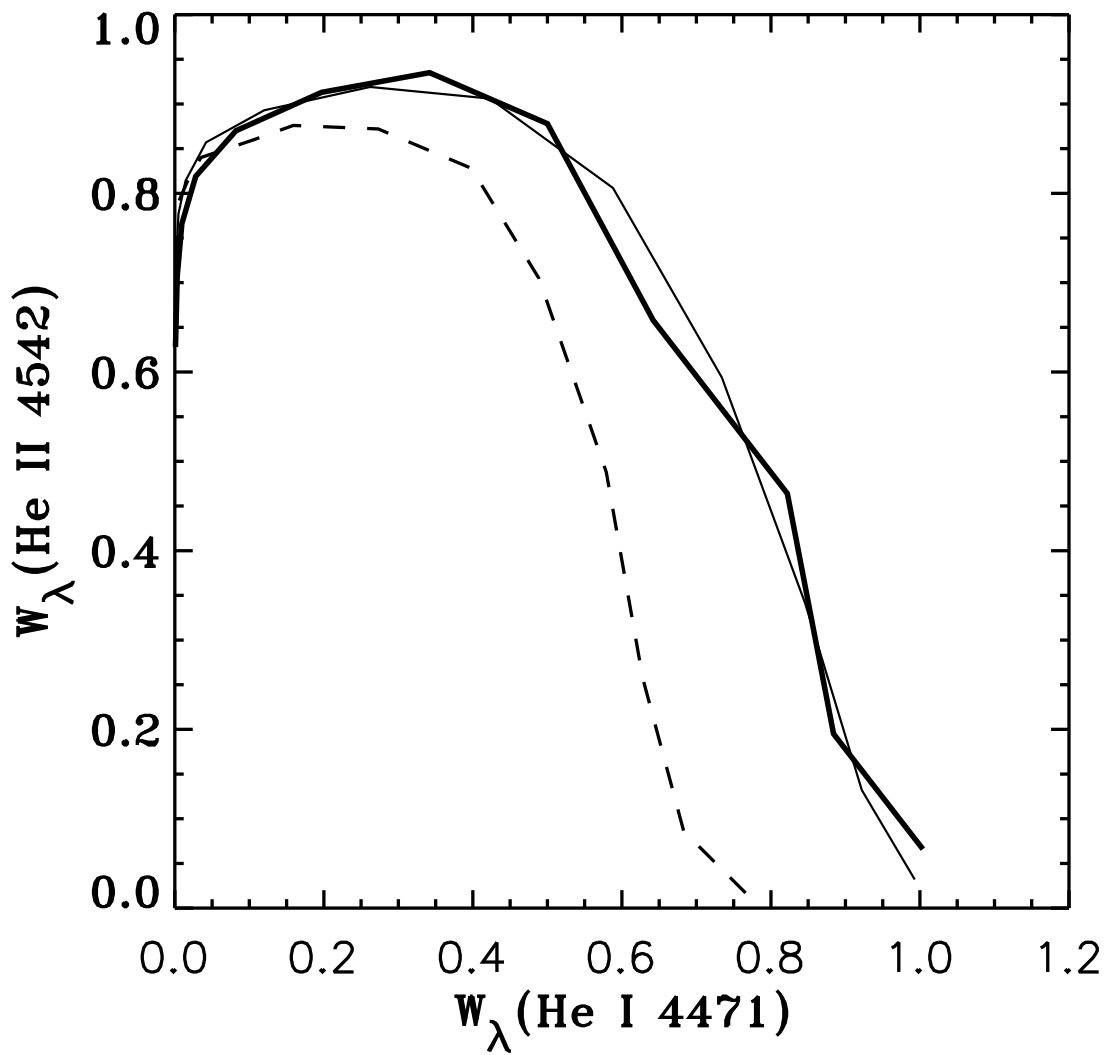} \figurenum{3}
\caption{Dependence of the derived helium abundance on metallicity. This figure shows
the computed equivalent widths of \ion{He}{1}\,$\lambda$4471 vs. \ion{He}{2}\,$\lambda$4542 
for line-blanketed models of solar-composition (bold), SMC metallicity (solid line)
and H-He (dashes) models.   The same surface gravity, $\log g = 4.0$ and helium abundance,
N(He)/N(H)=0.10, was assumed for all models. The H-He models, which by definition have no 
metal lines, yield smaller helium line strengths. Such models would lead to overestimates of the
helium abundance even for low-metallicity stars.}
\label{WHE2fig}
\end{figure}

Metal line-blanketing also affects the derived helium abundance. As shown in
Fig.~\ref{WHE2fig}, both \ion{He}{1} and \ion{He}{2} lines are stronger in
a line-blanketed model than in a pure H-He model having the same
parameters, $T_{\mathrm{eff}}$\ and $\log g$. To match the observed
strengths of \ion{He}{1}\,$\lambda$4471 and \ion{He}{2}\,$\lambda$4542 
with a H-He model would require a higher gravity or a higher helium abundance. 
Since the strength and
profile of the gravity indicator, H\,$\gamma$, is unaffected by line
blanketing, the higher-gravity solution is ruled out, and one would adopt an
enhanced helium abundance. There is therefore good reason to be skeptical
about the extent of enhancement in helium abundances in O-type stars if
derived from pure H-He model atmospheres \citep{herrero92}. And in fact, we
derive normal helium abundances for all the program stars, even for those
that are highly enriched in nitrogen.


\section{SPECTRAL DIAGNOSTICS}
\label{SpCalib}

\subsection{Atmospheric Parameters from the Hydrogen and Helium Lines}
\label{SpClass}

\begin{figure}[tbp]
\epsscale{.9} \plotone{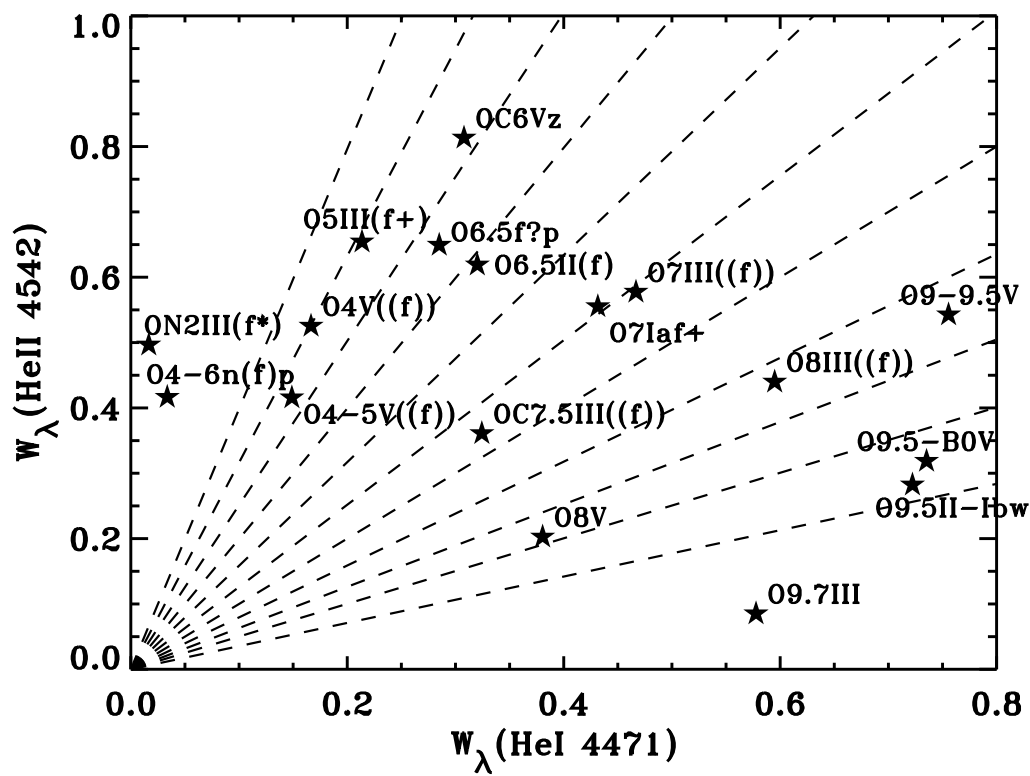} \figurenum{4}
\caption{Quantitative vs. qualitative spectral classification. The measured 
equivalent widths of \ion{He}{1}\,$\lambda$4471 and \ion{He}{2}\,$\lambda$4542 
in the program stars are shown by the star symbols. Each stellar point is labeled with 
spectral type as estimated visually by Walborn. 
The beam of dashed lines show Conti's boundaries
between spectral classes, from early to late O types (O3, O4, O5, O5.5, O6, O6.5, ... , O9.5) going
clockwise. The qualitative and quantitative classification systems usually agree to one 
sub-type, but see text for exceptions.}
\label{SPCLSfig}
\end{figure}

There are two types of
spectral classification systems in use for O-type stars: a quantitative scheme
devised by \citet{petrie47} and \citet{conti71} based on the equivalent 
widths of H\,$\gamma$, \ion{He}{1}\,$\lambda$4471, 
and \ion{He}{2}\,$\lambda$4542;
and a qualitative scheme developed by \citet{walborn71, walborn72}, based on
visual inspection of the H, He, and metal lines. Both schemes estimate
spectral type by the ratio of \ion{He}{2} and \ion{He}{1} line strengths, and
luminosity class by the strength of Stark-broadened wings of H\,$\gamma$
(Walborn also makes use of various He lines as well). In this study, we 
adopt Walborn et al.'s (2000) classifications for the program stars. 
In any case, the two schemes usually agree to better than a spectral type, as shown in
Fig.~\ref{SPCLSfig}. The two exceptions are: AV80 (O4-6n(f)p), a
rapidly rotating star whose line strengths are difficult to measure; and NGC~346
MPG~682 (O9-O9.5V), which has also been classified as O8V, in which case
there is no disagreement.

\begin{figure}[tbp]
\plotone{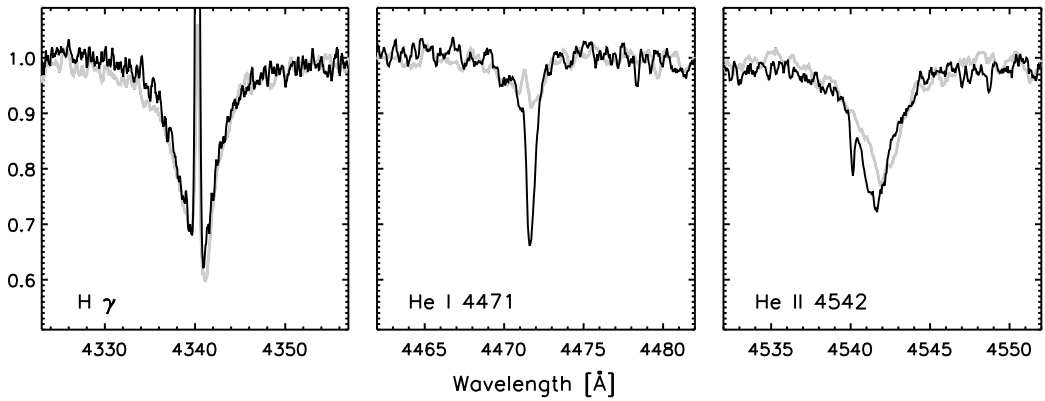} \figurenum{5}
\caption{Observational difficulties in spectral classification. The figure
shows the main classification lines in the optical spectrum of NGC~346
MPG~324 (O4V((f)); dashed line) and MPG~113 (OC6Vz; solid line). The two stars
have similar fundamental parameters, but nebular emission fills the 
\ion{He}{1}\,$\lambda$4471~line of MPG~324, resulting in an earlier
(spurious) spectral type. Metal lines must be used as independent checks on
the spectral classification and fundamental properties indicated by the H
and He lines. }
\label{NEBfig}
\end{figure}

Spectral classification becomes more difficult for stars embedded in
nebulosity because of contamination from nebular emission lines. This is the
case for some stars in NGC 346. Figure~\ref{NEBfig}, which compares the optical
spectra of two stars in NGC~346, illustrates the problem. \citet{walborn00}
classified the spectrum of NGC~346 MPG~324 as O4V because of its weak 
\ion{He}{1} absorption lines. However, the 
\ion{He}{1} lines may be weak because they are filled in by
nebular emission. In cases such as this, we used the UV
spectrum to check for consistency with the optical spectral type. 

\begin{figure}[tbp]
\plotone{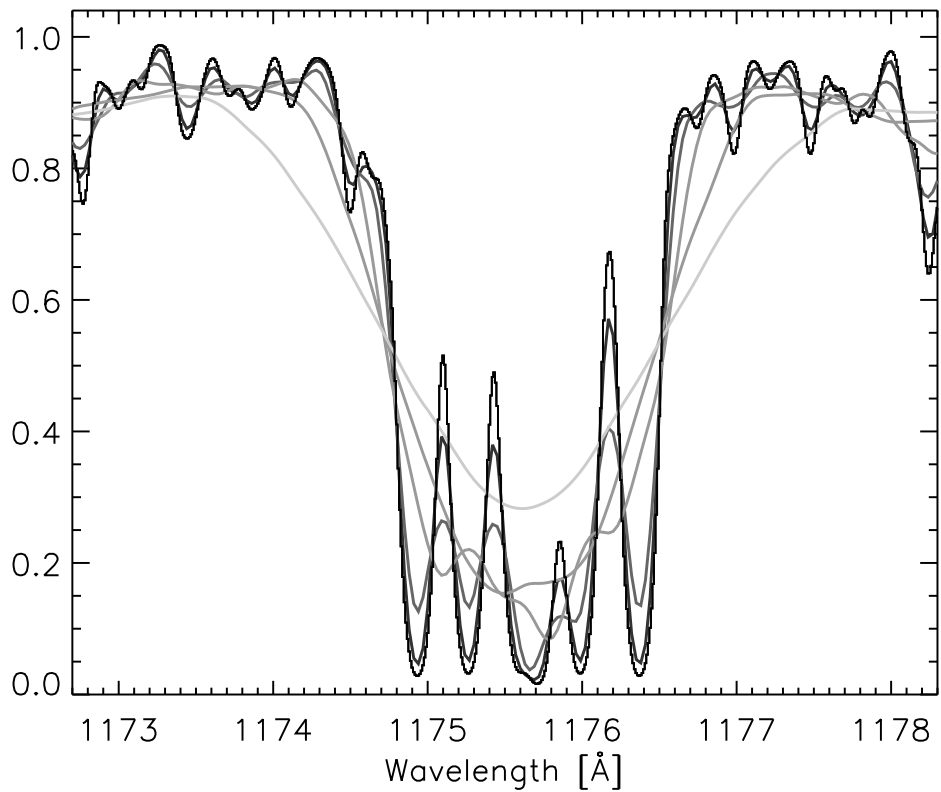} \figurenum{6}
\caption{Estimation of the projected rotational velocity from the normalized profile of the
\ion{C}{3}\,$\lambda$1176 multiplet in the case of a model with $T_{\mathrm{eff}}$~= 35,000\,K, 
$\log g$~= 4.0, $Z =0.2\,Z_\sun$. The figure compares the profile of a non-rotating model 
(histogram style) with those of rotating models ($V \sin i$~= 20, 40,
80, 160, 320\,km\,s$^{-1}$). The profile of the multiplet becomes progressively smoother at 
higher rotational velocities, but individual components remain visible at $V \sin i
\leq$ 40\,km\,s$^{-1}$. }
\label{VROTfig}
\end{figure}

If nebular emission is significant, we can only analyze the wings of 
the H and He line profiles. Without
reliable equivalent widths, there is some uncertainty regarding rotational
vs. pressure (Stark) broadening of the H and He lines. This ambiguity is not
serious in most cases since the rotational velocity, $V\sin i$, can usually
be estimated from the profile of the \ion{C}{3}\,$\lambda$1176
multiplet. As shown in Fig.~\ref{VROTfig}, individual components of the
multiplet start to merge as $V\sin i$ increases. The profile of 
\ion{C}{3}\,$\lambda$1176 is a useful diagnostic at all rotational 
velocities seen in
our sample, from $V\sin i=25$\,km\,s$^{-1}$\  or
less (NGC~346 MPG~487) to $V\sin i=300$\,km\,s$^{-1}$\  or
more (AV 80).

\begin{figure}[tbp]
\epsscale{.9} \plotone{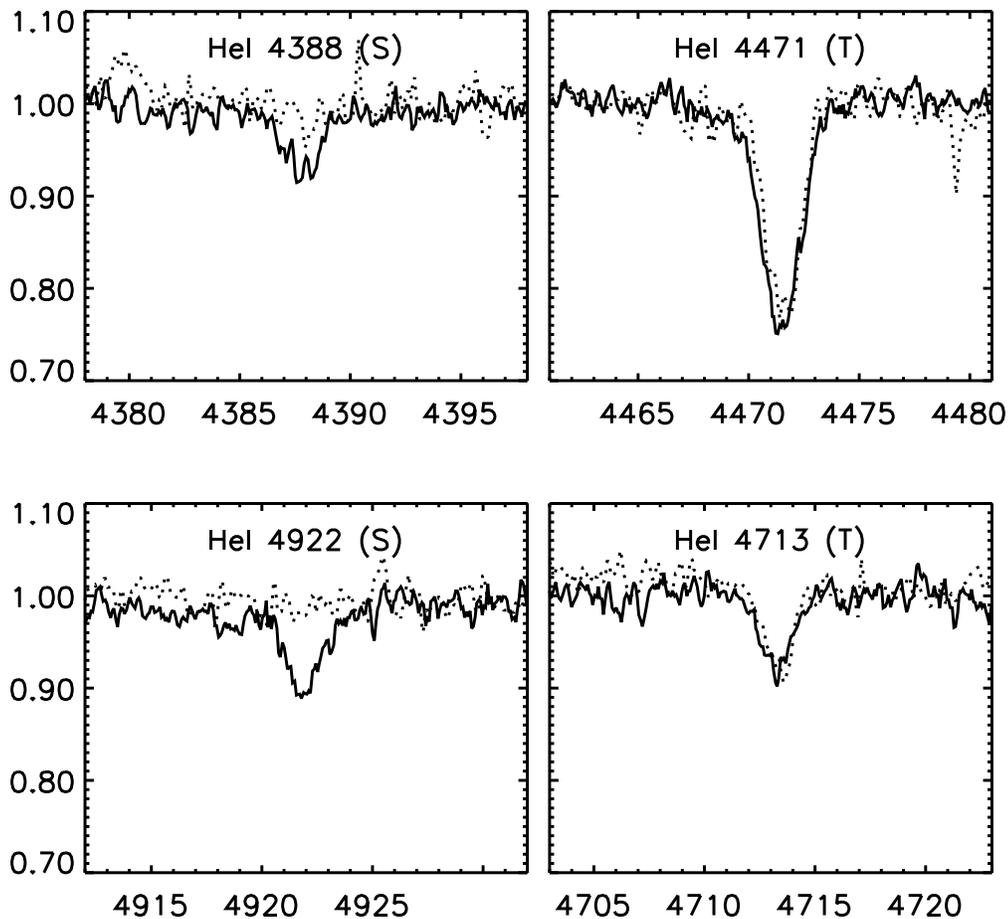} \figurenum{7}
\caption{Dependence of the \ion{He}{1} singlet (S) and triplet (T) line-strengths to gravity.
The figure compares the optical spectrum of the O7 Iaf supergiant, AV~83
(dashed), to that of the OC7.5 III giant AV~69. Although the two stars have the same
temperature, T$_{\rm eff}$=35,000~K,  AV~69  has a higher gravity.
The \ion{He}{1} triplets are totally insensitive to gravity, while the \ion{He}{1} singlets are stronger
in AV~69. The different behavior of the \ion{He}{1} singlets and triplets is due to non-LTE
effects on the \ion{He}{1} level populations in the two stars as explained in the text.   }
\label{AVHEfig}
\end{figure}

Other reasons to expand the list of lines for spectral analysis from
just the three strong H and He lines are to check for consistency and 
to make finer distinctions. Fig.~\ref {AVHEfig} compares the helium
spectrum of AV83 (O7Iaf+) and AV69
(OC7.5III). Despite their different spectral classifications, the line strengths 
of \ion{He}{1}\thinspace $\lambda $4471 (a triplet) and \ion{He}{2}\thinspace $\lambda $4542 
are identical, and our models indicate that the two stars have similar effective 
temperatures, $T_{\mathrm{eff}}$~$\approx ~$35,000\thinspace K. 
However, the \ion{He}{1} singlet lines are different in the two spectra: 
they are strong in AV69 but almost absent in AV83. Our models, which match both the 
singlet and triplet lines of \ion{He}{1}, indicate that this difference is due to the 
higher surface gravity of AV69 ($\log g=3.4$) than in AV83 ($\log g=3.25$).  
\ion{He}{1} singlets respond differently from \ion{He}{1} triplets to 
non-LTE effects on the level populations because of the different structure of 
their energy levels
(singlet lines are connected to the ground level of helium, whereas the
lowest triplet level is metastable). Because of its lower gravity (lower
atmospheric density), non-LTE effects are stronger in the spectrum of AV83.
The weaker Stark-broadened wings of the Balmer lines and the strong 
emission of \ion{He}{2}\thinspace $\lambda $4686 provide consistency checks
for the lower gravity of AV83.

\subsection{Temperature Estimation from Metal Lines}
\label{TeffEstim}

As discussed above, we made full use of hydrogen and helium lines in the
optical spectrum for calibration. However, because of problems with
nebular masking of  \ion{He}{1} lines (Fig.~\ref{NEBfig}), we also use lines
of heavier elements for estimating temperature. Below, we describe how
optical and UV metal lines can be used as supporting evidence on $T_{\mathrm{eff}}$.

\begin{figure}[tbp]
\epsscale{.99} \plotone{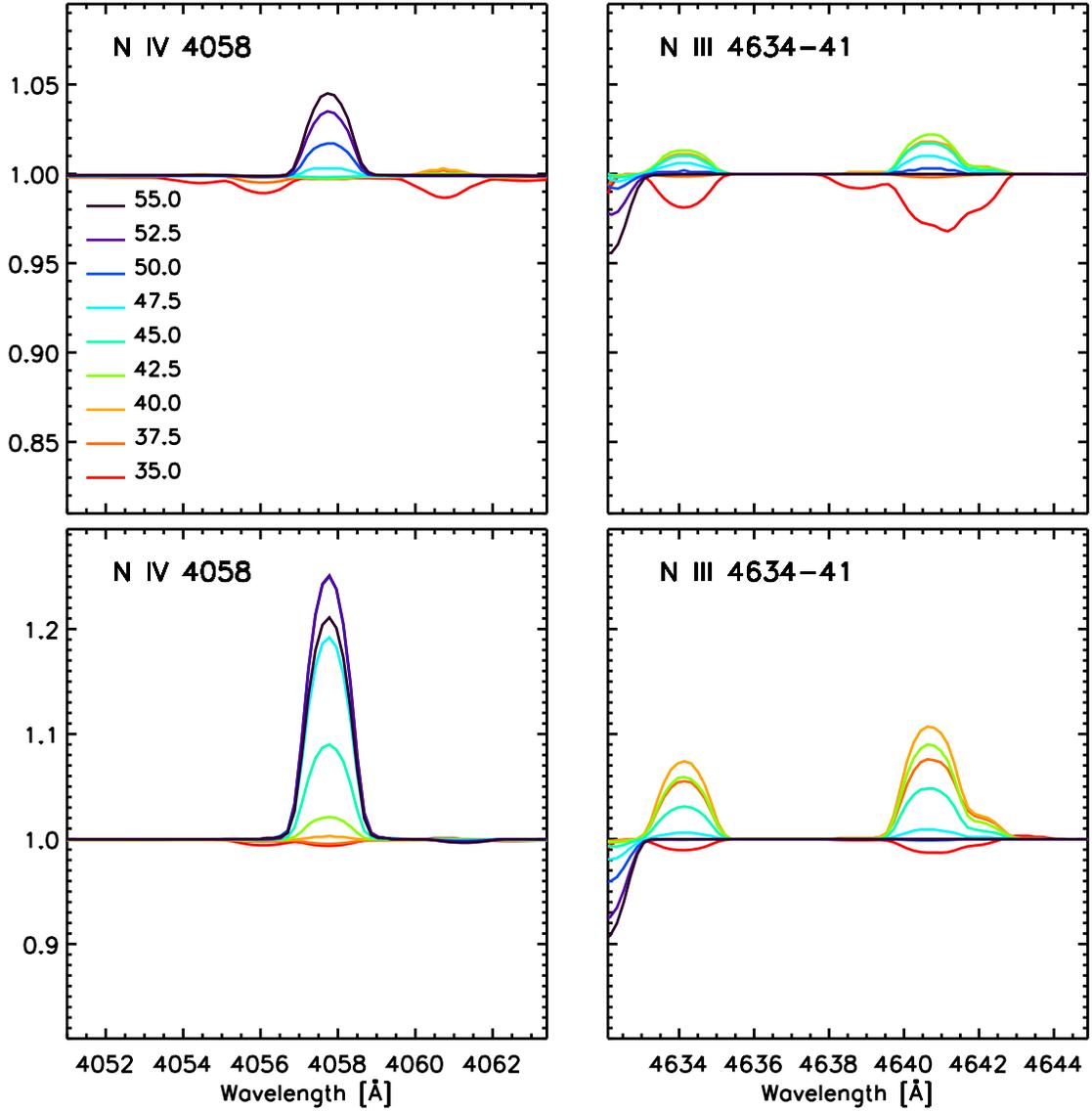} \figurenum{8}
\caption{Sensitivity of \ion{N}{3}\,$\lambda\lambda$4634-41 (right panels) and \ion{N}{4}\,$%
\lambda$4058 (left panels) to effective temperature for a fixed $\log g=4.0$, and $V_{\mathrm{t}%
}=10$\,km\,s$^{-1}$. The top panels assume $N = 0.2~N_\sun$, the
bottom panels, $N = N_\sun$. \ion{N}{4} emission
is stronger than the \ion{N}{3} emission at $T_{\mathrm{eff}}$$\geq$50000\,K, 
which is one criterion defining the earliest O spectral type, O2. }
\label{N34Tfig}
\end{figure}

Most metal lines in the optical spectral region can appear in either
absorption or emission. These ``Of emission lines'' include \ion{N}{4}%
\thinspace $\lambda $4058, \ion{Si}{4}\thinspace $\lambda $4088, 4116, %
\ion{N}{3}\thinspace $\lambda $4634-41, \ion{C}{3}\thinspace $\lambda $%
4647-51, and \ion{S}{4}\thinspace $\lambda $4486, 4503. There is
observational and theoretical evidence that the emission is due to non-LTE
effects (overpopulation of the upper level of a transition) rather than to
emission in an extended atmosphere. As shown in Fig.~\ref{N34Tfig},
our plane-parallel models with $\log
g=3.5$ show noticeable \ion{N}{4}\thinspace $\lambda $4058 emission at
$T_{\mathrm{eff}}>37,500$\,K if N=N$_{\sun}$; \ion{N}{3}\thinspace $%
\lambda $4634-41 emission occurs at $T_{\mathrm{eff}}\geq 35,000$%
\thinspace K if N$\geq 0.2$ N$_{\sun}$; but \ion{Si}{4}\thinspace $\lambda $%
4088, 4116 emission occurs only at very high temperatures ($T_{\mathrm{eff}}
\geq 50,000$\thinspace K) and then, only weakly. However, before we can use
these lines to make reliable estimates of $T_{\mathrm{eff}}$\ and
abundances, we must first improve the treatment of dielectronic
recombination as well as the model atoms, because these lines involve
highly-excited energy levels.

We therefore directed our attention to the UV metal lines, which provide
useful supporting material for temperature estimation, especially for those
stars whose \ion{H}{1} and \ion{He}{1} lines are masked by nebular emission.
Fig.~\ref{C34fig} illustrates the case of the carbon ionization balance. 
While both \ion{C}{3}\thinspace $\lambda $1175 and \ion{C}{4}\thinspace 
$\lambda $1169 grow stronger with increased gravity, higher carbon abundance
and microturbulent velocity, the ratio of line strengths is most sensitive
to temperature, so the effective temperature can be estimated with
reasonable accuracy, although there is a small offset toward higher
temperatures ($<2000$\thinspace K) if a nebular carbon abundance is assumed. 
Despite its successes,
UV metal lines cannot be the primary and only means of estimating effective
temperature because the ionization level of an element depends on gravity as
well as temperature ($\S 4.4$). Contrary to the optical spectrum, gravity cannot be
derived in the UV from the \ion{H}{1} Lyman lines because they are masked
by interstellar absorption.

\subsection{Sensitivity of Photospheric Spectrum to Fundamental Parameters}
\label{Sensitivity}

The main objection to photospheric analysis is that photospheric lines
strong enough to be measured may already be saturated. To investigate this
possibility, we computed a set of model UV spectra over the temperature
range, $T_{\mathrm{eff}}$ = 30,000 -- 50,000\thinspace K, for $\log g=4.0$
and $V_{\mathrm{t\ }}=10$\thinspace km\thinspace s$^{-1}$. These spectra are
shown in Fig.~10 (only Fig.~10a is shown in the paper edition).
Table~\ref{LineTab} gives a summary description of the key UV lines 
that are sensitive to temperature but not significantly contaminated
by interstellar contributions. The range
of $T_{\mathrm{eff}}$\ over which a spectral is sensitive to
temperature is listed for models with $\log g=4.0$. At lower gravity, the
temperatures are about 2000\thinspace K higher. As expected, resonance lines
of abundant elements like \ion{C}{4}\thinspace $\lambda $1549 are
insensitive to temperature because they are saturated. Some subordinate
lines, such as \ion{C}{4}\thinspace $\lambda $1169 and \ion{S}{5}\thinspace $%
\lambda $1502, are not useful temperature indicators by themselves because
their strength does not change monotonically with $T_{\mathrm{eff}}$,
although they can be used for ionization balance (see Fig.~\ref{C34fig}).
Most of the lines that are sensitive temperature indicators arise from ions
that are in trace ionization states.
Below 35,000\,K, most lines in Table~\ref{LineTab} are insensitive to  temperature.
There are other, lower-ionization lines that
could, but they are usually contaminated by interstellar contributions. 
Excited lines of these ions such as \ion{Si}{3}\,$\lambda\lambda$1294-1303,
are a good temperature indicators for late O and early B stars (see Fig.~10j). Fortunately,
the optical \ion{He}{1} and \ion{He}{2} lines work very well to determine 
$T_{\mathrm{eff}}$\ of late O-type stars. For very early O stars, we have the
opposite problem: helium ionization balance does not work because \ion{He}{1}
lines are no longer present. Likewise, \ion{C}{3} lines at 1176 and
1247\,\AA\ have also disappeared. New criteria for spectral classification
and analysis must be used. \citet{walborn02} define O2-type stars as those
in which \ion{N}{4}\,$\lambda$4058 emission is stronger than \ion{N}{3}\,$%
\lambda$4634-41 emission lines. As shown in Fig.~\ref{N34Tfig}, such
stars must have $T_{\mathrm{eff}}$ $\geq 50,000$\,K. We find that \ion{O}{5}%
\,$\lambda$1371 and the \ion{Fe}{6}\,$\lambda$1260-1280 complex present
adequate criteria for very hot stars in the UV as shown in Fig.~10i and 10r.

\subsection{Ionization Balance and Surface Gravity}
\label{FeGrav}

It is well known from the Saha equation that the level of ionization depends
not only on temperature but also on the atmospheric density, or
surface gravity. Departures from LTE amplify this
dependence, because NLTE ``overionizes'' a species 
relative to LTE, and this effect is larger at lower densities.
As discussed in connection with Fig.~\ref{AVHEfig}, 
the differences in \ion{He}{1} line strengths are mainly the result of 
different surface gravities. Fortunately, the \ion{C}{4}/\ion{C}{3} ratio
is nearly independent of gravity (Fig.~\ref{C34fig}).

The iron ionization level is also a good indicator of temperature of O-type
stars: \ion{Fe}{4} lines are strong in late O stars, \ion{Fe}{5} dominates
in mid-O-type spectra, and \ion{Fe}{6} lines are present in early-O 
spectra (see Fig.~9 of Lanz \& Hubeny 2003). 
To investigate the effect of gravity on the iron ionization
balance, we take advantage of the numerous \ion{Fe}{4} and \ion{Fe}{5}
lines in the ultraviolet to construct an ionization index,

\[
{\bf WFE54} = \frac{3. \times W_\lambda(1360-1380)}{2. \times W_\lambda(1600-1630)} .
\]

\noindent By using equivalent widths, we account
automatically for differences in the continuum of different models. We
display this ionization index for the `S' models in Fig.~\ref{WFE45fig}. The
index increases with $T_{\mathrm{eff}}$\ as ionization increases, but the
index is also sensitive to gravity, except around $T_{\mathrm{eff}}\approx 32,000$\,K
(at high temperatures, \ion{Fe}{4} lines weakened and the index also measures the
contribution of other lines in the selected range). If $T_{\mathrm{eff}}$\ can be determined
from the carbon or helium lines, then we can determine the surface gravity of O stars 
from UV iron lines if  the metallicity and microturbulent
velocity are known. In practice, we used the iron ionization index as  
supporting evidence for the parameters derived from the He and C lines ($T_{%
\mathrm{eff}}$) and from the Balmer line wings ($\log g$), searching for a
consistent match of the complete spectrum.

The idea of using the iron ionization balance to determine surface gravities
can be extended to early B/late O stars using the \ion{Fe}{4} to \ion{Fe}{3}
balance and to early-O stars by using the \ion{Fe}{6} to \ion{Fe}{5}
balance. The range of gravities in early-O stars is however limited;
moreover, the integrated contribution of \ion{Fe}{6} lines in the
$\lambda\lambda$1250-1300 range remains small (see Fig.~10r) making the definition
of a sensitive index difficult.

The case of iron ionization illustrates well the potential of using lines
from different species to determine the stellar parameters. For instance,
the \ion{Si}{4} resonance lines are also known to be sensitive to luminosity
\citep{walborn84, IUE89}. Matching the full UV and optical
spectrum provides additional checks and reliability that analyses based on H
and He lines only lack. A crucial ingredient, however, is an extensive
treatment of NLTE line blanketing -- as incorporated in TLUSTY and CMFGEN
models -- to ensure that ionization balance of heavier elements is
predicted accurately.

\subsection{Microturbulence and Abundances}
\label{Microturb}

\begin{figure}[tbp]
\epsscale{.6} \plotone{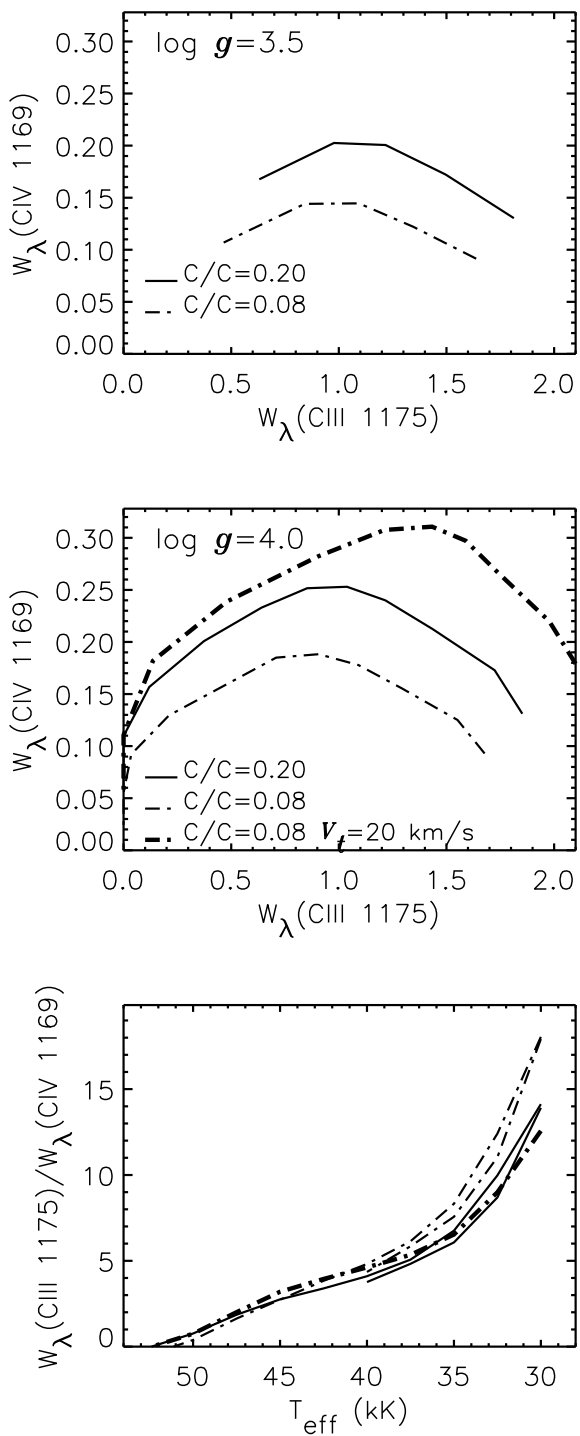} \figurenum{9}
\caption{Dependence of the UV carbon lines on effective temperature, gravity,
metallicity, and turbulent velocity. The top two panels, shows the absolute 
strengths of \ion{C}{3}\,$\lambda$1175 and \ion{C}{4}\,$\lambda$1169 over the range of 
of temperatures, T$_{\rm eff}$=30,000 -- 50,000~K (temperature increases to the left). 
Although the line strengths are sensitive 
to all four parameters, the ratio of line strengths is sensitive mainly to
effective temperature (bottom panel). }
\label{C34fig}
\end{figure}

``Microturbulence'' was first invoked by \citet{struve32} to explain the
large Doppler widths of lines seen in supergiant spectra. Although a useful
device for matching observed line strengths, microturbulence plays havoc
with abundance determinations. As \citet{rpk92} warned, photospheric models
ignoring the presence of an atmospheric velocity gradient may require
significantly higher microturbulent velocities to achieve the same
equivalent widths. Velocity fields are essential for determining abundances
of weak `pseudophotospheric' metal lines in relatively dense winds in which
these lines are formed around the sonic point. In less dense stellar winds,
as those observed in SMC O stars, metal lines are formed deeper in the
quasi-static photosphere. Using CMFGEN wind models, and thus accounting for
the velocity field, we had nevertheless to adopt microturbulent velocities
between 5 and 20\,km\,s$^{-1}$\ to match the observed strength of UV metal
lines \citep{AV83, NGC346}. The effect of microturbulence can be seen in the
left panel of Fig.~\ref{C34fig}, where an increase in $V_{\mathrm{t\ }}$
from 10 to 20\,km\,s$^{-1}$\ is more than enough to compensate for a factor
of 2.3 lower carbon abundance. No model to date, either photospheric or
unified model, has evaluated the depth-dependent velocity fields in the
subsonic regions of an atmosphere. The main point for this study is that the
temperature derived from the \ion{C}{4}/\ion{C}{3} ratio (Fig.~\ref{C34fig},
right panel) 
is changed in only a minor way by the assumed microturbulent velocity.

The abundance of oxygen in the \ion{H}{2} regions in a galaxy is often taken
as an indicator of the overall metallicity of the stellar population. Recent
studies of \ion{H}{2} regions in the SMC concur on an oxygen abundance%
\footnote{%
Throughout this paper, abundances are given in logarithm of the number
density, in a scale where the hydrogen abundance is [H]=12.0}, [O]= 8.1
(e.g., Garnett et~al. 1995, Kurt et~al. 1999), that is, one fifth of the
solar value. Stellar studies in the SMC of F-K supergiants \citep{spite89},
of A supergiants \citep{venn99}, and of O stars \citep{haser98, AV83, NGC346}
yield metallicities consistent with the nebular value. These same studies,
however, show that the N/C abundance ratio is not solar. Table~\ref{AbunTab}
provides a compilation of SMC abundances derived from analyses of \ion{H}{2}
regions and B and A-type stars. In our spectral analysis (\S\ref{Analysis}),
we started out using the `S' grid of model spectra supplied by \citet{OS02}.
In the S grid, metal abundances are uniformly set to one fifth of their
solar values. In agreement with previous abundance studies, we soon found
that we had to modify the abundance mix and/or the microturbulent velocity
to achieve a match with observations. The nebular analyses suggest that the
S grid of model spectra should be adequate for elements other than nitrogen
and carbon. However, the S models overestimate both the nitrogen and carbon
abundances if the stellar surface abundances still reflect the original,
nebular abundances. We therefore re-computed the spectra based on the S grid
of model atmospheres, but this time with SMC nebular abundances for CNO and
with the SMC A-star abundance for Si and Fe. We refer to these model spectra
as the `NEB' grid of spectra. Evolutionary models of rotating SMC stars show
that there can be considerable enhancement of the surface abundance of
nitrogen in the main-sequence phase \citep{MM01}, supported by our analysis
of NGC~346 stars \citep{NGC346}. We therefore calculated another series of
spectra, this time with solar nitrogen abundance, which we call `N5X'
spectra. Thus, the nitrogen abundance in the various spectral grids range
over a factor over 30 (N/N$_\sun = 0.033 - 1.0$). \citet{MM01} predict that
helium enrichment occur at much later phases. We therefore employed a solar
helium abundance for all the spectral grids. In our analyses, we did not
find any evidence to the contrary, although \citet{AV83} report that helium
may be enhanced in the atmosphere of AV83. Table~\ref{AbunTab} summarizes
also the abundance mix of the various models.

The results of our sensitivity study of previous sections indicate that the
S models can be used for a first estimate temperature from the \ion{He}{2}/%
\ion{He}{1} and the \ion{C}{4}/\ion{C}{3} line strength ratios. However, the
carbon-based temperature so derived can be up to $\approx 2000$\,K too low
if the microturbulent velocity is higher than assumed in the S model ($V_{%
\mathrm{t\ }} = 10$\,km\,s$^{-1}$) and/or if the carbon abundance is lower
than assumed in the S model (C/C$_\sun$ = 0.2). A knowledge of $V_{\mathrm{%
t\ }}$ can be found from the strengths of the \ion{O}{4}\,$\lambda$1340
triplet and \ion{S}{5}\,$\lambda$1502 line, which are assumed to have
abundances close to one-fifth solar.


\section{SPECTRAL ANALYSIS}
\label{Analysis}

For consistency, we used as a starting point the same
spectral-classification lines of hydrogen and helium to determine effective
temperature and surface gravity. In addition, we took advantage of
high-resolution UV spectra from STIS and FUSE, which enabled us to extend
the principle of ionization balance to other elements, such as carbon, iron,
and sulphur. In all cases, we assumed that program stars are single and that
rotation works only to broaden the spectral lines. Our model grid samples
the parameter space as in \citet{OS02}, with 2500\,K steps in temperature
and 0.25~dex in gravity. In several cases, it was obvious that the best
match to the observations could be achieved with a model between grid
points. As discussed by \citet{OS02}, we interpolated the atmospheric
structure and the NLTE level populations for the adopted temperature and
gravity, before recomputing the spectrum.

We estimated errors on the derived parameters by comparing observations and
models with parameters close to those adopted. Models that can be clearly
excluded thus provide an upper limit to uncertainties. Conservatively, we
estimated errors to be $\pm$2000\,K in $T_{\mathrm{eff}}$, $\pm$0.2\,dex in $%
\log g$, $\pm$5\,km\,s$^{-1}$\ in $V_{\mathrm{t}}$, and less than a factor
of 2 in abundances. The accuracy might be better in some cases. The results
of our analysis are summarized in Table~\ref{ResuTab}. We discuss our
results for each individual object in the following sections.

\subsection{O-type Dwarfs}

All six O-type dwarfs in the sample are members of the young cluster,
NGC~346. Together, they cover the full range of O spectral types, from O4V
to 09.5-B0V, so they are useful in establishing the age of the cluster. We
refer to them by their MPG number \citep{massey89}. We also include here
NGC~346 MPG~355 (ON2III; Walborn et al. 2002), as part of the dwarf sequence
because of its proximity to the zero-age main sequence. MPG~355 and MPG~12
(O9.5-B0V (N strong)) have anomalously strong nitrogen features in their
spectra, while MPG~113 (OC6 Vz) has anomalously strong carbon lines in its
spectrum according to \citet{walborn02}. This disparity in nitrogen or
carbon predominance implies that evolutionary abundance changes have already
taken place in some members of NGC~346. Our approach as described below was
to analyze the spectrum of each star independently and only then to attempt
to discern evolutionary abundance patterns. Unless noted otherwise, we make
use of the `S' model spectra ($Z/Z_\sun = 0.2$, $V_{\mathrm{t\ }}$=10\,km\,s$%
^{-1}$) for analysis.

\citet{NGC346} already reported on the NGC~346 stars, based on an analysis
conducted fully independently to the present study. They used both TLUSTY
and CMFGEN models to match photospheric and wind lines. They obtained an
excellent agreement between the two codes demonstrating that most lines are
indeed formed in a quasi-static photosphere. The results of the two studies
are consistent.

\textit{NGC 346 MPG 355} (ON2III(f$^{*}$)).--- Earlier studies of this star %
\citep{rpk89, haser98} estimated the stellar parameters, $T_{\mathrm{eff}}$
= 55,000$\pm$5000\,K, $\log g$=3.9$\pm$0.1. We also find a very high
temperature, $T_{\mathrm{eff}}$ = 52,500$\pm$2000\,K, and $\log g=4.0$. A
high temperature is inferred from the strength of \ion{N}{5}$\lambda\lambda$%
4603, 4619 relative to \ion{N}{4}$\lambda$1718, the strength of \ion{Fe}{6}
relative to \ion{Fe}{5} lines in the UV, and the emission strength of %
\ion{N}{4}$\lambda\lambda$4058, 7109, 7123 and of \ion{C}{4}$\lambda\lambda$%
5801, 5812. An unusually high microturbulent velocity, $V_{\mathrm{t\ }}$ =
25\,km\,s$^{-1}$, is required to match the strength of metal lines in the
UV. The atmosphere appears to be greatly enhanced in nitrogen to solar or
super-solar values. The \ion{C}{4}$\lambda\lambda$5801, 5812 emission lines
are well fit by a model with the nebular carbon abundance. We were not able
to analyze \ion{O}{5}$\lambda$1371, \ion{N}{4}$\lambda$1718, or \ion{He}{2}$%
\lambda$1640, because these lines are asymmetric and presumably arise in the
wind. (See Bouret et~al. 2003 for a discussion of the wind properties).

\textit{NGC 346 MPG 324} (O4V((f))).--- Visible structure in the \ion{C}{3}$%
\lambda$1175 feature and the sharpness of UV Fe and Ni lines imply a
rotational velocity, $V \sin i = 60 - 80$\,km\,s$^{-1}$. At this $V \sin i$, %
\ion{He}{1}$\lambda$4471 is relatively narrow, so most of the line profile
is contaminated by nebular emission, and the line cannot be used for helium
ionization balance. In the UV, the ionization balances of \ion{C}{4}/\ion{C}{3}
and \ion{Fe}{5}/\ion{Fe}{4} point toward a temperature slightly hotter than 
$T_{\mathrm{eff}}$ = 40,000\,K with $\log g$=4.0. The optical \ion{H}{1}, %
\ion{He}{1} and \ion{He}{2} lines are consistent with this result. \ion{N}{4}%
$\lambda$1718 and the \ion{Si}{4}$\lambda$1400 doublet are well fit by the
model. The strengths of \ion{O}{4}$\lambda$1340 triplet and \ion{S}{5}$%
\lambda$1502 imply a high microturbulent velocity, $V_{\mathrm{t\ }}$ =
15\,km\,s$^{-1}$. Nitrogen is enhanced but not to the extent of the
enrichment found in MPG~355. The absence of \ion{C}{4}$\lambda\lambda$5801,
5812 indicates a low carbon abundance.

\textit{NGC346 MPG 368} (O4-5V((f))).--- The optical spectrum of this star
is quite noisy and heavily contaminated by nebular emission. In addition,
the severe disagreement in the radial velocity of the star measured from the
optical (130\,km\,s$^{-1}$) and from the UV (200\,km\,s$^{-1}$) spectrum
suggests either that MPG~368 is a previously undetected spectroscopic binary
or that different stars were observed by HST and the AAT. Nevertheless, a
model with $T_{\mathrm{eff}}$ = 40,000\,K, $\log g=3.75$, $V_{\mathrm{t\ }}$
= 15\,km\,s$^{-1}$, and enhanced nitrogen and low carbon abundances,
produces an acceptable fit to both optical and UV CNO lines (e.g., \ion{N}{3}%
$\lambda\lambda$1182, 1184, 4097, 4631, 4634; \ion{N}{4}$\lambda$1718). This
same model produces a poor fit to the blue spectrum: the model hydrogen and
helium lines are all too strong.

\textit{NGC346 MPG 113} (OC6Vz).--- In the spectrum of this star, the
components of the \ion{C}{3}$\lambda$1175 multiplet are nicely resolved,
implying a low rotational velocity, $V \sin i\la 35$\,km\,s$^{-1}$. The
best-fit model in the grid has the parameters, $T_{\mathrm{eff}}$ =
40,000\,K and $\log g = 4.0$. The nitrogen and carbon lines are best fit
with a SMC nebular abundance.

\textit{NGC346 MPG 487} (O8V).--- Like MPG 113, this star has a very
sharp-line spectrum ($V \sin i\la 20$\,km\,s$^{-1}$). The H and He lines
indicate $T_{\mathrm{eff}}$ = 35,000\,K and $\log g = 4.0$. However, most of
the model UV metal lines are too strong even if lower metal abundances are
assumed. Lowering the microturbulence to $V_{\mathrm{t\ }}=2-5$\,km\,s$^{-1}$%
\ greatly ameliorates the problem. The carbon and nitrogen lines favor a low
(nebular) abundance.

\textit{NGC346 MPG 682} (O9-O9.5V).--- Because of a failed target
acquisition by STIS, we cannot be sure that the target observed was really
MPG~682. We therefore restricted our analysis to the optical spectrum only.
The best fit model is $T_{\mathrm{eff}}$ = 32,500\,K and $\log g=3.75$, with
low (nebular) carbon and nitrogen abundances.

\textit{NGC346 MPG 12} (O9.5-B0V).--- Two models, $T_{\mathrm{eff}}$ =
32,500\,K and $\log g=3.75$, and $T_{\mathrm{eff}}$ = 30,000\,K and $\log
g=3.5$, fit equally well the UV and the optical spectrum. So, we adopted
intermediate parameters, $T_{\mathrm{eff}}$ = 31,000\,K and $\log g=3.6$.
The \ion{Si}{4}$\lambda$1400/\ion{Si}{3}$\lambda$1296 balance constrains
well the parameters. Our best fit requires that $V_{\mathrm{t\ }}=5$\,km\,s$%
^{-1}$. Even with a relatively low microturbulent velocity, we derived a low
carbon abundance to fit \ion{C}{3}$\lambda\lambda$1175, 1247, 4650. The
nitrogen abundance had to be enhanced to nearly solar value in order to
match the observed \ion{N}{3}$\lambda\lambda$4097, 4634, 4641 and \ion{N}{4}$%
\lambda$1718.

\subsection{O-type Giants}

With the exception of AV80, all the O giants in the sample have been
observed by FUSE as well as STIS. The extension of the observed spectra into
the FUSE spectral range (904 -- 1187\,\AA) provides additional elements for
ionization balance, such as \ion{N}{4}$\lambda$1718/\ion{N}{3}$\lambda$980
(if \ion{N}{4}$\lambda$1718 was included in a STIS spectrogram) and %
\ion{S}{5}$\lambda$1502/\ion{S}{4}$\lambda$1073, as well as giving
additional evidence on elemental abundances.

\textit{AV80} (O4-6n(f)p).--- This star is a hotter analogue of a Be star in
showing rotationally broadened lines and double-peaked emission of %
\ion{He}{2}$\lambda$4686. Because of the high rotational velocity ($V \sin
i\simeq 325$\,km\,s$^{-1}$), only the strongest lines, such as the H and He
lines and \ion{C}{3}$\lambda$1175, are useful for analysis. A model with $T_{%
\mathrm{eff}}$ = 38,000\,K and $\log g=3.5$ produces an acceptable fit to
the observations. The strength of \ion{N}{3}$\lambda$4097 (blended with H$%
\delta$), moderately strong emission of \ion{N}{4}$\lambda$4058 (unmatched
by the model), and a strong \ion{N}{5}$\lambda$1240 P~Cygni feature indicate
that nitrogen is enhanced, but the \ion{N}{3} line cannot be used to derive
a quantitative estimate and the model does not reproduce the \ion{N}{4} and %
\ion{N}{5} lines formed in the wind. The UV carbon lines indicate a low
carbon abundance. In the UV, the iron lines, \ion{O}{4}$\lambda$1340 and %
\ion{S}{5}$\lambda$1502 indicate a high microturbulence, $V_{\mathrm{t}}=20$%
\,km\,s$^{-1}$.

\textit{AV75} (O5III(f+)).--- The H and He lines indicate $T_{\mathrm{eff}}$
= 38,500\,K and $\log g=3.5$ or lower. The moderately strong \ion{N}{3}$%
\lambda\lambda$4634, 4641 emission and strong \ion{N}{3}$\lambda$980
absorption together with the absence of \ion{N}{4}$\lambda$4058 emission are
compatible with this estimate, as are the \ion{C}{4}/\ion{C}{3}, \ion{S}{5}/%
\ion{S}{4}, and \ion{Fe}{5}/\ion{Fe}{4} ion ratios. There are indications of
a strong nitrogen enhancement and a low (nebular) carbon abundance. Spectral
lines of \ion{P}{5}, \ion{S}{5}, \ion{Si}{4} -- elements not involved in
nuclear processing -- are all stronger than predicted unless the
microturbulence is increased to $V_{\mathrm{t\ }}\simeq 20$\,km\,s$^{-1}$.

\textit{AV 220} (O6.5f?p).--- The question mark in the spectral
classification does not signify a question of whether the star is an Of-type
star. The spectrum of AV~220 shows very strong \ion{He}{2}$\lambda$4686
emission, the defining characteristic of Of-type spectra. Instead, the
question mark denotes doubt that it is a normal Of supergiant because the %
\ion{C}{3}$\lambda$4650 emission lines are as strong as the \ion{N}{3}$%
\lambda\lambda$4634, 4641 emission lines. \citet{walborn00} assert that such
spectral peculiarities involve ``ejected shell phenomena''. Our models are
inappropriate for such stars, so our results are uncertain. Although the
spectrum is rather noisy, it is clear that the UV lines are quite sharp,
indicating a rotational velocity, $V\sin i\la 30$\,km\,s$^{-1}$. Whether
these sharp lines are shell lines or true photospheric lines is an open
question. It is interesting to note that the FUSE spectrum reveals many
narrow features, like the excicted \ion{S}{4}$\lambda$1100 lines, along with
broad features, as the resonance \ion{S}{4}$\lambda\lambda$1063, 1073 or %
\ion{P}{5}$\lambda$1118 lines. The \ion{C}{3}$\lambda$1175 also show
remarkable changes between the FUSE (broad saturated line) and STIS (weaker
line with narrow components visible) spectra, thus providing a direct
evidence of changes in the circumstellar environment of AV~220. An in-depth
analysis of AV~220 remains beyond the scope of this paper. Ionization
balance of carbon and iron lines in the UV implies $T_{\mathrm{eff}}$ =
37,500\,K and $\log g=3.75$. Based on the relatively high gravity, we list
AV~220 along with the O giants rather than with supergiants. The observed
strengths of \ion{N}{3}$\lambda\lambda$980, 4097 indicate that nitrogen is
highly enhanced. Our model with a solar nitrogen abundance matches well the %
\ion{N}{3}$\lambda\lambda$4634, 4641 emission lines, although it fails to
reproduce the \ion{C}{3}$\lambda$4650 emission lines. The strengths of the %
\ion{O}{4}$\lambda$1340 triplet and of the \ion{S}{5}$\lambda$1502 line
imply that $V_{\mathrm{t\ }}=10$\,km\,s$^{-1}$.

\textit{AV 15} (O6.5II(f)).--- We derive the parameters, $T_{\mathrm{eff}}$
= 37,000\,K, $\log g=3.5$, with the following adjustments. A model with $V_{%
\mathrm{t\ }}=20$\,km\,s$^{-1}$\ is needed to match the strength of the
numerous \ion{Fe}{5} lines, the \ion{O}{4}$\lambda$1340 triplet and the %
\ion{S}{5}$\lambda$1502 line. The observed strengths of \ion{N}{3}$%
\lambda\lambda$980, 4097, 4634, 4641 imply an enhanced nitrogen abundance,
while the UV carbon lines imply a low (nebular) carbon abundance.

\textit{AV 95} (O7III((f))).--- A model with $T_{\mathrm{eff}}$ = 35,000\,K
and $\log g=3.4$ results in a good fit to the optical spectrum, as well as
to the UV carbon, iron, and sulphur lines. The observed strength of %
\ion{N}{3}$\lambda$4097 and the \ion{N}{3} emission lines imply an enhanced
nitrogen abundance, while the UV carbon lines imply a low (nebular) carbon
abundance.

\textit{AV 69} (OC7.5III((f))).--- \citet{walborn76} and \citet{walborn00}
suggest that the OC stars are actually normal stars in having CNO abundances
that reflect the material out of which the stars formed. We find a very good
fit to the FUSE+STIS\ spectrum of AV~69 for a `NEB' model with $T_{\mathrm{%
eff}}$ = 35,000\,K, $\log g$=3.4, and $V_{\mathrm{t\ }}=15$\,km\,s$^{-1}$,
in agreement with the detailed study of \citet{AV83}. Our final adopted
abundances are slightly higher than those of the `NEB' models, yielding an
excellent fit to the \ion{C}{3}$\lambda\lambda$1176, 1247 and \ion{C}{4}$%
\lambda$1169 as well as \ion{N}{3}$\lambda$980. Based on the latter, our
nitrogen abundance is up to a factor of 3 higher than the very low value
adopted by \citet{AV83}.

\textit{AV 47} (O8III((f)).--- \citet{walborn00} call the spectrum of AV~47
``entirely normal for its (SMC) type'', and is quite similar to the spectrum
of AV~69 (in particular the UV Fe lines). The Balmer lines are, however,
broader. Our best model, $T_{\mathrm{eff}}$ = 35,000\,K and $\log g=3.75$
produces an adequate fit to the UV and optical spectrum, although the UV %
\ion{Fe}{4} and some optical \ion{He}{2} lines are predicted too strong. The
adopted $T_{\mathrm{eff}}$\ is the best compromise between a higher
temperature indicated by the \ion{Fe}{4} lines and a lower value by the %
\ion{He}{2} lines. Additionally, we needed to adopt low (nebular) abundances
for carbon. The nitrogen abundance is difficult to determine accurately, but %
\ion{N}{3}$\lambda\lambda$1182, 1184, and 4097 are fitted with an abundance
one fifth solar.

\textit{AV 170} (O9.7III).--- The absence of \ion{C}{4}$\lambda$1169, the
strong \ion{Si}{3}$\lambda$1296, and the strong \ion{Fe}{4} lines indicate a
low $T_{\mathrm{eff}}$. The best fit model has $T_{\mathrm{eff}}$ =
30,000\,K, $\log g=3.5$. Nitrogen is clearly enhanced over that of a `S'
model, and we consistently match \ion{N}{3}$\lambda\lambda$980, 1182, 1184,
4097 with a solar abundance. Moreover, we derived a low carbon abundance
from \ion{C}{3}$\lambda\lambda$1175, 1247. To reproduce the iron line
strengths, we had to adopt a low microturbulence velocity, $V_{\mathrm{t}}=5$%
\,km\,s$^{-1}$.

\subsection{O-type Supergiants}

\textit{AV 83} (O7Iaf+).--- We derived the same temperature ($T_{\mathrm{eff}%
}$ = 35,000\,K) as for AV~69 (OC7.5III((f))), but a lower gravity, $\log
g=3.25$. As previously noted by \citet{AV83}, the two stars have virtually
identical optical spectra in some respects, such as the strengths of %
\ion{He}{1}$\lambda\lambda$4471, 4713 and \ion{He}{2}$\lambda\lambda$4542,
4200. However, a lower surface gravity of AV~83 is clearly indicated by (1)
the weaker wings of the \ion{H}{1} Balmer lines, (2) the strong emission of %
\ion{He}{2}$\lambda$4686, and (3) the lack of \ion{He}{1} singlets at 4143,
4387, 4922\,\AA\ (they appear in absorption in the spectrum of AV~69). The
weakening of the \ion{He}{1} singlets is used as an indicator of
luminosity-class \citep{walborn71}, but our models show that they disappear
fully only at temperatures of 35,000\,K and above. The UV spectral region
indicates a strong wind \citep{AV83}, so that many of the diagnostic lines
such as \ion{C}{3}$\lambda$1175 appear as P~Cygni lines and cannot be
reproduced by our static, plane-parallel models. However, ionization balance
of the iron lines (\ion{Fe}{5} \textit{vs.} \ion{Fe}{4}) is still possible
and is consistent with a temperature of 35,000\,K. All the photospheric
lines in the UV spectrum indicate a high microturbulent velocity, $V_{%
\mathrm{t\ }}=20$\,km\,s$^{-1}$.

\textit{AV 327} (O9.5II-Ibw).--- In addition to the \ion{H}{1} Balmer lines,
the silicon and the iron ionization balance indicate a low gravity. We
obtained a best match to the spectrum with $T_{\mathrm{eff}}$ = 30,000\,K
and $\log g=3.25$. Nitrogen is highly enriched, and the carbon abundance is
very low (lower than the neb model) as indicated by \ion{C}{3}$%
\lambda\lambda $1247, 4650. We adopted a microturbulent velocity, $V_{%
\mathrm{t\ }}=15$\,km\,s$^{-1}$.


\section{RESULTS AND DISCUSSION}
\label{Results}

Table~\ref{ResuTab} summarizes the results of our analysis and readily
reveals several trends. First, there is a good correspondence (obviously
expected!) between spectral class and fundamental parameters. The only
two exceptions are  AV~220 and AV~47, which have somewhat higher
gravities than expected from their assigned luminosity class. 
Second, the
microturbulent velocity decreases from early to late O stars, as previously
suggested by \citet{NGC346}. Large microturbulent velocities are a common
feature of giants and supergiants. The late-O
supergiant, AV~327, has a large microturbulent velocity similar to those 
found in early B supergiants (e.g., Trundle et al. 2004). The high levels of  
microturbulence derived in our analysis are not a spurious result of static
models. Similar large values of the microturbulence are derived from wind models 
such as CMFGEN \citep{NGC346} and FASTWIND \citep{trundle04}. Third,
all supergiants and all giants except AV~69 show significantly enhanced nitrogen. 
Nitrogen enrichment is also
found in the early-O dwarfs, whereas late-O dwarfs show nebular abundances
reflecting their original composition. The only exception is MPG~12 which is
actually near to the terminal-age main sequence (TAMS). Surface abundances are
further discussed in \S\ref{CNSect}.

\subsection{Fundamental Stellar Parameters}
\label{FundamSect}

We calculated the radius of each star from the basic formula:

\[
\frac{f_{\mathrm{v}}}{\mathcal{F}_{\mathrm{v}}}=\frac{R^{2}}{d^{2}}\ 
\]
where $f_{\mathrm{v}}$ is the observed visual flux of the star corrected for
reddening, $\mathcal{F}_{\mathrm{v}}$ is the intrinsic visual flux, i.e.
luminosity per square centimeter at the stellar surface, and $d$ is the
distance to the SMC. We used an IDL version of Kurucz' program, UBVBuser, to
calculate TLUSTY model colors $(B-V)$. We then derived color excesses and
visual extinction assuming the standard ratio $R = A_V/E(B-V) = 3.1$. All
stars have color excesses smaller than 0.15\,mag; hence, the reddening
correction should not introduce significant uncertainties. We assumed a
distance modulus, $\Delta M=19.0\pm 0.1$\,mag ($d=63\pm 3$\,kpc), for the
SMC.\footnote{%
In previous related papers \citep{walborn00, AV83, NGC346}, we assumed $%
\Delta M=19.1$\,mag; recent studies favor a shorter distance, $\Delta M=18.9$%
\,mag \citep{harries03}.} For O-type stars, the visual flux scales almost
linearly with temperature. The radius is relatively well determined, since
both the temperature and distance are known to about 5\%.

Given the stellar radius, we then calculated other fundamental stellar
parameters, 
\[
L=R^{2}(T_{\mathrm{eff}}/5780)^{4} 
\]
\[
M_{\mathrm{sp}}=R^{2}10^{(\log g-4.44)} 
\]
where the bolometric luminosity ($L$), radius ($R$), and spectroscopic mass (%
$M_{\mathrm{sp}}$) are in solar units. A good consistency check is obtained
in using the visual absolute magnitudes and the bolometric corrections
extracted from TLUSTY models \citep{OS02}, thus indicating that
uncertainties on $\log L$ are at most $\pm 0.1$\,dex. Table~\ref{StelParTab}
shows the results for the 17 program stars. The stars are displayed on a H-R
diagram (Fig.~\ref{HRfig}) together with Geneva evolutionary models
appropriate for SMC metallicity ($Z/Z_\sun = 0.2$) from which we
interpolated evolutionary masses $M_{\mathrm{ev}}$ and stellar ages. We use
here evolutionary tracks of non-rotating models, but tracks of rotating
models are not very different according to \cite{MM01}. Although the program
stars cover the full range in spectral type and luminosity class, there are
holes in the distribution, particularly between 41,500 and 52,500\,K. The
ON2III star in the sample, MPG~355, is found to be very hot, whereas the one
O4 star, MPG~324, is found to be no hotter than 41,500\,K. Under the %
\citet{Vacca96} spectral calibration, MPG~324 would be assigned a
temperature of nearly 49,000\,K. The observing program did not probe below $%
\log L_{\mathrm{bol}}$=4.9, which corresponds to a mass of about 20\,$M_\sun$%
.. All stars but MPG~355 appear well evolved from the zero-age main sequence
(ZAMS). This may be a selection effect, in particular for the late O dwarfs.
Because the brightest stars of a given spectral type are usually selected
for study, there is a risk that they turn out to be binaries. While a
significant fraction of Galactic O stars are known binaries (e.g.,
Ma\'iz-Apell\'aniz et al. 2004), little is currently known for SMC O stars
on this topic.


\subsection{Luminosities}
\label{LumSect}

Figure~\ref{LMTfig} illustrates that
program stars of similar luminosity class show a considerable scatter in
bolometric luminosity. For example, the four most luminous stars are giants (luminosity
class III or II), not supergiants. Of-type stars are generally regarded as
supergiants, but AV 220, one of two Of-type stars in the program, is no more
luminous than an O8V star! There appears to be little correspondence between
luminosity class and luminosity. Instead, luminosity class is better
correlated with the ratio of luminosity to mass. As shown in Fig.~\ref{LMTfig},
O-type dwarfs (luminosity class V) generally have $L/M_{\mathrm{sp}}<7,000$
(in solar units); giants (luminosity class III) have $L/M_{\mathrm{sp}%
}=12,000-20,000$; and supergiants (luminosity class I and II) have $L/M_{%
\mathrm{sp}}>20,000$. None of the program stars has $L/M_{\mathrm{sp}%
}>25,000 $.

The correspondence between luminosity class and $L/M_{\mathrm{sp}}$ is not
surprising. It has been known for more than a half century that 
some central stars of planetary nuclei ($M\approx
0.6\,M_\odot$, $L\approx 5,000\,L_\odot$) also have O or Of-type spectra
(e.g., Heap 1977). The degeneracy of the optical spectrum among stars whose
luminosities differ by a factor of 100 demonstrates that the physical basis
of luminosity class cannot possibly be luminosity \textit{per se}. Instead,
the spectral characteristics defining luminosity class are responsive to the
effective gravity ($g_{\mathrm{eff}}$) of the atmosphere, which is lower
than the Newtonian gravity ($g$) because of the outward-directed force of
radiation pressure: 
\[
g_{\mathrm{eff}} = (1-\Gamma _{\mathrm{rad}}) g. 
\]
The Eddington factor, $\Gamma _{\mathrm{rad}}$, which is the ratio of the
luminosity to the Eddington luminosity is related to the luminosity-to-mass
ratio by \citep{SA78}: 
\[
\Gamma _{\mathrm{rad}}\equiv \frac{L}{L_{\mathrm{Edd}}}=2.6\times 10^{-5}%
\frac{L}{M_{\mathrm{sp}}} 
\]
where both luminosity and mass are in solar units. The fact that most
program stars show a good correspondence between luminosity class and $L/M_{%
\mathrm{sp}}$ ratio suggests that radiation pressure ($\Gamma _{\mathrm{rad}%
} $) is the primary means of lowering the effective gravity. However, 
no star has a $L/M_{\mathrm{sp}}$ close to the 
predicted limit of about 38,000, which suggests that some other mechanism(s)
besides radiation pressure is at work. That other mechanism is almost
certainly centrifugal
acceleration. \cite{MM01} defined a factor, $\Gamma _{\mathrm{rot}}$,
analogous  to the Eddington factor; hence, accounting for the levitating
effect of rotation, the effective gravity becomes: 
\[
g_{\mathrm{eff}} = (1-\Gamma _{\mathrm{rad}}-\Gamma _{\mathrm{rot}}) g, 
\]
where 
\[
\Gamma _{\mathrm{rot}} = \frac{\Omega^2}{2\pi G\rho_{\mathrm{m}}} 
\]
with $\Omega$ the angular velocity and $\rho_{\mathrm{m}}$ the average
density internal to the surface equipotential. We interpret 
the relatively high $L/M_{\mathrm{sp}}$ ratio of
AV~83, AV~327, and AV~170 as the effect of rapid rotation. 
This interpretation is bolstered by their high
nitrogen abundance caused by rotationally induced mixing of
nuclear-processed material to the surface ($\S 6.6$); moreover, the apparently slow
acceleration of the wind of AV~83 suggests that it is enhanced in the
equatorial plane and that the star is seen pole-on \citep{AV83}. For AV~47,
we derive a surface gravity that is more typical of main sequence stars in
agreement with its low $L/M_{\mathrm{sp}}$ ratio. On the other hand, the
high value for MPG~368 is not surprising as we argued that it might be a
binary. Finally, the peculiar spectral type of AV~220 makes it difficult to
reach a conclusion.


\subsection{Effective Temperature Scale}
\label{TeffSect}

The relation between our effective temperatures and spectral type is shown
in Fig.~\ref{Tefig}. For earlier-type O stars, the relation depends totally
on the ON2~III(f*) star MPG~355. For a given spectral type, the more
luminous stars have a lower effective temperature, a well-known result for
Galactic O stars. The full lines in Fig.~\ref{Tefig} indicate the new
relations for low metallicity O stars. The calibrations connect
straightforwardly our values, without attempting a least-square fit because
of the size of the stellar sample.

Our effective temperatures are lower than the ``standard'' calibration
established by \cite{Vacca96} from results based on NLTE H-He model
atmosphere analyses. Similarly all recent studies based on NLTE metal
line-blanketed model atmospheres found lower $T_{\mathrm{eff}}$, starting
with \cite{HHL98}, \cite{martins02}, \cite{crowther02}, and \cite{HL02a}.
Our temperatures are typically 10\% lower than Vacca's for mid to late-O
stars. The difference increases to 15-20\% at spectral type O4 and O5.

All other recent $T_{\mathrm{eff}}$\ determinations for O stars are based on
NLTE unified model atmospheres calculated with three codes, CMFGEN %
\citep{CMFGEN, TuebJDH}, FASTWIND \citep{santolaya97}, and WM-basic %
\citep{pauldrach01}. CMFGEN incorporates metal line-blanketing in a
detailed, similar way to TLUSTY, and we showed that an excellent agreement
is reached when analyzing O stars with TLUSTY and CMFGEN \citep{NGC346}. On
the other hand, for the sake of faster computation, FASTWIND and WM-basic
treat metal line blanketing in an approximate way. Moreover, our study, as
well as analyses conducted with CMFGEN, benefit from analyzing the complete
far-UV and optical spectrum, while FASTWIND analyses \citep{herrero02,
repolust04, massey04} are based solely on the optical hydrogen Balmer and
helium lines, and WM-basic models have been used to match only the UV metal
lines without considering the optical spectrum \citep{bianchi02}.

We have extracted over 60 new $T_{\mathrm{eff}}$\ determinations of Galactic
and SMC O stars based on these recent analyses, and we compare them in Fig.~%
\ref{Tcompfig}. All new values are lower than Vacca's. Fig.~\ref{Tcompfig}
shows, however, a relatively large scatter for any given spectral type. We
discuss these determinations for each spectral type.

\textit{O2--O3}.--- \cite{walborn02} splitted the original O3 spectral type
into 3 types, O2, O3, and O3.5, depending on the relative strength of
nitrogen emission lines in the optical spectrum. \cite{walborn04} support
our high effective temperature for the ON2~III(f*) star MPG~355. They derive
similar high temperatures (52,500 to 55,000\,K) for 4 other O2 stars in the
LMC. On the other hand, \cite{repolust04} derived a low temperature for the
Galactic O2 supergiant HD93129A, $T_{\mathrm{eff}}$ = 42,500\,K, admitting
however that this value might be spuriously low because of spectrum
contamination by a companion. As indicated by their different spectral
morphology, O3 stars have lower temperatures, in the range from 46,000 to
48,000\,K \citep{martins02, repolust04}.

\textit{O4--O5.5}.--- There are about 20 new determinations in this spectral
range, with all $T_{\mathrm{eff}}$\ clustering around 40,000\,K with a
scatter of $\pm$2,000\,K. There are three notable exceptions. First, \cite{massey04}
derived high values for the O5~V star AV~14 (44,000\,K), and for the
O5~V((f)) star AV~377 (45,500\,K). Nebular \ion{He}{1} line filling might
the culprit; they also derived a surprisingly large helium enrichment
(He/H=0.35) for the latter star. Second, \cite{bianchi02} obtained a low
temperature for the O5-6~III(f) star HD~93843 (34,000\,K).

\textit{O6--O6.5}.--- The 6 values obtained with TLUSTY and FASTWIND yield a
typical $T_{\mathrm{eff}}$\ of 38,000\,K with a scatter of $\pm$2,000\,K.
Within the O6 type, \cite{bianchi02} analyzed 4 Galactic stars covering all
luminosity classes with WM-basic and consistently found markedly lower
temperatures, $T_{\mathrm{eff}}$ = 33,000$\pm$1,000\,K.

\textit{O7--O7.5}.--- The 15 new $T_{\mathrm{eff}}$\ generally cluster
around 35,000\,K with a range of 1000\,K. However, lower temperatures are
obtained for two extreme O7Iaf+ SMC supergiants, 32,000\,K for AV~232 %
\citep{crowther02} and 33,000\,K for its twin AV~83 \citep{AV83}. Note that
our photospheric analysis of AV~83 favors $T_{\mathrm{eff}}$ = 35,000\,K.
Although the low temperatures derived for AV~232 and AV~83 agree with the
value derived for the Galactic O7~Ib(f) star (33,000\,K; Bianchi~\& Garcia
2002), we note that the spectral type of the latter indicate that it is not
a supergiant as extreme as the two SMC stars.

\textit{O8}.--- Much fewer determinations are available at spectral type O8,
indicating a typical value around 34,000\,K with a scatter of 1000\,K.

\textit{O9}.--- The 6 values derived with TLUSTY, CMFGEN, and FASTWIND
cluster in a limited range between 31,500 and 33,000\,K, leading to assign a
typical temperature of 32,500\,K for O9 stars.

\textit{O9.5}.--- The 9 values span a range between 29,000 and 32,000\,K
with no obvious outlier. We assign therefore $T_{\mathrm{eff}}$ = 30,500$\pm$%
1,500\,K to the O9.5 type.

In summary, we generally find a good agreement between results based on
TLUSTY, CMFGEN, and FASTWIND models. However, temperatures derived from the
UV metal line spectrum with WM-basic by \cite{bianchi02} are systematically
lower by 2,000 to 6,000\,K compared to other results. The typical error
quoted on individual results are $\pm$1000\,K in favorable cases to more
typically $\pm$2000\,K, which means that the error bars cover the span of
temperatures found at a given spectral type.

Because of the large scatter in current results, it remains difficult to
characterize empirically the dependence of the temperature scale on metallicity.
In past works, as well as in the recent studies based on FASTWIND models, classification
and calibration of O-type spectra rely almost exclusively on the hydrogen
and helium lines, so there is no direct dependence of the temperature scale
on metallicity. Some dependence is, however, expected, because of weaker
metal line blanketing and weaker wind feedback onto the photosphere. Along
these lines, based on FASTWIND modeling of hydrogen and helium lines, \cite
{massey04} argued that early to mid O stars in the SMC have temperatures
higher by up to 4000~K compared to similar Galactic stars, while the
differences are smaller for later O stars. On the other hand, results from
CMFGEN and TLUSTY models using additionally UV metal lines indicates cooler
temperatures in some SMC stars than in Galactic stars. We argue that the
high temperatures derived in a few cases by \cite{massey04} might be a
spurious consequence of \ion{He}{1} line contamination by nebular emission.
Indeed, the nebular contribution to \ion{He}{1} lines cannot be corrected as
well for SMC stars as for Galactic stars because of the larger distance,
hence the smaller apparent size of the nebulosity, thus resulting in
possible overestimation of $T_{\mathrm{eff}}$\ in SMC stars. We therefore
stress that additional temperature indicators beyond the helium lines are
extremely helpful.

As illustrated in Fig.~\ref{Tcompfig}, the current scatter in the results makes
hardly possible to reach a conclusion about a systematic difference between
Galactic (filled symbols) and SMC (open symbols) stars and, therefore, to
offer a quantitative relation tying the temperature scale to metallicity ---
which would be a great help for studies of the first generations of massive
stars. We are extending our analysis based on CMFGEN and TLUSTY models to a
significant sample of Galactic stars for providing the adequate comparison
sample in order to arrive to a fully satisfactory conclusion. Until then,
and because of the consistency of the results derived from CMFGEN wind
models and TLUSTY photospheric models, as well as our comprehensive analysis
of the various spectral diagnoses, we recommend to use the present results
as the temperature calibration for low metallicity O stars (Table~\ref{Calib1Tab}).
Alternatively, the calibration based on all recent studies of O stars can be
used as a general $T_{\mathrm{eff}}$-spectral type relation 
(bottom panel of Fig.~\ref{Tcompfig}).


\subsection{Ionizing Luminosities}
\label{BCQSect}

For each star, the adopted TLUSTY model provides the bolometric correction
and ionizing fluxes, $q_0$ and $q_1$, in the hydrogen Lyman ($\lambda\le 912$\,\AA)
and neutral helium ($\lambda\le 504$\,\AA) continua, respectively. The ionizing
flux in the \ion{He}{2} continnum ($\lambda\le 228$\,\AA) is highly sensitive 
to emission from shocks in the stellar wind and, therefore, our photospheric models
cannot provide relevant values. Using previously derived stellar radii, we
then obtain the ionizing luminosities, $Q_0$ and $Q_1$. Fig.~\ref{BCQfig} displays
effective temperatures, bolometric corrections and ionizing luminosities together in
relation with spectral type. Following lower $T_{\mathrm{eff}}$, we obtain lower
bolometric corrections and lower luminosities, compared to previous standard
calibrations \citep{SK82, Vacca96}.

For O dwarfs, our estimates of the ionizing luminosities are on average a factor of
2 to 3 smaller than values derived from \cite{Vacca96}, 
$\Delta\log Q_0 = 0.44\pm 0.16$ and $\Delta\log Q_1 = 0.34\pm 0.04$. For giants,
the differences are even larger, though we found a notably larger scatter. On
average, the differences are
$\Delta\log Q_0 = 0.55\pm 0.23$ and $\Delta\log Q_1 = 0.66\pm 0.32$.
Although Vacca et al.'s calibrations were established for Galactic stars, we
compare them to our results for low metallicity stars. The OSTAR2002
models, however, indicate that the effect of metallicity on ionizing
fluxes is quite limited, decreasing $q_0$ by 0.02~dex but increasing $q_1$
by less than 0.1~dex. Therefore, our results imply a substantial downward
revision of ionizing luminosities of O stars. Consequences for models of \ion{H}{2}
regions and for the ionization of the interstellar and intergalactic medium need
to be evaluated.

Table~\ref{Calib2Tab} provides our new calibration of bolometric correction and
ionizing luminosities with spectral type.


\subsection{Surface Abundances}
\label{CNSect}

Our previous study of the spectra of O stars in the SMC \citep{walborn00}
raised several important questions concerning the abundances of the elements
and their relation to evolutionary properties of the stars. In NGC~346, the
nitrogen surface abundance is enhanced in all but one of the 7 program stars %
\citep{NGC346}, while AV~83 and AV~69 exhibit striking differences in
surface composition even though they have near-identical stellar parameters %
\citep{AV83}. With our larger sample, we are now in position to revisit the
question of the surface enrichment of massive stars during their evolution,
how it increases with mass, extent of evolution from the ZAMS and/or
rotational velocity, and finally to compare our results with the predictions
of stellar evolution models.

As outlined in \S\ref{Microturb}, the results of chemical abundance
analyses are tied to the microturbulent velocity, which we derived by
simultaneously matching \ion{O}{4}$\lambda$$\lambda$1338-1342, \ion{S}{5}$%
\lambda$1502, and a multitude of iron lines. We list the adopted values in
Table~\ref{ResuTab}. In most cases, we derived a microturbulent velocity, $%
V_{\mathrm{t}}\ge 10$\,km\,s$^{-1}$, consistent with the large values
obtained from the CMFGEN wind models \citep{NGC346}, which are therefore not
an artefact of static models. As hinted by Bouret et~al., our results
suggest that microturbulence decreases towards late-O spectral types,
particularly in the case of the O dwarfs. Stellar metallicities, as
derived from iron and oxygen lines, are given in Table ~\ref{ResuTab}.
Except for MPG~487 which has an anomalously weak line spectrum, we
consistently get values that are one fifth of the solar value. Because of
this agreement within a relatively large sample, we conclude that this value
($Z = 0.2\,Z_\odot$) is well representative of the current metallicity of
the SMC, in agreement with metallicities measured in earlier studies limited
to fewer stars \citep{venn99, NGC346}. In all cases, we could match the %
\ion{He}{1} and \ion{He}{2} lines using the solar helium abundance and the
adopted microturbulent velocities. No indications of surface helium
enrichment are found in this sample, agreeing with the predictions of
stellar evolution models that account for the effects of rotation %
\citep{MM01}. On the other hand, only three of the program stars show
nitrogen abundances as low as the SMC nebular value, N/N$_\odot = 0.03$. We
adopt this value as the initial abundance relative to which we evaluate the
nitrogen enrichment of the other stars. These three stars have a carbon
abundance equal to the nebular abundance, C/C$_\odot = 0.08\pm 0.02$, that we
adopt as the initial carbon abundance. The N/C abundance ratios discussed
below (and in Table~\ref{ResuTab}) refer to these adopted values.

Fig.~\ref{HRNCfig} shows the program stars on a HR diagram with symbols
denoting their measured N/C abundance ratios.
It also shows the ZAMS and the TAMS for non-rotating
models \citep{Geneva3}. \cite{MM01} showed that models with rotation have
longer main-sequence phase, with a shift of the TAMS to lower temperatures.
There is no clear evidence that the N/C ratio increases with
evolution away from the ZAMS, or with effective temperature or luminosity.
Conversely, stars with near-identical parameters may have very different
surface compositions. For example, the three stars, AV~83, AV~69, and AV~47,
have similar parameters but show a wide range of N/C ratios (Table 4).
Hence, evolution alone cannot explain the observed N/C ratios and, indeed,
classical (non-rotating) stellar evolution models do not predict any
nitrogen enrichment during the main-sequence stage. On the other hand,
nitrogen enrichment $is$ predicted in models with rotationally-induced
mixing. Furthermore, \cite{AV83} convincingly argued that AV~83 is a fast rotator
while AV~69 is most likely a slow rotator. 

We find N/C enhancement ratios (relative to the SMC nebular value) ranging from 1 to
30. These observed enhancement factors are larger than predicted factors of 5-10 for
stars  with an initial rotational velocity, $V=300$\,km\,s$^{-1}$. \cite{MM01}
 say that increasing the initial velocity from 300 to 
 400\,km\,s$^{-1}$\ will not make signifcant changes. 
 Our results may indicate that current rotating models may underestimate 
 the mixing efficiency, although some of the difference
may come from Maeder~\& Meynet's adoption of a higher initial nitrogen abundance  
in SMC stars (hence, decreasing the factor of enhancement). 

Contrary to the case of nitrogen, all the program stars have a carbon
abundance that is consistent with the SMC nebular abundance within the
observational uncertainties. Only AV~327
may have a slightly lower abundance. Similarly, the oxygen abundance O/O$_\odot = 0.2$,
that we derived from the relatively weak \ion{O}{3}$\lambda$1149-54 triplet (the %
\ion{O}{4}$\lambda$$\lambda$1338-1342 lines are saturated and \ion{O}{5}$%
\lambda$1371 is sensitive to the wind properties) is   
consistent with the nebular value. We further discuss the behavior of the
CNO surface abundances in relation to the evolutionary status in Sect.~\ref
{AgeSect}.

Given these results, we are now in a poistion to explain the the spectral morphology 
of O-type stars. Among the program stars, MPG~355 was classified as an ON
star \citep{walborn02}, and two, MPG~113 (OC6 Vz) and AV~69 (OC7.5 III((f))), 
were classified stars as OC stars \citep{walborn00}. Our results indicate that a
three-fourths of stars in our sample are nitrogen-rich, even though they are not
classified as ON stars. Hence, an ON classification may indicate only the most
extreme enhancements. On the
other hand, an OC classification may simply indicate that the surface abundances
have not been altered. In both
MPG~113 and AV~69, the optical spectra show emission near \ion{C}{3}$\lambda$%
4650 without any \ion{N}{3}$\lambda$$\lambda$4634-4640 emission, and %
\ion{N}{3}$\lambda$4097 is absent. \cite{walborn00} suggested that these two
stars may actually reflect the composition of the nebular material from
which they formed, whereas other supposedly normal stars have undergone
rotational mixing. This suggestion is indeed confirmed by our quantitative
analysis, showing that the OC classification may be used to identify O stars
which are not yet nitrogen-rich, and are thus presumably intrinsically slow
rotators.


\subsection{The Mass Discrepancy and Rotation}
\label{MassSect}

We derived ``spectroscopic'' masses from the surface gravities and stellar radii
(\S\ref{FundamSect}), and ``evolutionary'' masses from Geneva evolutionary tracks for single,
non-rotating stars (see Fig.~\ref{HRfig}). For stars more massive than 20\,$%
M_\odot$, tracks of models with rotation are not very different according to 
\cite{MM01}. In Fig.~\ref{HRfig}, tracks are labeled by the initial mass of
the model. The evolutionary masses remain close to their initial masses ($M_{%
\mathrm{MS}}\ga 0.92\,M_{\mathrm{ZAMS}}$) throughout their main sequence
lifetimes because of low mass loss rates. We compare the spectroscopic and
evolutionary masses in Fig.~\ref{Massfig}. The figure shows that only three stars
(MPG~324, MPG~487, and AV~47) have spectroscopic masses as high or higher
than their evolutionary masses. The estimates actually agree within the
analysis uncertainties ($\log g$ would only need to be lower by 0.1\,dex or
less to reconcile the spectroscopic and evolutionary masses). However,
systematically, all the other stars have a spectroscopic mass lower than the
evolutionary mass. This disagreement is known as the ``mass discrepancy''
and has been explored in detail observationally by \cite{herrero92,
herrero02} in their studies of Galactic stars.

Spectroscopic and evolutionary mass estimates 
actually measure different things. Evolutionary masses follow from the
mass-luminosity relation (tracks are almost horizontal in Fig.~\ref{HRfig})
which is hardly affected by rotation at $V \sin i = 300$\,km\,s$^{-1}$. 
Spectroscopic masses derive from the stellar radius and effective
surface gravity. As discussed in \S\ref{LumSect}, the Newtonian gravity is
lowered by radiation pressure and centrifugal acceleration. As \cite{MM01} 
suggested, rotation is the root cause of the mass discrepancy.
 Model atmospheres in hydrostatic equilibrium only account
for radiation pressure plus gas pressure to balance gravity, so 
spectroscopic masses are correct only if the centrifugal acceleration is
negligible, that is, only for slow rotators. Otherwise, spectrum analyses
yield underestimates of stellar mass.

Since the surfaces of rapidly rotating stars are enriched in nitrogen, 
we should expect the largest nitrogen enhancements in stars
with the most severe mass discrepancies. In Fig.~\ref{Massfig}, we plotted
stars with different symbols depending on the measured nitrogen enrichment.
However, the success of this test is limited, in part because of the large
error bars on the spectroscopic masses (typically, $\pm 10\,M_\odot$) and of
the small number of stars showing the original composition. Revisiting this
expected relation using a larger stellar sample should be very helpful to
improve our understanding of the effects of rotation.

We cannot correct the spectroscopic masses for rotation, because we do not
know the actual rotation rates (we only have access to $V \sin i$). However,
the problem may be turned around and we can use the mass discrepancy as a
way to estimate the rotation rates. Using the relations in \S\ref{LumSect}
and requiring the two mass estimates to be identical, we derive rotational
velocities in the range from 200 to 400\,km\,s$^{-1}$. All but one star in
our sample have low apparent rotational velocities ($V \sin i \le 120$\,km\,s%
$^{-1}$, see Table~\ref{ResuTab}). Preferentially, stars in our sample are
thus viewed pole-on. It remains, however, unclear at this point if this is
or not a significant bias. For instance, we do not know if the polar and
equatorial regions are homogeneously enriched. However, the derived rotation
rates are not close enough to the break-up velocity to significantly affect
the shape of the stars and, consequently, introduce a bias in the
determination of effective temperatures.


\subsection{Evolutionary Status and Ages}
\label{AgeSect}

The program stars cover the full range of luminosity classes, from class V
stars to extreme supergiants (AV~83, O7Iaf+). Nevertheless, Fig.~\ref
{HRNCfig} shows that all but the two coolest stars (AV~327, AV~170) fall
between the ZAMS and the TAMS in the HR diagram. Allowing for rotation
(cooler TAMS), all stars are thus in the stage of core hydrogen burning. A
timescale argument excludes all stars to be at a later evolutionary stage
(looping back to the blue). In particular, because 75\% of the program stars
show a moderate or strong nitrogen enhancement, this enrichment needs to
occur during the main sequence phase and cannot be delayed to post-main
sequence phases. As discussed above, models with rotation indeed predict
nitrogen enrichment during the main sequence phase.

Stellar ages have been determined by interpolating between isochrones (Fig.~%
\ref{HRfig}) and are listed in Table~\ref{StelParTab}. Seven of the program
stars are members of NGC~346, the largest \ion{H}{2} region in the SMC. Our
age estimates are consistent with Bouret et~al.'s (2003) and show a spread
between 1 and 8~Myr. We note that this spread is increased by our lower
effective temperatures (using Vacca et~al.'s $T_{\mathrm{eff}}$\ scale, the
age of MPG~355 would not change while the age of the later O stars would be
smaller). Bouret et~al. concluded that NGC~346 is about 3~Myr old. They
interpreted the position of MPG~355 close to the ZAMS as resulting from
homogeneous evolution due to rotation close to break-up velocity %
\citep{maeder87}. They argued following \cite{walborn00} that the old MPG~12
is not a coeval member of NGC~346, and attributed the larger age of MPG~487
to possible binarity (i.e., the star appears too luminous in the HR
diagram). To their study, we only add one more star, MPG~682, which however
also turn to be relatively old (about 7~Myr). In NGC~346, we now have a very
young star, 3 stars with ages clustered around 3~Myr, and 3 ``old'' stars (6
to 8~Myr), in a sequence of decreasing masses. We may uphold Bouret et~al.'s
conclusion, and MPG~682 would also need not be coeval. This trend, however,
suggests an alternate scenario, that is that the most massive stars of the
cluster formed last, and once having formed, they put a halt to further star
formation.

AV~327 is the only star exhibiting a hint of carbon deficiency, and may thus
be viewed as the ''most evolved'' star of our sample. AV~327 is about 6~Myr
old. We derive slightly greater ages only for three other stars (AV~170,
MPG~12, and MPG~682), but they are significantly less massive ($M\approx
20~M_\odot$ compared to $35~M_\odot$ for AV~327). As discussed in \S\ref
{CNSect}, all other stars show their original carbon, oxygen, and helium
composition. As pointed out by \cite{MM01}, models with rotation naturally
explain N-rich stars with normal helium. A similar conclusion holds for
carbon and oxygen. Indeed, small differences between the predicted N/C and
N/O ratios at the end of hydrogen burning indicate that the C and O
depletion remains limited in this phase. During the main sequence phase, the
primary change in surface composition of rapidly-rotating stars is nitrogen
enhacement. Strong carbon and oxygen depletion only occurs at later phases
(helium burning) similarly to the non-rotating case. The model predictions,
therefore, agree with our findings.


\section{Conclusions}
\label{Conclusion}

We have presented a comprehensive analysis of the far-UV and optical,
high-resolution spectrum of 17 O stars in the SMC based on TLUSTY NLTE metal
line blanketed model atmospheres. We have determined the basic stellar
parameters, effective temperatures, surface gravities, stellar radii,
masses, luminosities, and ages, while consistently deriving the surface
chemical compositions. Wind properties have been discussed in two companion
papers for the first half of the sample \citep{AV83, NGC346}, or will be
presented in a future paper (Koesterke et~al., in prep.). We summarize the
main findings of our analysis.

Supporting the conclusion of several recent studies of a limited number of
Galactic and Magellanic Clouds O stars, we find that the effective
temperature scale of O stars needs to be revised downward relative to \cite
{Vacca96} calibration. Our temperatures are 15 to 20\% cooler for O4-O5
stars, but the difference decreases to about 10\% for mid to late O stars.
Our analysis benefits from using various spectral diagnoses from the optical
and the far-UV spectrum. Moreover, as demonstrated by \cite{NGC346},
excellent agreement in effective temperatures derived from TLUSTY and CMFGEN
models is achieved. Our new temperatures therefore form a sound basis for a
new $T_{\mathrm{eff}}$-spectral type calibration of low metallicity O stars.

Apart from a few exceptions, we also find a good agreement between
determinations based on TLUSTY, CMFGEN, and FASTWIND models for stars of
same spectral subtype, though the scatter may typically be as large as $\pm
2000$\,K. On the other hand, temperatures derived from UV metal lines with
WM-basic \citep{bianchi02} are systematically cooler by 2000 to 6000\,K
compared to all other recent studies. Given the differences between models,
analysis methodology, and the scatter at given spectral types between
current results, we believe that a reliable quantitative estimate of the
effect of metallicity on the effective temperature calibration still
requires additional work. In particular, we have started a similar analysis
of a large sample of Galactic stars with TLUSTY and CMFGEN. This study as
well as Massey et~al.'s analysis of their complete sample currently under
way will form appropriate bases for comparison with the results presented in
this paper.

As a consequence of lower effective temperatures, we derive
lower ionizing luminosities (typically by a factor of 3), implying
a significant downward revision of previous relations of ionizing
luminosities with spectral type. We provide a new calibration of
the ionizing luminosities of O stars, which is only
weakly dependent on metallicity.

The surface composition of these 17 O stars is the second main result of our
study. We found that a large fraction of the stars (about 75\% of our
sample) show surface nitrogen enrichment during their lifetime on the main
sequence. Furthermore, there is no enrichment in helium and no depletion in
carbon or oxygen. These results support \cite{MM01} predictions of stellar
evolution models accounting for rotationally-induced mixing. We find,
however, N/C enrichment factors of 10 to 40 that are significantly larger
than the theoretical predictions of factors of 5 to 10.

Additionally, we derive for all stars (but one) a present metallicity of one
fifth the solar value, in good agreement with other recent studies. In the
few cases where nitrogen surface abundance has not yet been affected by
mixing, we also obtain CNO abundances consistent with earlier nebular
studies. Finally, our analysis of UV metal lines indicate that NLTE metal
line blanketed model atmospheres still require a significant microturbulence
to match the strength of the observed UV lines. The required microturbulent
velocity might decrease from early to late O stars. We stress that CMFGEN
wind models also require similar microturbulent velocities, which thus are
not a spurious result from static model atmospheres.

Finally, we find that most stars in our sample show the so-called ``mass
discrepancy'' between spectroscopic and evolutionary mass estimates. Along
with nitrogen enrichment, we interpret that discrepancy as a result of rapid
rotation that lower the measured effective gravities. Therefore, most stars
in our sample are fast rotators with $V_{\mathrm{rot}} > 200$\,km\,s$^{-1}$,
and the low measured $V \sin i$\ thus imply that most stars are viewed
almost pole-on.

To conclude, our results emphasize the importance of rotation in our
understanding of the properties of massive stars as worked out by \cite
{heger00} and \cite{maeder00, MM01}. They also provide a framework for
understanding populations of massive stars in low metallicity environments
at low and high redshifts.

\acknowledgments

We thank Danny Lennon for making available the optical spectra.
This work was supported by several NASA grants from the Space Telescope
Science Institute (programs GO~7437, AR~7985), which is operated by the
Association of Universities for Research in Astronomy, Inc., under NASA
contract NAS5-26555; from the \textsl{FUSE\/}\ program B134 (NAG5-10895);
and from the NASA Astrophysics Data Program (NRA-01-01-ADP-099, NAG5-13187).


\clearpage

\begin{deluxetable}{llccc}
\tabletypesize{\small}
\tablewidth{0 pt}
\tablecaption{Program Stars. \label{StarTab}}
\tablehead{
\colhead{ID} & \colhead{Spectral Type}   & \colhead{Optical}  & \colhead{UV}  &
 \colhead{Far-UV} }
\startdata
 NGC 346 MPG 355   & ON2 III(f$^{*}$)  & AAT  &  STIS  &  \tablenotemark{a} \\
 NGC 346 MPG 324   & O4 V((f))         & AAT  &  STIS  &  \tablenotemark{a} \\
 NGC 346 MPG 368   & O4-5 V((f))       & AAT  &  STIS  &  \nodata \\
 NGC 346 MPG 113   & OC6 Vz            & ESO  &  STIS  &  \nodata \\
 NGC 346 MPG 487   & O8 V              & AAT  &  STIS  &  \nodata \\
 NGC 346 MPG 682   & O9-O9.5 V         & ESO  &  \tablenotemark{b}  &  \nodata \\
 NGC 346 MPG 12    & O9.5-B0 V (N str) & ESO  &  STIS  &  FUSE \\ [2mm]
 AV 80             & O4-6n(f)p         & AAT  &  STIS  &  \nodata \\
 AV 75             & O5 III(f+)        & ESO  &  STIS  &  FUSE \\
 AV 15             & O6.5 II(f)        & ESO  &  STIS  &  FUSE \\
 AV 95             & O7 III((f))       & ESO  &  STIS  &  FUSE \\
 AV 69             & OC7.5 III((f))    & AAT  &  STIS  &  FUSE \\
 AV 47             & O8 III((f))       & ESO  &  STIS  &  FUSE \\
 AV 170            & O9.7 III          & AAT  &  STIS  &  FUSE \\ [2mm]
 AV 220            & O6.5 f?p          & ESO  &  STIS  &  FUSE \\
 AV 83             & O7 Iaf+           & ESO  &  STIS  &  FUSE \\
 AV 327            & O9.5 II-Ibw       & ESO  &  STIS  &  FUSE \\
\enddata
\tablenotetext{a}{Composite of many stars in NGC 346}
\tablenotetext{b}{Target acquisition failed; wrong target}
\end{deluxetable}

\clearpage

\begin{deluxetable}{lll}
\tabletypesize{\small}
\tablewidth{0 pt}
\tablecaption{Temperature sensitivity of UV spectral features ($\log g = 4.0$).
   \label{LineTab}}
\tablehead{
\colhead{Line ID} & \colhead{T sensitive at:}   & \colhead{SI} }
\startdata
\ion{C}{3}\,$\lambda$1176    & 32.5 -- 50.0 kK                  & STIS, FUSE   \\ 
\ion{C}{3}\,$\lambda$1247    & 32.5 -- 47.5 kK                  & STIS \\ 
\ion{C}{4}\,$\lambda$1169    & peak strength at $\approx$ 41 kK & STIS, FUSE   \\ 
\ion{C}{4}\,$\lambda$1549    & saturated                        & STIS \\ 
\ion{N}{3}\,$\lambda$980     & 35.0 -- 50.0 kK                  & FUSE \\ 
\ion{N}{3}\,$\lambda$1184    & 37.5 -- 45.0 kK                  & STIS, FUSE \\ 
\ion{N}{4}\,$\lambda$1718    & blend with \ion{Fe}{4}           & STIS \\ 
\ion{N}{5}\,$\lambda$1240    & saturated / shocks               & STIS \\ 
\ion{O}{3}\,$\lambda$1150    & 37.5 -- 50.0 kK                 & FUSE \\ 
\ion{O}{4}\,$\lambda$1340    & 32.5 -- 40.0 kK                  & STIS \\ 
\ion{O}{5}\,$\lambda$1371    & $>$37.5 kK                       & STIS \\
\ion{O}{6}\,$\lambda$1036    & saturated / shocks               & FUSE \\ 
\ion{Si}{3}\,$\lambda$1298   & $<$35 kK                          & STIS \\ 
\ion{Si}{4}\,$\lambda$1126   & 32.5 -- 50.0 kK                  & FUSE \\ 
\ion{Si}{4}\,$\lambda$1398   & 35.0 -- 50.0 kK                  & STIS \\ 
\ion{S}{4}\,$\lambda$1073    & 35.0 -- 45.0 kK                  & FUSE \\ 
\ion{S}{4}\,$\lambda$1100    & 37.5 -- 42.5 kK                  & FUSE \\ 
\ion{S}{5}\,$\lambda$1502    & peak strength at $\approx$ 41 kK & STIS \\ 
\ion{Fe}{4}\,$\lambda\approx$1610 & 37.5 -- 45.0 kK             & STIS \\
\ion{Fe}{5}\,$\lambda\approx$1375 & $\leq $35.0 vs. $>$35.0 kK  & STIS \\ 
\ion{Fe}{6}\,$\lambda\approx$1270 & $>$45 kK                     & STIS 
\enddata
\end{deluxetable}

\clearpage

\begin{deluxetable}{cccccccccc}
\tabletypesize{\small}
\tablewidth{0 pt}
\tablecaption{SMC and model abundances. \label{AbunTab}}
\tablehead{
\colhead{Element} & & \colhead{SMC} &  & & \colhead{Sun\tablenotemark{a}} && & \colhead{Models}&  \\
    & \colhead{\ion{H}{2} regions\tablenotemark{b}} & \colhead{B stars\tablenotemark{c}} & \colhead{A stars\tablenotemark{d}}& &&& \colhead{S}& \colhead{NEB}&\colhead{N5X} }
\startdata
 \mbox{[C]}  & 7.45  &  7.4     & \nodata & \phm{xxx}&8.52&\phm{xxx} & 7.82 & 7.45 & 7.82 \\ 
 \mbox{[N]}  & 6.45  &  $<$6.7  & 7.33 & &7.92 && 7.22 & 6.45 & 7.92 \\ 
 \mbox{[O]}  & 8.10  &  8.1     & 8.14 & &8.82 && 8.12 & 8.10 & 8.12 \\ 
 \mbox{[Si]} & \nodata  &  \nodata  & 6.97 & &7.55 && 6.85 & 6.97 & 6.85 \\ 
 \mbox{[Fe]} & \nodata  &  \nodata  & 6.70 & &7.50 && 6.80 & 6.70 & 6.80 \\ 
\enddata
\tablenotetext{a}{\citet{Sun98}}
\tablenotetext{b}{\citet{garnett95}, \citet{kurt99}}
\tablenotetext{c}{\citet{rolleston93}}
\tablenotetext{d}{\citet{venn99}}
\end{deluxetable}

\clearpage

\begin{deluxetable}{llllrrllrl}
\tablecolumns{10}
\tabletypesize{\small}
\tablewidth{0 pt}
\rotate
\tablecaption{Results of the spectral analysis. 
  \label{ResuTab}}
\tablehead{
\colhead{ID} & \colhead{Spectral Type}   & \colhead{\teff}  & \colhead{$\log g$}  &
 \colhead{$V_{\rm t}$} &  \colhead{$V \sin i$} & \colhead{C/H\tablenotemark{a}} & \colhead{N/H\tablenotemark{a}} &
\colhead{N/C\tablenotemark{b}} & \colhead{M/H\tablenotemark{a}} \\
 & & \colhead{[K]} &  & \colhead{[\kms]}  & \colhead{[\kms]} & & & &  }
\startdata
 AV 83             & O7 Iaf+           & 35000  &  3.25  &  20 & 80 & 0.08 & 1.0 & 30 & 0.2 \\
 AV 327            & O9.5 II-Ibw       & 30000  &  3.25  &  15 & 60 & 0.04 & 0.2 & 12 & 0.2 \\ [3mm]
 AV 80             & O4-6n(f)p         & 38000  &  3.5   &  20 &325 & 0.08 & \tablenotemark{c} & $\gg$1 & 0.2 \\
 AV 75             & O5 III(f+)        & 38500  &  3.5   &  20 &120 & 0.08 & 1.0 & 30 & 0.2 \\
 AV 220            & O6.5 f?p          & 37500  &  3.75  &  10 & 30 & 0.08 & 1.0 & 30 & 0.2 \\
 AV 15             & O6.5 II(f)        & 37000  &  3.5   &  20 &100 & 0.08 & 1.0 & 30 & 0.2 \\
 AV 95             & O7 III((f))       & 35000  &  3.4   &  20 & 80 & 0.08 & 0.5 & 15 & 0.2 \\
 AV 69             & OC7.5 III((f))    & 35000  &  3.4   &  15 & 70 & 0.1  & 0.06&  2 & 0.2 \\
 AV 47             & O8 III((f))       & 35000  &  3.75  &  10 & 60 & 0.08 & 0.2 &  6 & 0.2 \\
 AV 170            & O9.7 III          & 30000  &  3.5   &   5 & 40 & 0.08 & 1.0 & 30 & 0.2 \\ [3mm]
 NGC 346 MPG 355   & ON2 III(f$^{*}$)  & 52500  &  4.0   &  25 &110 & 0.08 & 1.0 & 30 & 0.2 \\
 NGC 346 MPG 324   & O4 V((f))         & 41500  &  4.0   &  15 & 70 & 0.06 & 0.2 &  8 & 0.2 \\
 NGC 346 MPG 368   & O4-5 V((f))       & 40000  &  3.75  &  15 & 60 & 0.06 & 0.6 & 25 & 0.2 \\
 NGC 346 MPG 113   & OC6 Vz            & 40000  &  4.0   &  10 & 35 & 0.08 & 0.03&  1 & 0.2 \\
 NGC 346 MPG 487   & O8 V              & 35000  &  4.0   &   2 & 20 & 0.1  & 0.03&  1 & 0.1 \\
 NGC 346 MPG 682   & O9-O9.5 V         & 32500  &  3.75  &  10 & 80 & 0.08 & 0.03&  1 & 0.2 \\
 NGC 346 MPG 12    & O9.5-B0 V (N str) & 31000  &  3.6   &   5 & 60 & 0.06 & 1.0 & 40 & 0.2
\enddata
\tablenotetext{a}{Abundances and metallicities are relative to the solar values.}
\tablenotetext{b}{The N/C abundance ratio is relative to the SMC nebular N/C abundance ratio.}
\tablenotetext{c}{N-rich, but the high $V \sin i$ value makes a quantitative estimate uncertain.}
\end{deluxetable}

\clearpage

\begin{deluxetable}{llllcccccl}
\tabletypesize{\footnotesize}
\tablewidth{0 pt}
\rotate
\tablecaption{Stellar parameters and ages. 
  \label{StelParTab}}
\tablehead{
\colhead{ID} & \colhead{Spectral Type}   & \colhead{\teff}  & \colhead{$\log g$}  &
 \colhead{$R$} &  \colhead{$\log L$} & \colhead{$L/M_{\rm sp}$}& \colhead{$M_{\rm sp}$} & \colhead{$M_{\rm ev}$} & 
 \colhead{Age} \\
 & & \colhead{[K]} &  & \colhead{[$R_\sun$]}  &  &\colhead{[$10^3~L_\sun/M_\sun$]} &\colhead{[$M_\sun$]} & \colhead{[$M_\sun$]}  & \colhead{[Myr]}
  }
\startdata
 AV 83             & O7 Iaf+           & 35000  &  3.25  &  17 & 5.6 & 20.8            & 18 & 39  &  4.5  \\
 AV 327            & O9.5 II-Ibw       & 30000  &  3.25  &  19 & 5.4 & 11.2            & 22 & 35  &  6    \\ [3mm]
 AV 80             & O4-6n(f)p         & 38000  &  3.5   &  18 & 5.8 & 16.3            & 35 & 51  &  3    \\
 AV 75             & O5 III(f+)        & 38500  &  3.5   &  22 & 6.0 & 17.1            & 54 & 65  &  3    \\
 AV 220            & O6.5 f?p          & 37500  &  3.75  &   9 & 5.2 & \phantom{1}8.7  & 17 & 30  &  4    \\
 AV 15             & O6.5 II(f)        & 37000  &  3.5   &  17 & 5.7 & 14.6            & 33 & 46  &  3.5   \\
 AV 95             & O7 III((f))       & 35000  &  3.4   &  12 & 5.3 & 14.7            & 13 & 30  &  5    \\
 AV 69             & OC7.5 III((f))    & 35000  &  3.4   &  16 & 5.5 & 14.7            & 24 & 38  &  4.5  \\
 AV 47             & O8 III((f))       & 35000  &  3.75  &  15 & 5.5 & \phantom{1}6.6  & 46 & 39  &  4.5    \\
 AV 170            & O9.7 III          & 30000  &  3.5   &  13 & 5.1 & \phantom{1}6.3  & 19 & 22  &  7.5  \\ [3mm]
 NGC 346 MPG 355   & ON2 III(f$^{*}$)  & 52500  &  4.0   &  12 & 6.0 & 18.7            & 55 & 85  &  $<1$ \\
 NGC 346 MPG 324   & O4 V((f))         & 41500  &  4.0   &  11 & 5.5 & \phantom{1}7.3  & 41 & 40  &  3    \\
 NGC 346 MPG 368   & O4-5 V((f))       & 40000  &  3.75  &  10 & 5.4 & 11.2            & 21 & 36  &  3    \\
 NGC 346 MPG 113   & OC6 Vz            & 40000  &  4.0   &   7 & 5.1 & \phantom{1}6.3  & 20 & 30  &  2.5  \\
 NGC 346 MPG 487   & O8 V              & 35000  &  4.0   &  10 & 5.1 & \phantom{1}3.7  & 33 & 25  &  6    \\
 NGC 346 MPG 682   & O9-O9.5 V         & 32500  &  3.75  &   9 & 4.9 & \phantom{1}4.9  & 18 & 21  &  7    \\
 NGC 346 MPG 12    & O9.5-B0 V (N str) & 31000  &  3.6   &  10 & 4.9 & \phantom{1}5.7  & 14 & 20  &  8
\enddata
\end{deluxetable}

\clearpage

\begin{deluxetable}{llllllr}
\tabletypesize{\footnotesize}
\tablewidth{0 pt}
\tablecaption{Effective temperature, stellar luminosity and mass calibrations
  of spectral type.   \label{Calib1Tab}}
\tablehead{
\colhead{Spectral Type}   & \colhead{\teff}  & \colhead{$\log L$} & \colhead{$M_{\rm ev}$} &
\colhead{\teff}  & \colhead{$\log L$} & \colhead{$M_{\rm ev}$}   \\
 & \colhead{[K]} & & \colhead{[$M_\sun$]} & \colhead{[K]} & & \colhead{[$M_\sun$]} \\
 & \multicolumn{3}{c}{This paper} & \multicolumn{3}{c}{Vacca et al. (1996)}
  }
\startdata
\multicolumn{7}{c}{Dwarfs}\\
O4 V((f))         & 41500  & 5.48 & 40 & \phantom{$>$}48670 & \phantom{$>$}5.88 & \phantom{$>$}69 \\
O4-5 V((f))       & 40000  & 5.38 & 36 & \phantom{$>$}47400 & \phantom{$>$}5.81 & \phantom{$>$}62 \\
OC6 Vz            & 40000  & 5.09 & 30 & \phantom{$>$}43560 & \phantom{$>$}5.57 & \phantom{$>$}45 \\
O8 V              & 35000  & 5.09 & 25 & \phantom{$>$}38450 & \phantom{$>$}5.24 & \phantom{$>$}31 \\
O9-O9.5 V         & 32500  & 4.94 & 21 & \phantom{$>$}35260 & \phantom{$>$}5.01 & \phantom{$>$}24 \\
O9.5-B0 V (N str) & 31000  & 4.89 & 20 & \phantom{$>$}33980 & \phantom{$>$}4.93 & \phantom{$>$}22 \\ [3mm]
\multicolumn{7}{c}{Giants}\\
ON2 III(f$^{*}$)  & 52500  & 6.01 & 85 & $>$51000 & $>$6.15 & $>$100 \\
O5 III(f+)        & 38500  & 5.97 & 65 & \phantom{$>$}45410 & \phantom{$>$}5.93 & \phantom{$>$}68 \\
O4-6n(f)p         & 38000  & 5.76 & 51 & \phantom{$>$}45410 & \phantom{$>$}5.93 & \phantom{$>$}68 \\
O6.5 f?p          & 37500  & 5.18 & 30 & \phantom{$>$}41250 & \phantom{$>$}5.76 & \phantom{$>$}52 \\
O6.5 II(f)        & 37000  & 5.68 & 46 & \phantom{$>$}41250 & \phantom{$>$}5.76 & \phantom{$>$}52 \\
O7 III((f))       & 35000  & 5.27 & 30 & \phantom{$>$}39860 & \phantom{$>$}5.70 & \phantom{$>$}47 \\
O7 Iaf+           & 35000  & 5.57 & 39 & \phantom{$>$}38720 & \phantom{$>$}5.98 & \phantom{$>$}64 \\
OC7.5 III((f))    & 35000  & 5.54 & 38 & \phantom{$>$}38480 & \phantom{$>$}5.63 & \phantom{$>$}43 \\
O8 III((f))       & 35000  & 5.48 & 39 & \phantom{$>$}37090 & \phantom{$>$}5.57 & \phantom{$>$}39 \\
O9.5 II-Ibw       & 30000  & 5.40 & 35 & \phantom{$>$}32085 & \phantom{$>$}5.55 & \phantom{$>$}37 \\
O9.7 III          & 30000  & 5.07 & 22 & \phantom{$>$}32235 & \phantom{$>$}5.32 & \phantom{$>$}29
\enddata
\end{deluxetable}

\clearpage

\begin{deluxetable}{lllllll}
\tabletypesize{\footnotesize}
\tablewidth{0 pt}
\tablecaption{Bolometric correction and ionization luminosities relations with
  spectral type.   \label{Calib2Tab}}
\tablehead{
\colhead{Spectral Type}   & \colhead{BC}  & \colhead{$\log Q_0$} & \colhead{$\log Q_1$} &
\colhead{BC}  & \colhead{$\log Q_0$} & \colhead{$\log Q_1$}   \\
 & \colhead{[mag]} & \colhead{[s$^{-1}$]} & \colhead{[s$^{-1}$]} & 
 \colhead{[mag]} & \colhead{[s$^{-1}$]} & \colhead{[s$^{-1}$]} \\
 & \multicolumn{3}{c}{This paper} & \multicolumn{3}{c}{Vacca et al. (1996)}
  }
\startdata
\multicolumn{7}{c}{Dwarfs}\\
O4 V((f))         & -3.80  & 49.20 & 48.54 & \phantom{$<$}-4.40 & \phantom{$>$}49.70 & \phantom{$>$}48.99 \\
O4-5 V((f))       & -3.71  & 49.10 & 48.41 & \phantom{$<$}-4.32 & \phantom{$>$}49.61 & \phantom{$>$}48.90 \\
OC6 Vz            & -3.69  & 48.78 & 48.08 & \phantom{$<$}-4.06 & \phantom{$>$}49.34 & \phantom{$>$}48.61 \\
O8 V              & -3.29  & 48.53 & 47.37 & \phantom{$<$}-3.68 & \phantom{$>$}48.87 & \phantom{$>$}47.92 \\
O9-O9.5 V         & -3.09  & 48.25 & 46.59 & \phantom{$<$}-3.42 & \phantom{$>$}48.47 & \phantom{$>$}47.01 \\
O9.5-B0 V (N str) & -3.01  & 48.08 & 46.09 & \phantom{$<$}-3.30 & \phantom{$>$}48.27 & \phantom{$>$}46.50 \\ [3mm]
\multicolumn{7}{c}{Giants}\\
ON2 III(f$^{*}$)  & -4.57  & 49.86 & 49.42 & $<$-4.55 & $>$49.99 & $>$49.30 \\
O5 III(f+)        & -3.63  & 49.71 & 48.97 & \phantom{$<$}-4.21 & \phantom{$>$}49.73 & \phantom{$>$}48.98 \\
O4-6n(f)p         & -3.58  & 49.49 & 48.73 & \phantom{$<$}-4.21 & \phantom{$>$}49.73 & \phantom{$>$}48.98 \\
O6.5 f?p          & -3.51  & 48.82 & 48.02 & \phantom{$<$}-3.92 & \phantom{$>$}49.50 & \phantom{$>$}48.63 \\
O6.5 II(f)        & -3.50  & 49.40 & 48.61 & \phantom{$<$}-3.92 & \phantom{$>$}49.50 & \phantom{$>$}48.63 \\
O7 III((f))       & -3.34  & 48.92 & 48.03 & \phantom{$<$}-3.82 & \phantom{$>$}49.41 & \phantom{$>$}48.46 \\
O7 Iaf+           & -3.38  & 49.28 & 48.43 & \phantom{$<$}-3.75 & \phantom{$>$}49.69 & \phantom{$>$}48.57 \\
OC7.5 III((f))    & -3.34  & 49.19 & 48.30 & \phantom{$<$}-3.71 & \phantom{$>$}49.32 & \phantom{$>$}48.26 \\
O8 III((f))       & -3.32  & 49.01 & 47.97 & \phantom{$<$}-3.60 & \phantom{$>$}49.22 & \phantom{$>$}48.01 \\
O9.5 II-Ibw       & -2.89  & 48.75 & 46.85 & \phantom{$<$}-3.18 & \phantom{$>$}48.98 & \phantom{$>$}47.07 \\
O9.7 III          & -2.90  & 48.21 & 45.94 & \phantom{$<$}-3.18 & \phantom{$>$}48.67 & \phantom{$>$}46.71
\enddata
\end{deluxetable}

\clearpage

\clearpage	

\clearpage

\clearpage

\clearpage

\clearpage

\clearpage

\clearpage

\clearpage

\clearpage

\begin{figure}[tbp]
\epsscale{.99} \plotone{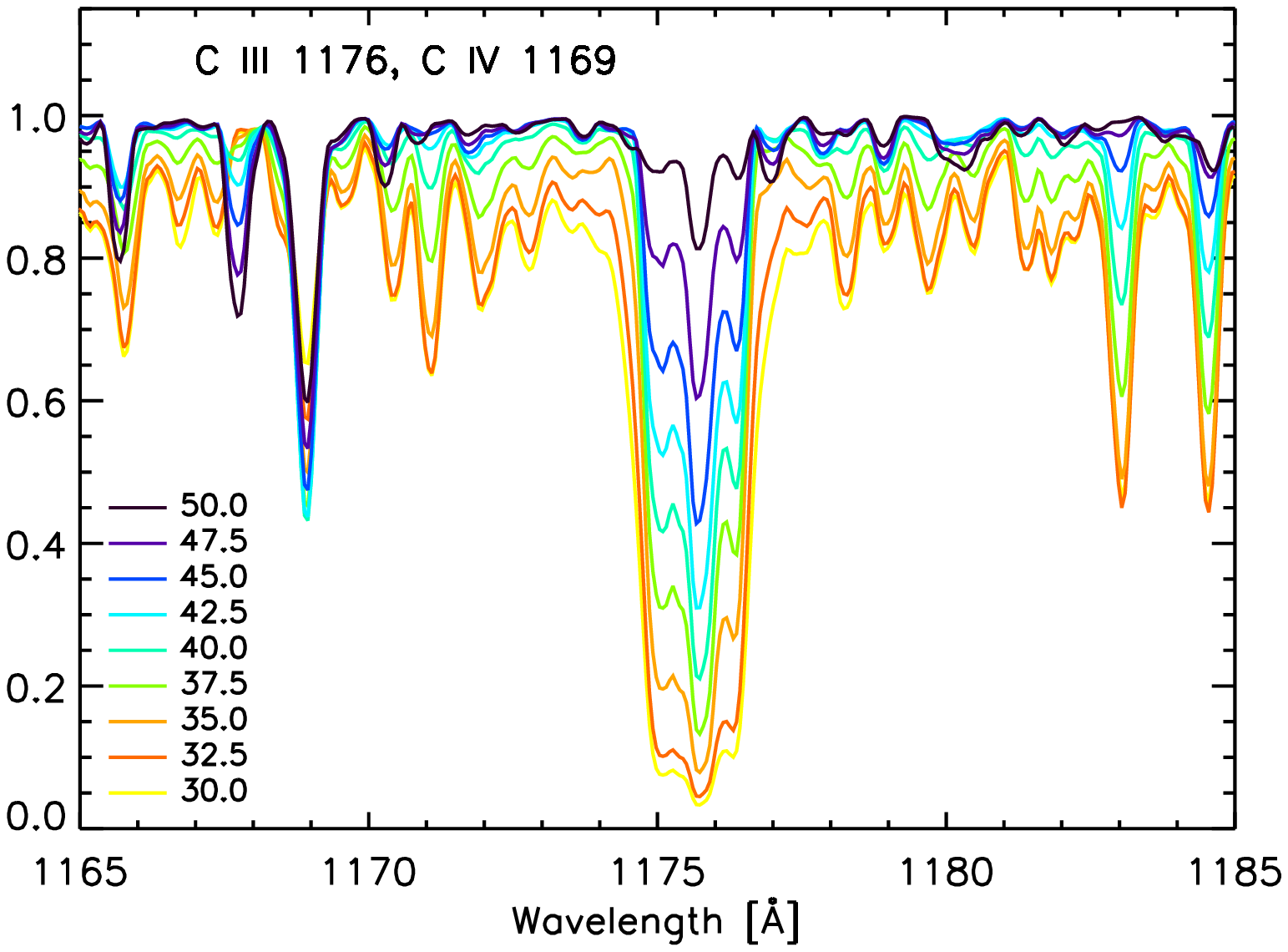} \figurenum{10a}
\caption{Sensitivity of \ion{C}{3}\,$\lambda$1176 and \ion{C}{4}\,$\lambda$%
1169 to effective temperature for a fixed $\log g=4.0$, $V_{\mathrm{t}}=10$%
\,km\,s$^{-1}$, $Z/Z_\sun=0.2$. Additional figures are shown
in the electronic edition of the Journal only for 
\ion{N}{5}\,$\lambda\lambda$1239,1243 + \ion{C}{3}\,$\lambda$1247(Fig. 10b), 
\ion{C}{4}\,$\lambda\lambda$1548, 1551 (Fig. 10c), 
\ion{C}{3}\,$\lambda$977, \ion{N}{3}\,$\lambda$980 (Fig. 10d), 
\ion{N}{3}\,$\lambda\lambda$1183, 1184 (Fig. 10e), 
\ion{N}{4}\,$\lambda$1718 (Fig. 10f), 
\ion{O}{3}\,$\lambda\lambda$1150, 1151, 1154 (Fig. 10g)
\ion{O}{4}\,$\lambda\lambda$1339, 1343, 1344 (Fig. 10h), 
\ion{O}{5}\,$\lambda$1371(Fig. 10i), 
\ion{Si}{3}\,$\lambda$1295-1303 (Fig. 10j), 
\ion{Si}{4}\,$\lambda\lambda$1123, 1128 (Fig. 10k), 
\ion{Si}{4}\,$\lambda\lambda$1394, 1403 (Fig. 10l), 
\ion{S}{4}\,$\lambda\lambda$1073, 1074 (Fig. 10m), 
\ion{S}{4}\,$\lambda$1098-1100 (Fig. 10n), 
\ion{S}{5}\,$\lambda$1502 (Fig. 10o), 
\ion{Fe}{4}\,$\lambda$1600-1630 (Fig. 10p), 
\ion{Fe}{5}\,$\lambda$1360-1380  (Fig. 10q), 
\ion{Fe}{6}\,$\lambda\lambda$1266, 1272, 1277, 1285 (Fig.~10r). }
\label{C34Tfig}
\end{figure}

\clearpage


\begin{figure}[tbp]
\epsscale{.699} \plotone{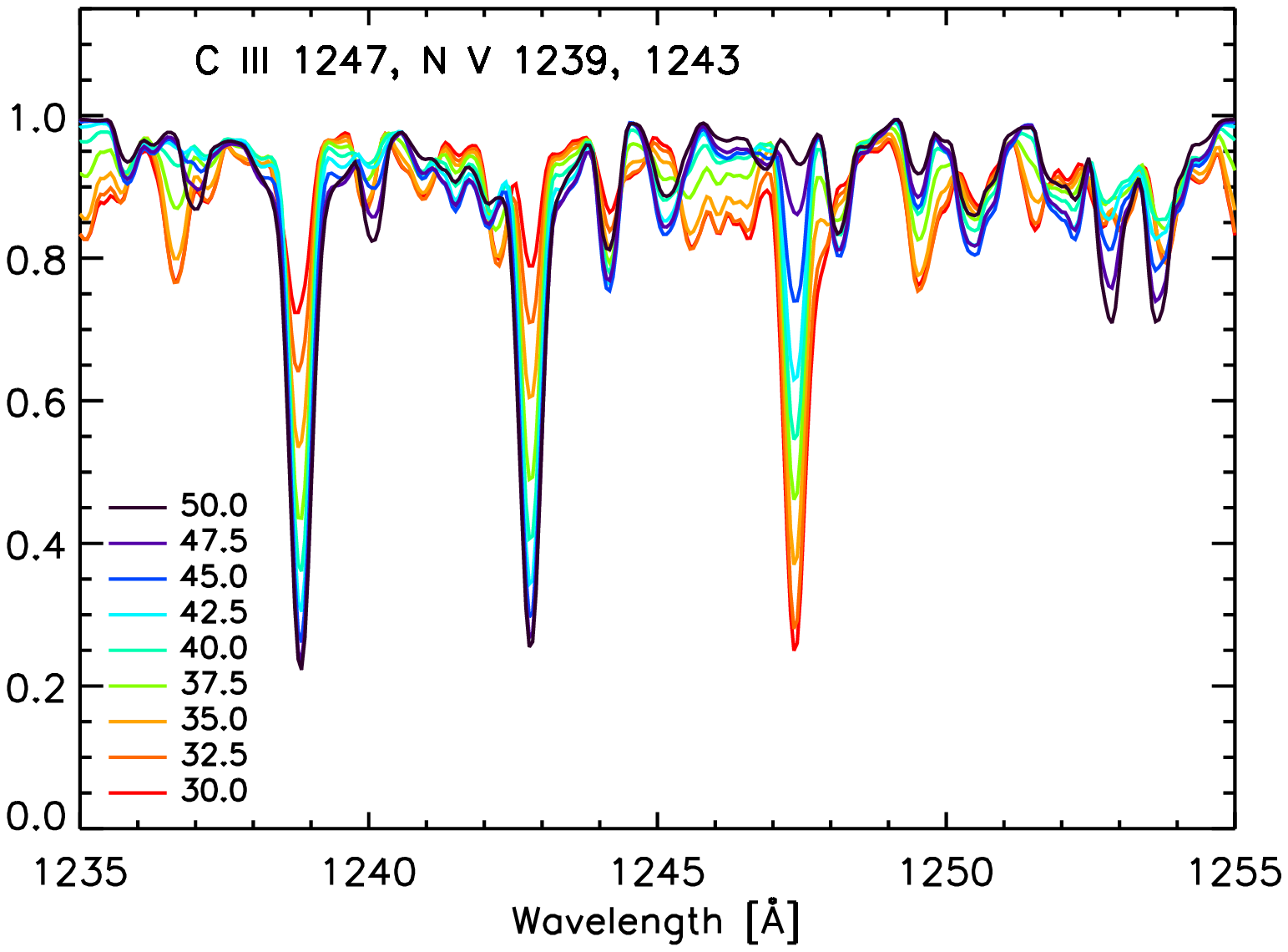} \figurenum{10b}
\caption{ONLY IN ELECTRONIC EDITION}
\end{figure}

\begin{figure}[tbp]
\epsscale{.699} \plotone{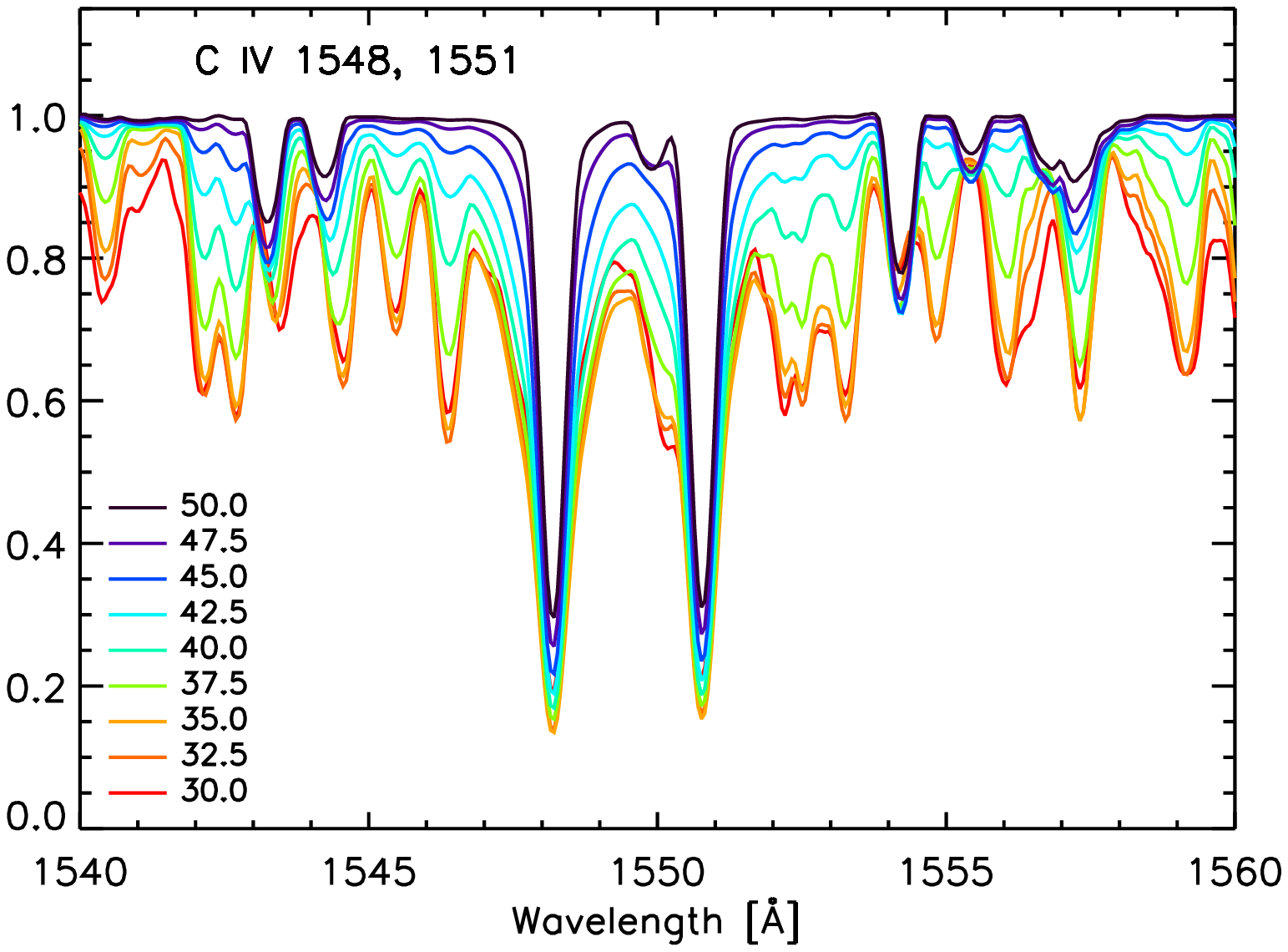} \figurenum{10c}
\caption{ONLY IN ELECTRONIC EDITION}
\end{figure}

\clearpage

\begin{figure}[tbp]
\epsscale{.699} \plotone{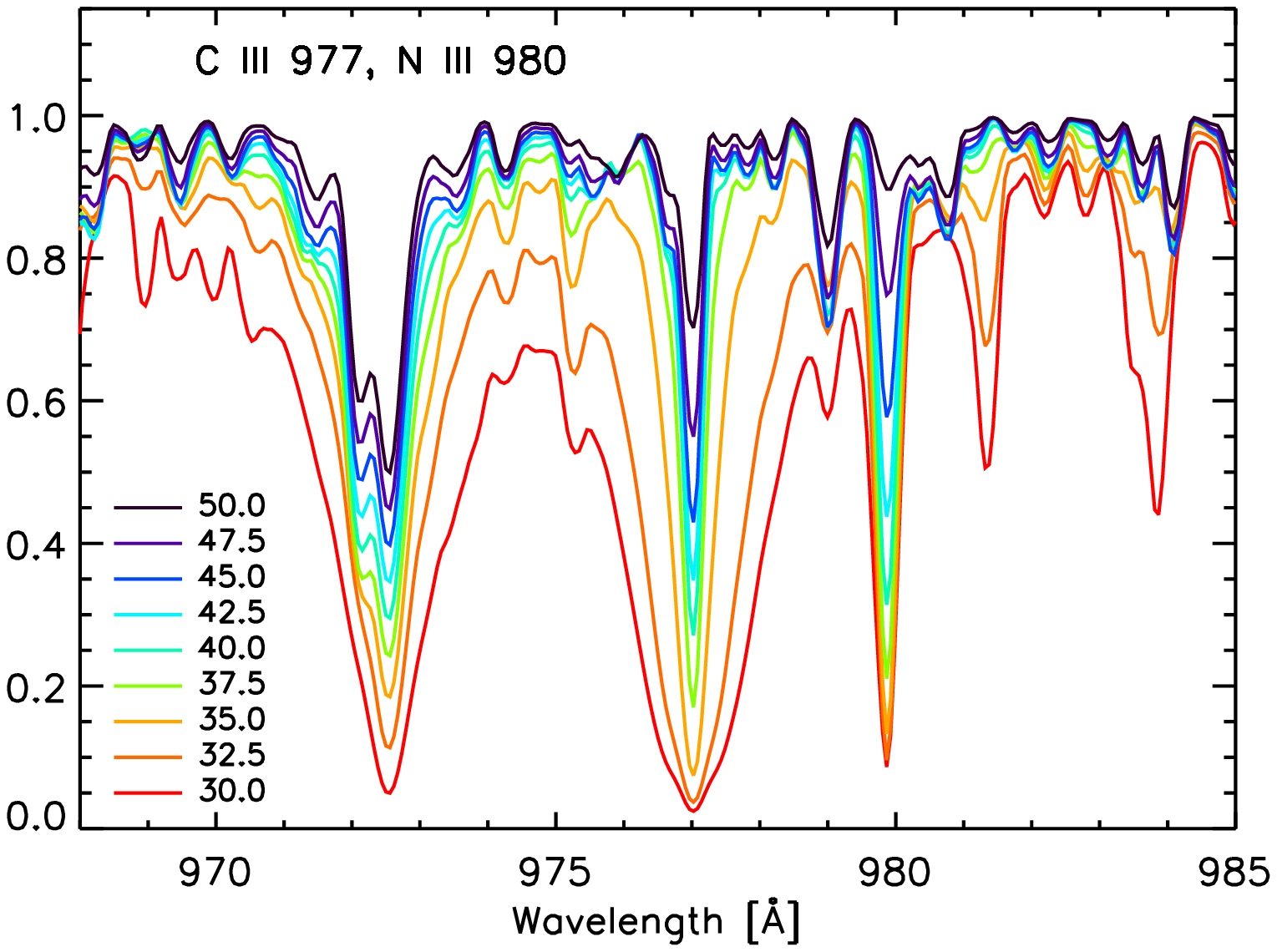} \figurenum{10d}
\caption{ONLY IN ELECTRONIC EDITION}
\end{figure}

\begin{figure}[tbp]
\epsscale{.699} \plotone{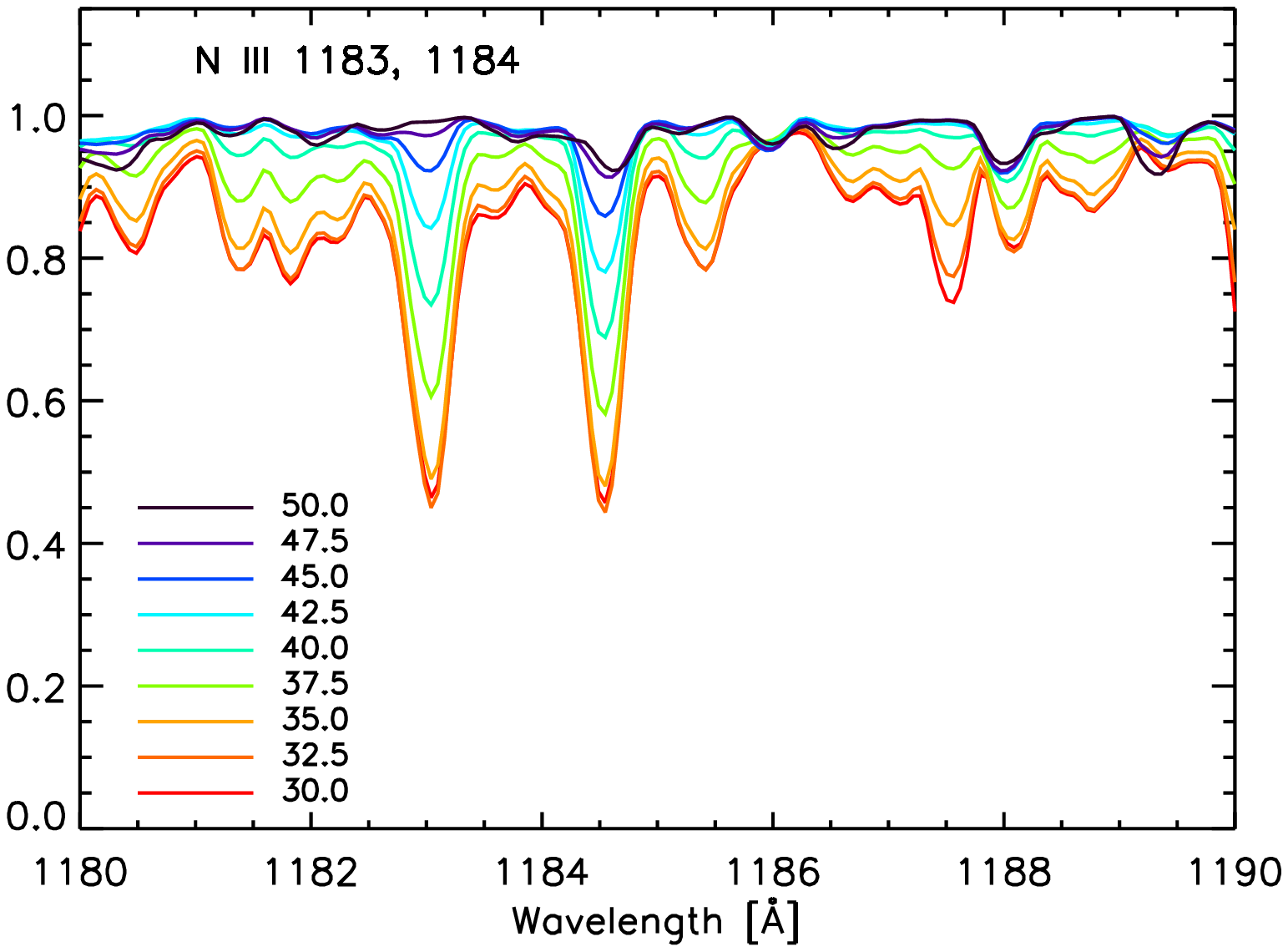} \figurenum{10e}
\caption{ONLY IN ELECTRONIC EDITION}
\end{figure}

\clearpage

\begin{figure}[tbp]
\epsscale{.699} \plotone{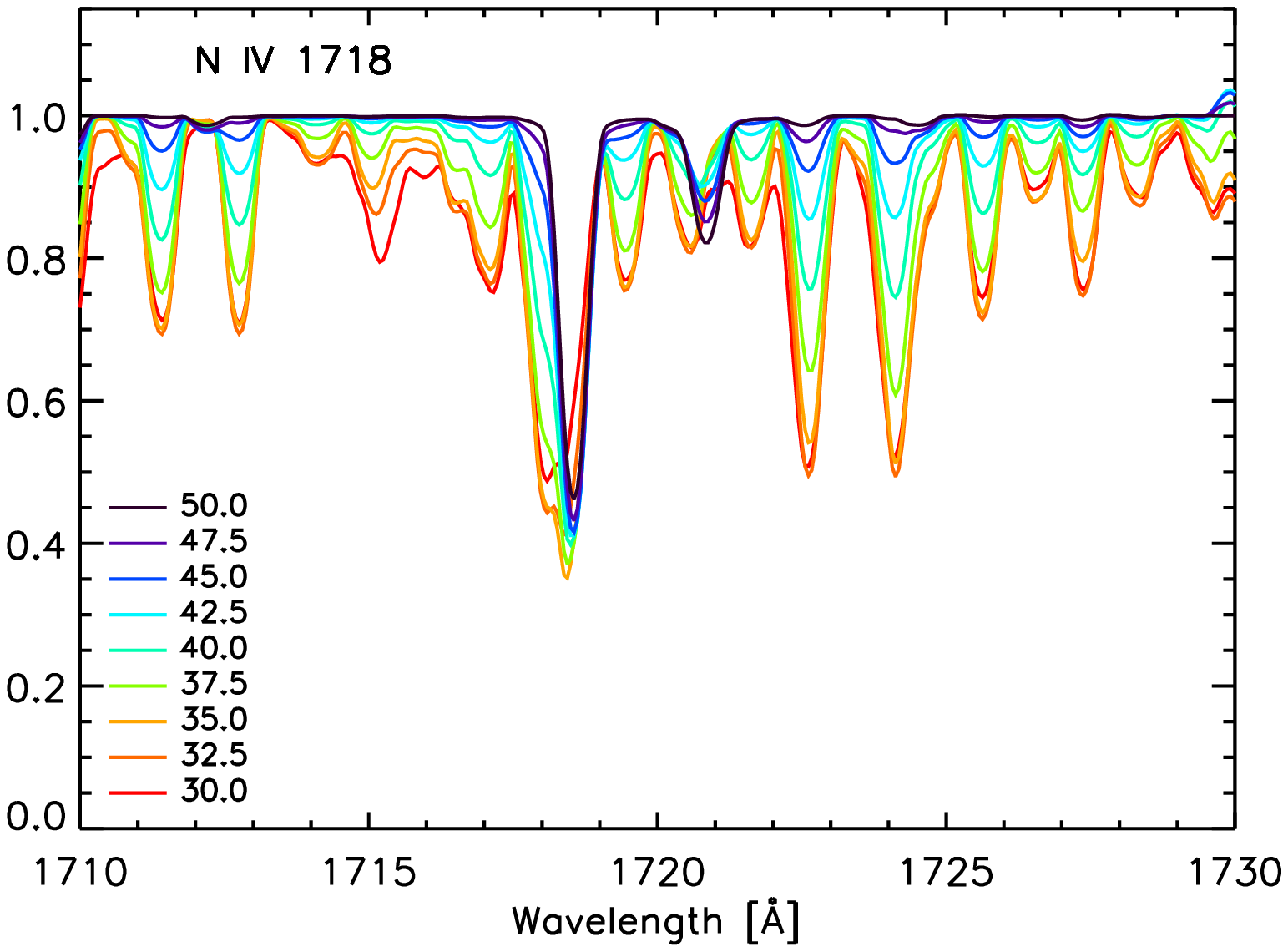} \figurenum{10f}
\caption{ONLY IN ELECTRONIC EDITION}
\end{figure}

\begin{figure}[tbp]
\epsscale{.699} \plotone{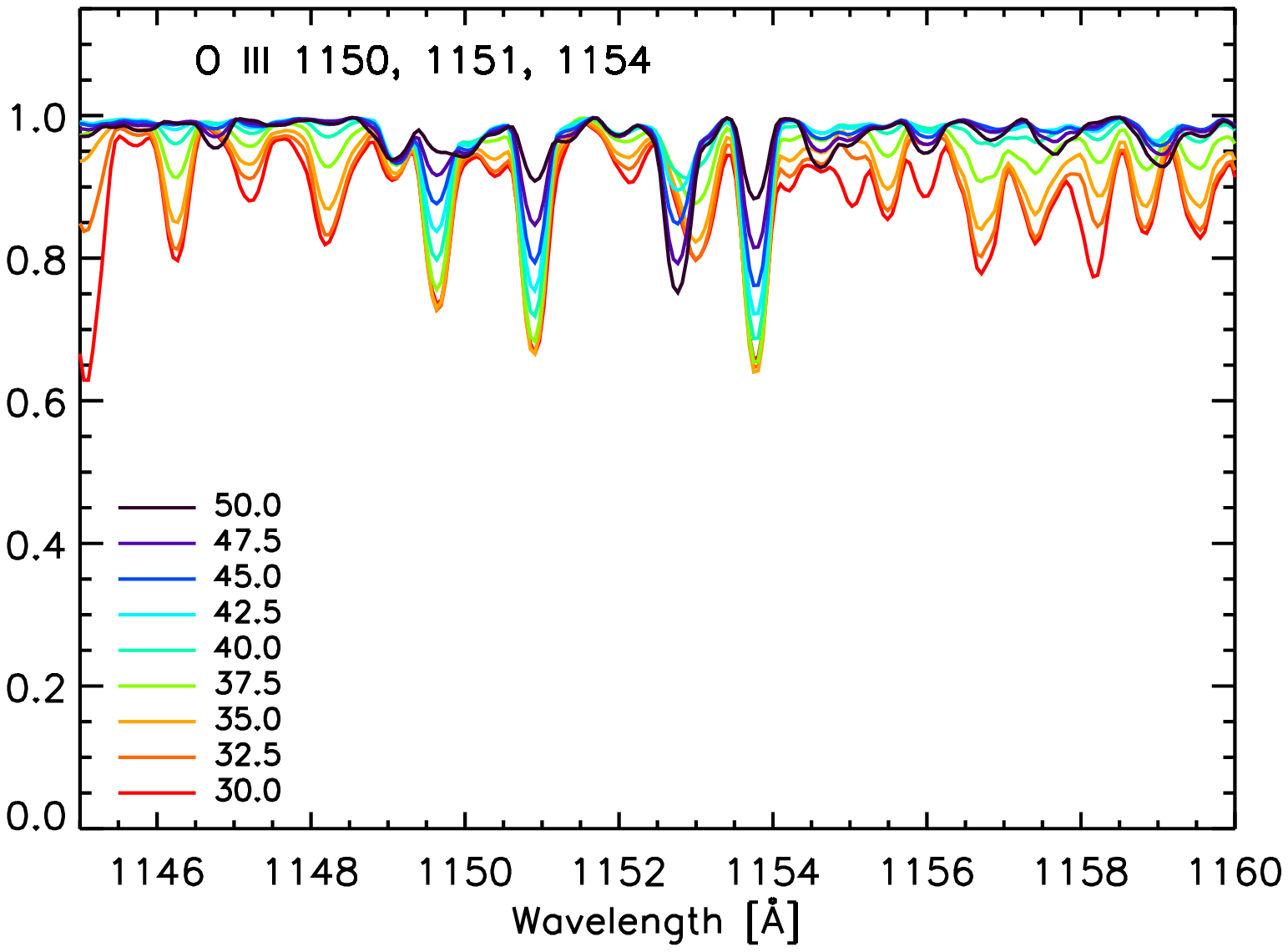} \figurenum{10g}
\caption{ONLY IN ELECTRONIC EDITION}
\end{figure}

\clearpage

\begin{figure}[tbp]
\epsscale{.699} \plotone{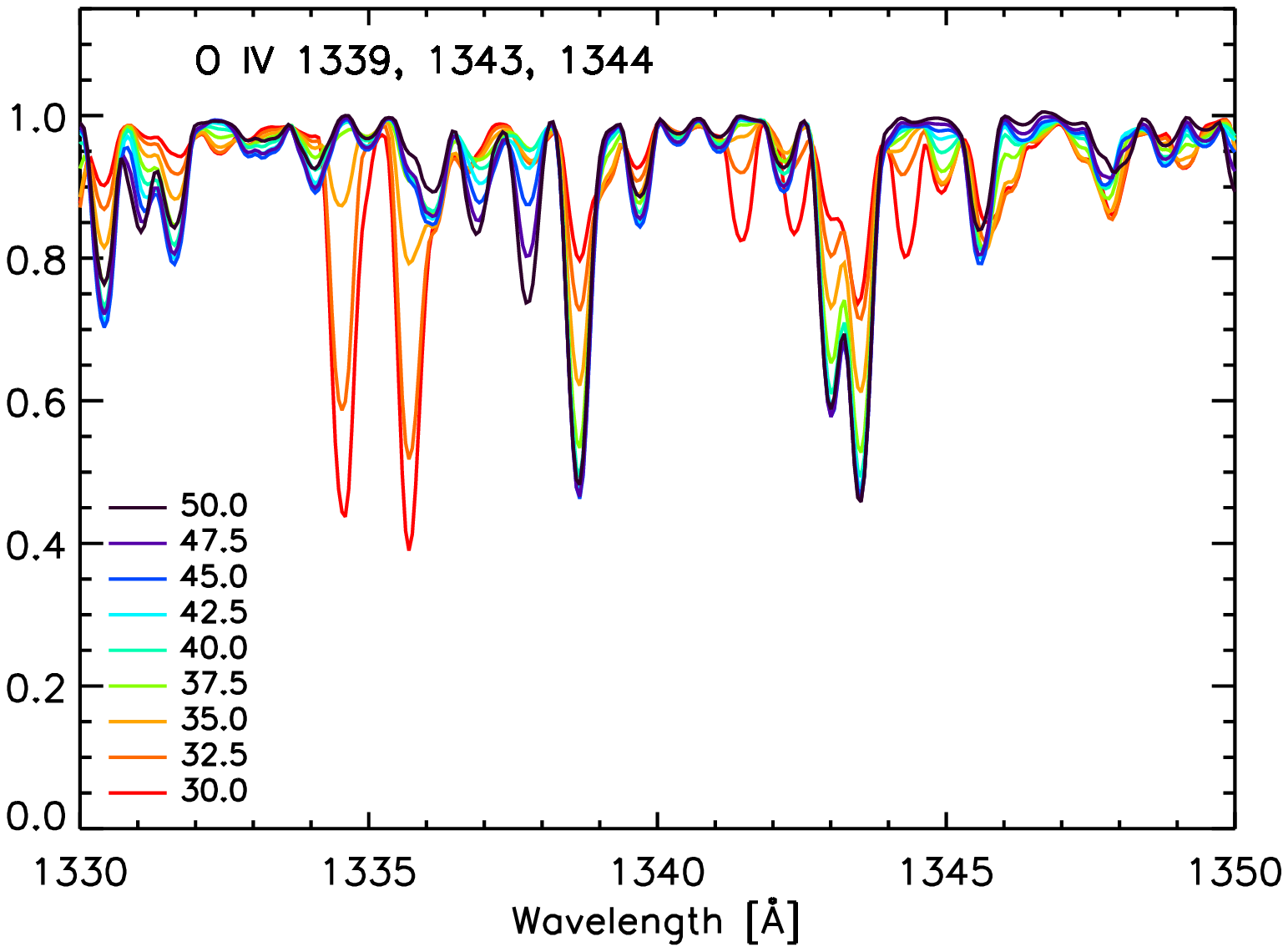} \figurenum{10h}
\caption{ONLY IN ELECTRONIC EDITION}
\end{figure}

\begin{figure}[tbp]
\epsscale{.699} \plotone{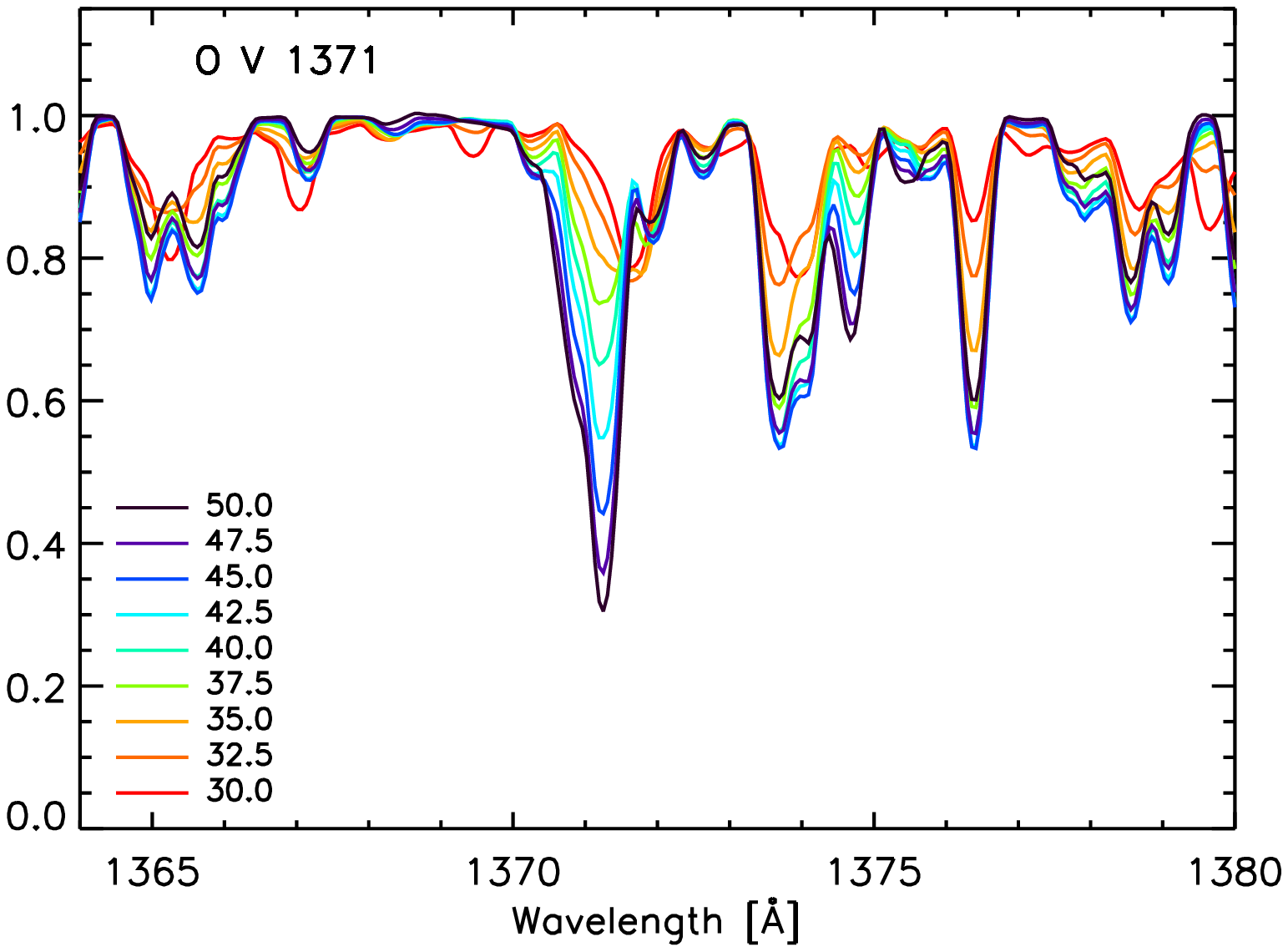} \figurenum{10i}
\caption{ONLY IN ELECTRONIC EDITION}
\end{figure}

\clearpage

\begin{figure}[tbp]
\epsscale{.699} \plotone{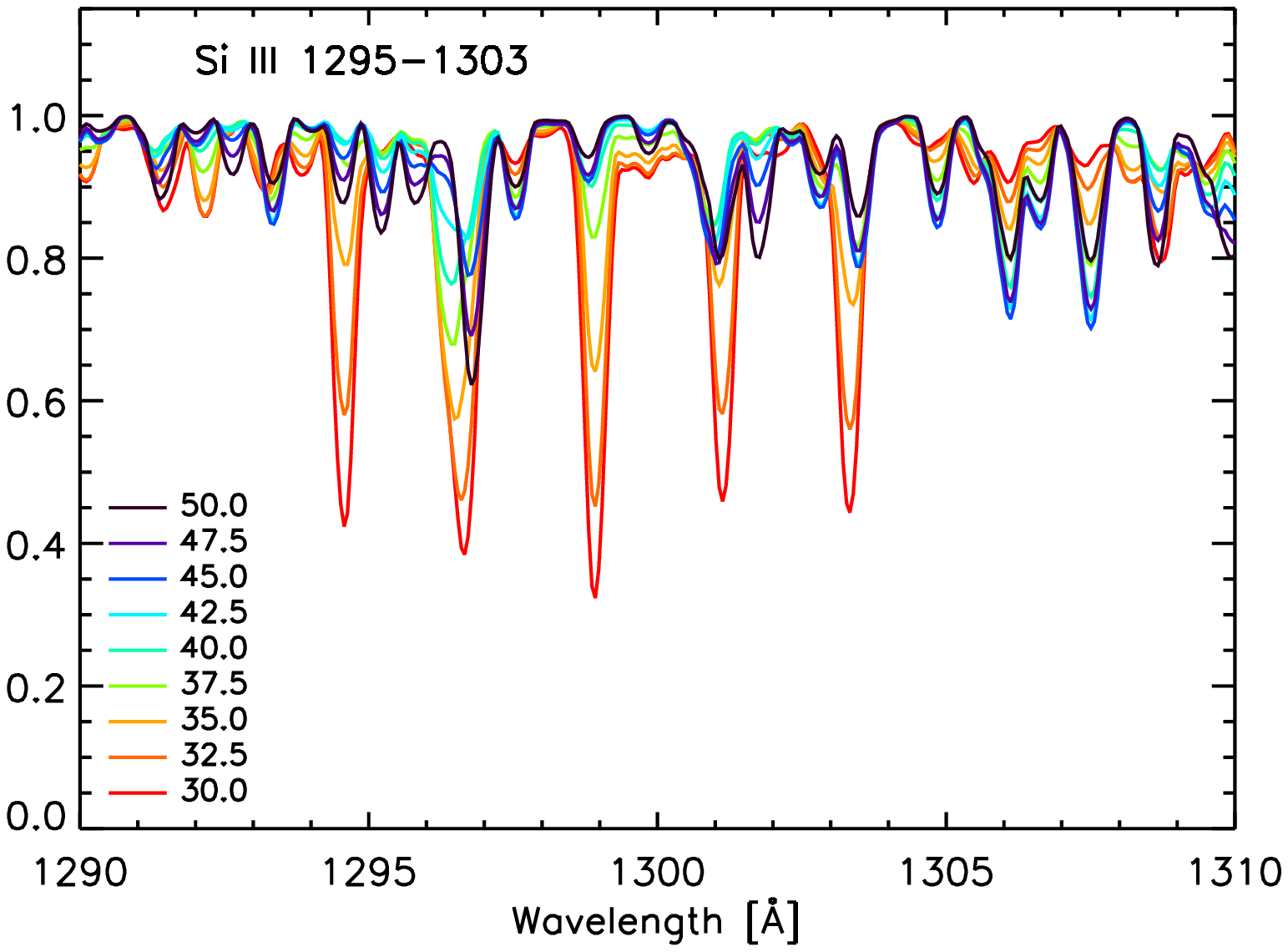} \figurenum{10j}
\caption{ONLY IN ELECTRONIC EDITION}
\end{figure}

\begin{figure}[tbp]
\epsscale{.699} \plotone{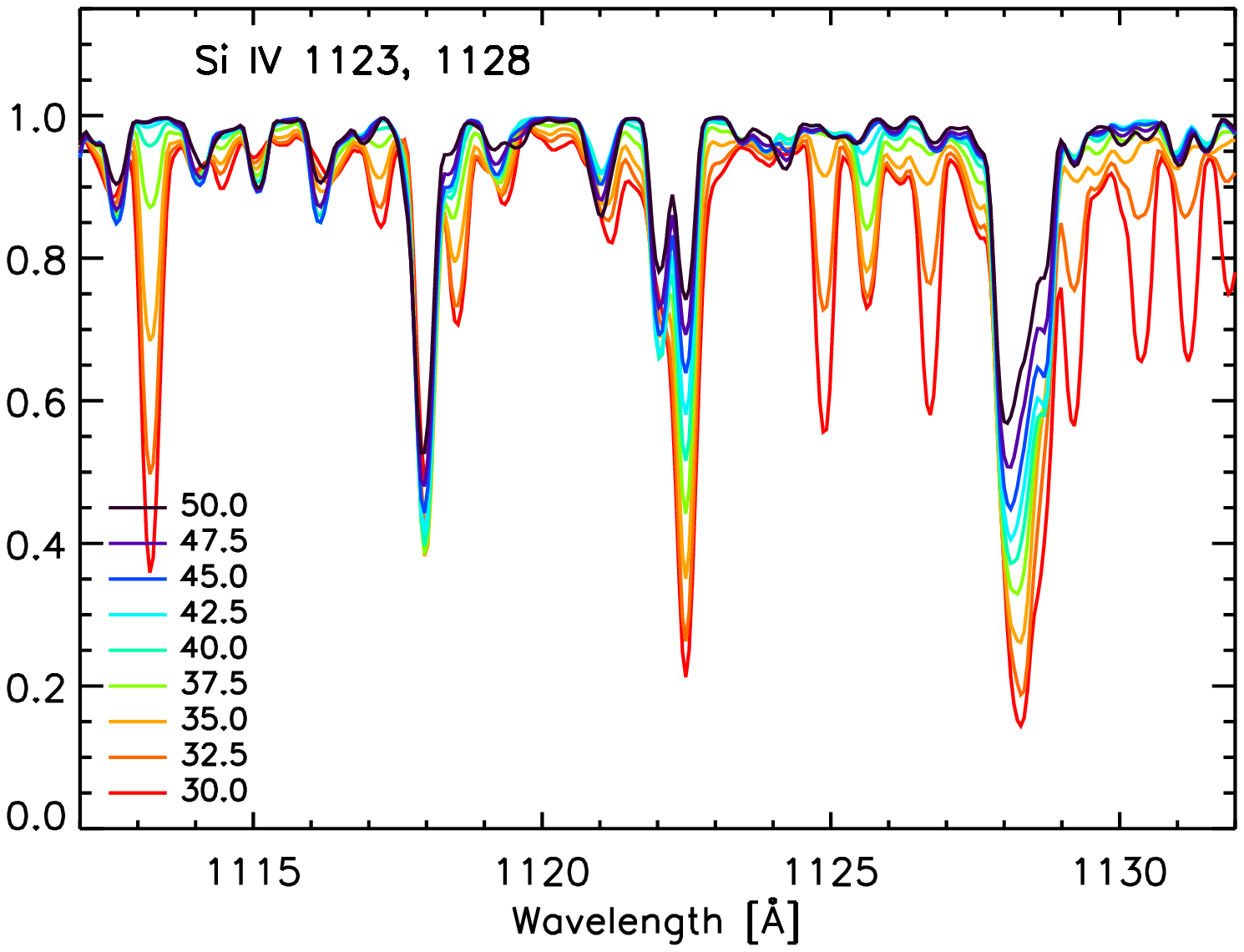} \figurenum{10k}
\caption{ONLY IN ELECTRONIC EDITION}
\end{figure}

\clearpage

\begin{figure}[tbp]
\epsscale{.699} \plotone{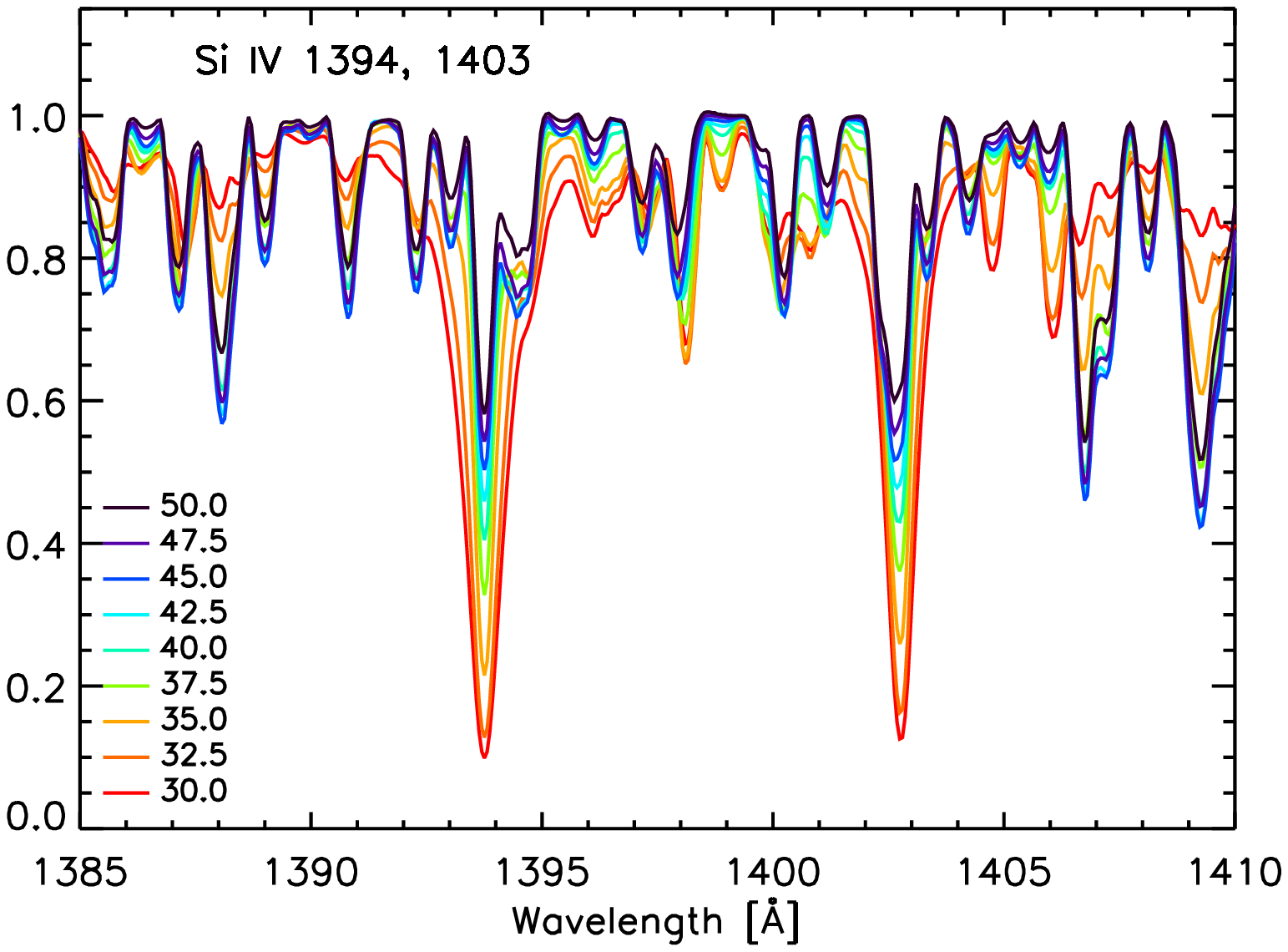} \figurenum{10l}
\caption{ONLY IN ELECTRONIC EDITION}
\end{figure}

\begin{figure}[tbp]
\epsscale{.699} \plotone{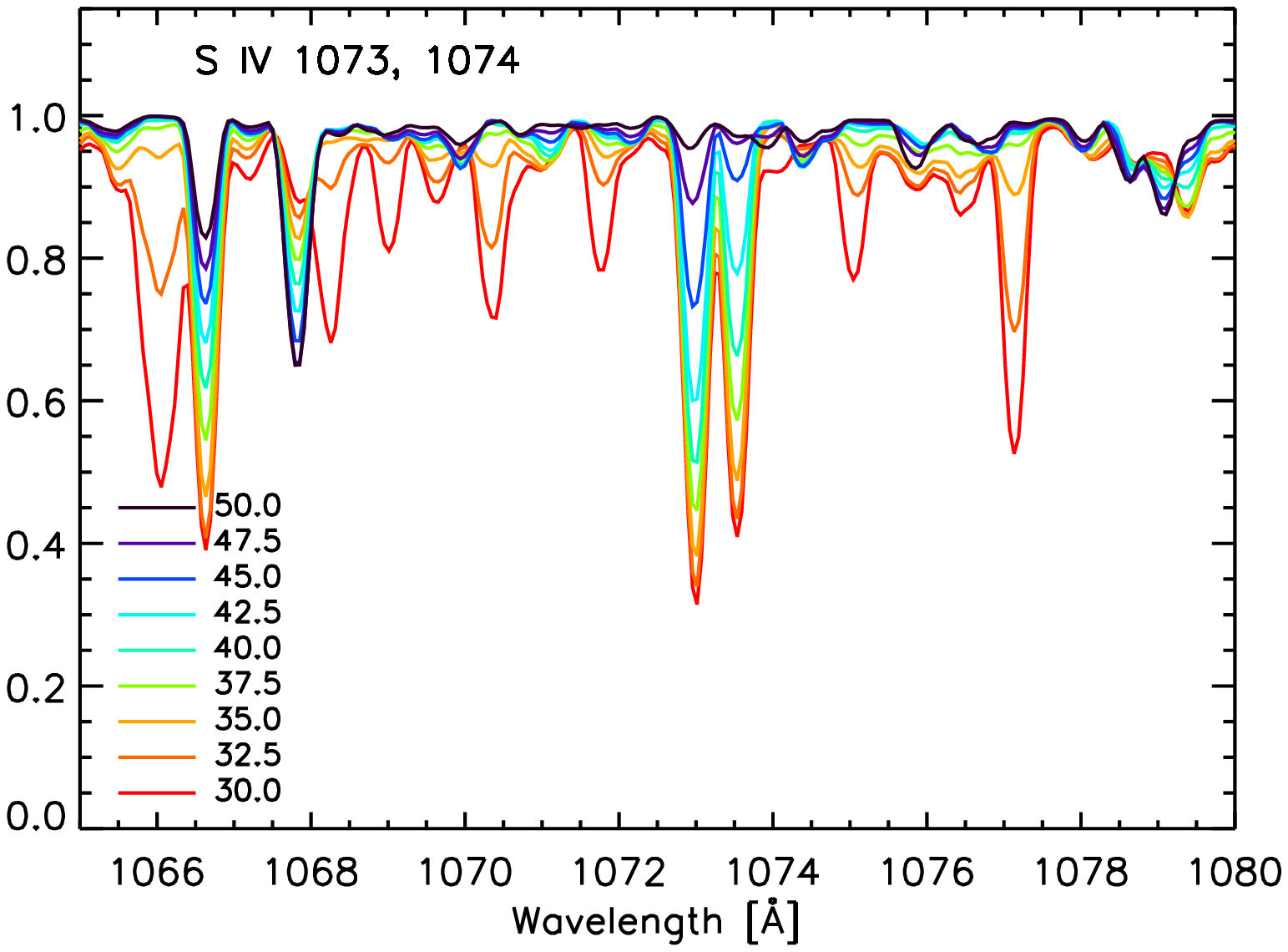} \figurenum{10m}
\caption{ONLY IN ELECTRONIC EDITION}
\end{figure}

\clearpage

\begin{figure}[tbp]
\epsscale{.699} \plotone{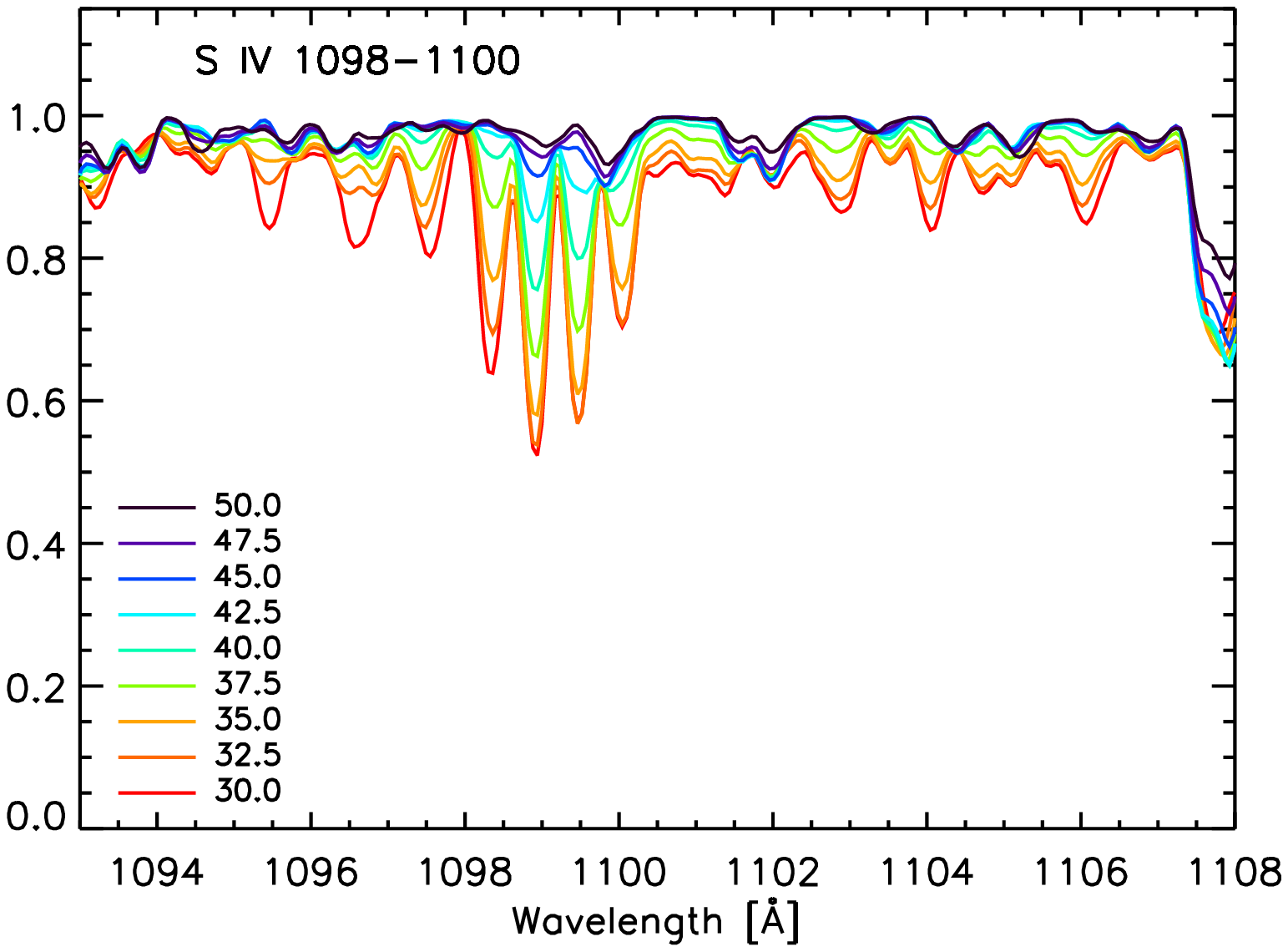} \figurenum{10n}
\caption{ONLY IN ELECTRONIC EDITION}
\end{figure}

\begin{figure}[tbp]
\epsscale{.699} \plotone{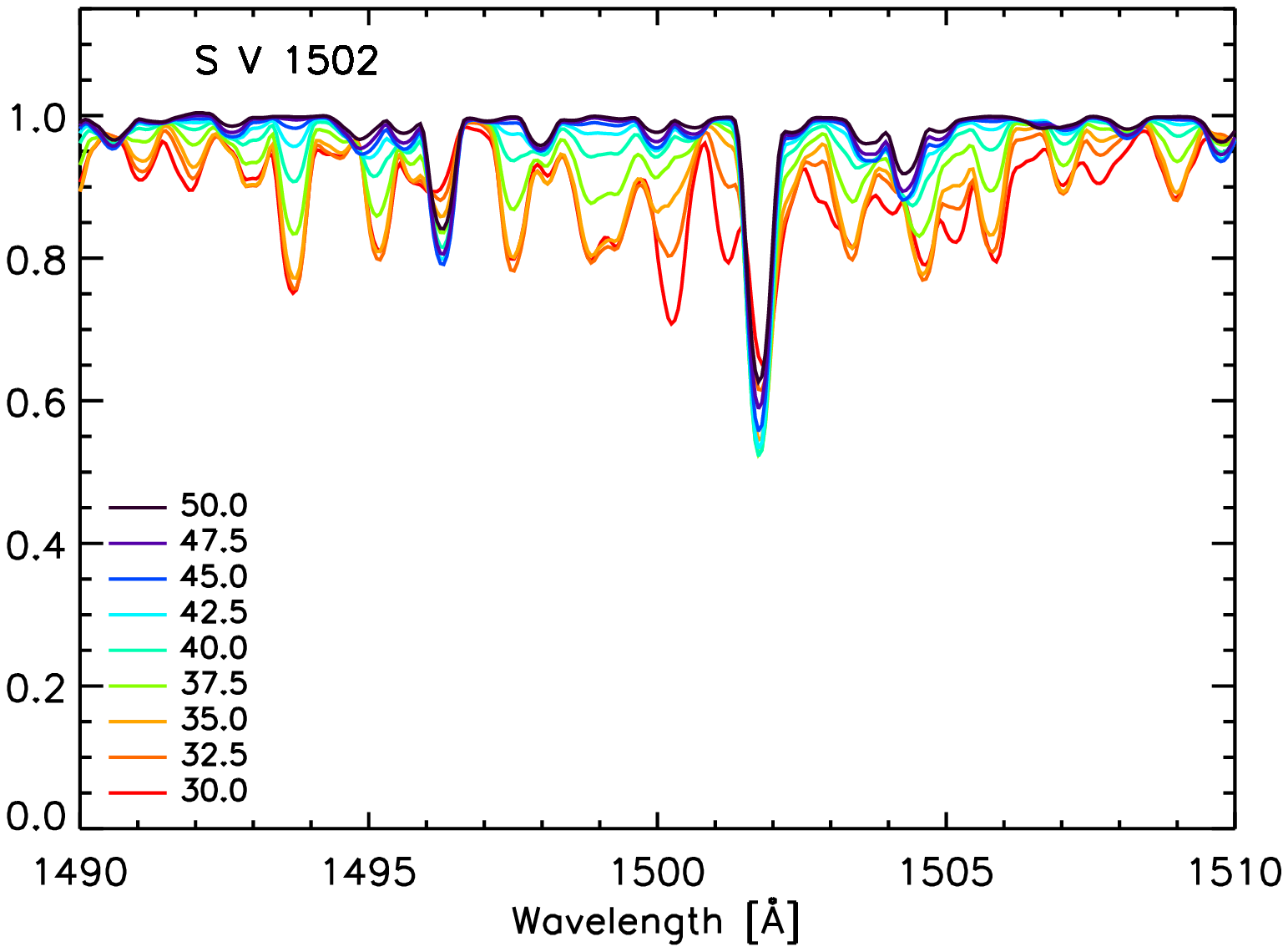} \figurenum{10o}
\caption{ONLY IN ELECTRONIC EDITION}
\end{figure}

\clearpage

\begin{figure}[tbp]
\epsscale{.699} \plotone{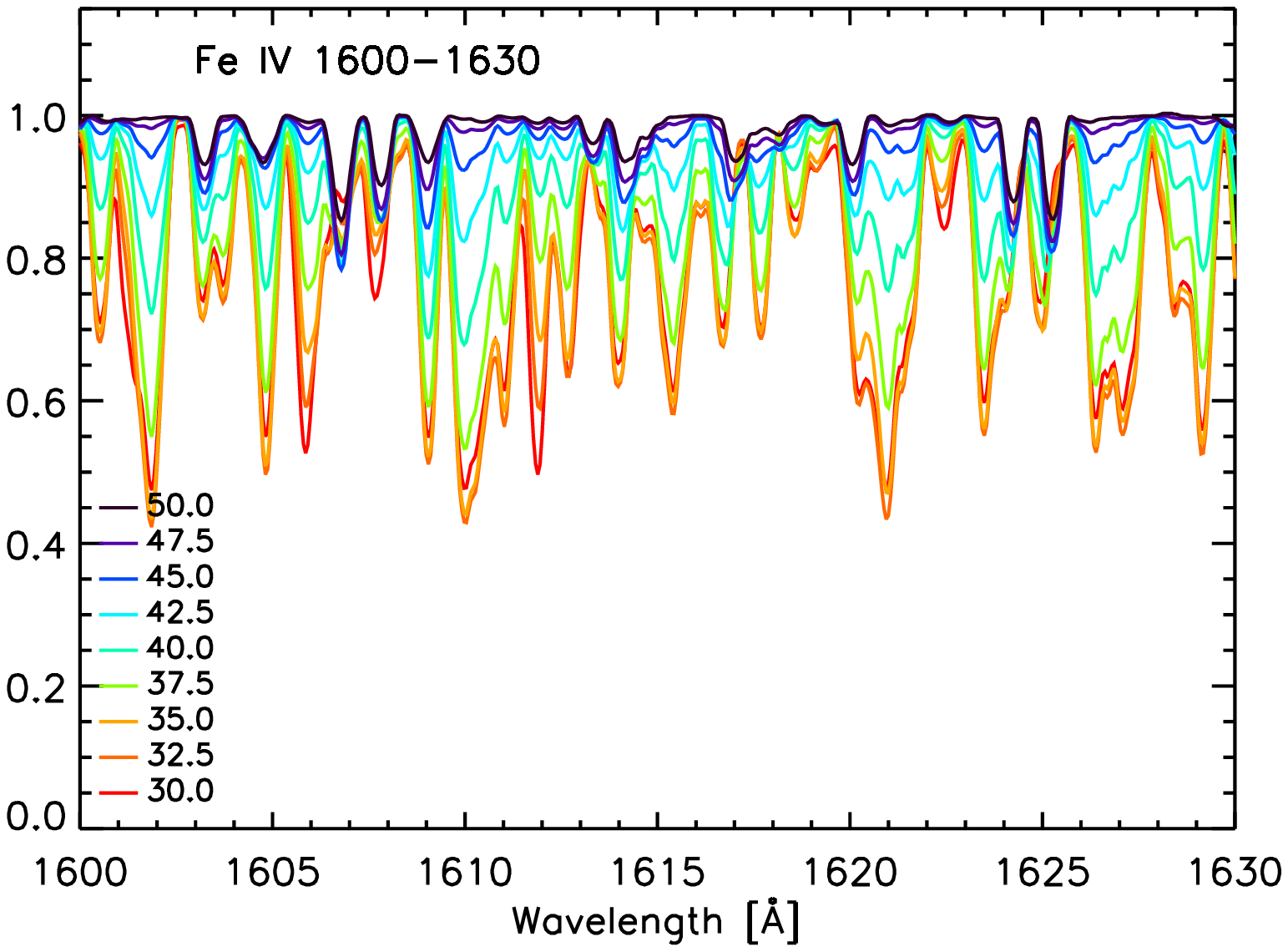} \figurenum{10p}
\caption{ONLY IN ELECTRONIC EDITION}
\end{figure}

\begin{figure}[tbp]
\epsscale{.699} \plotone{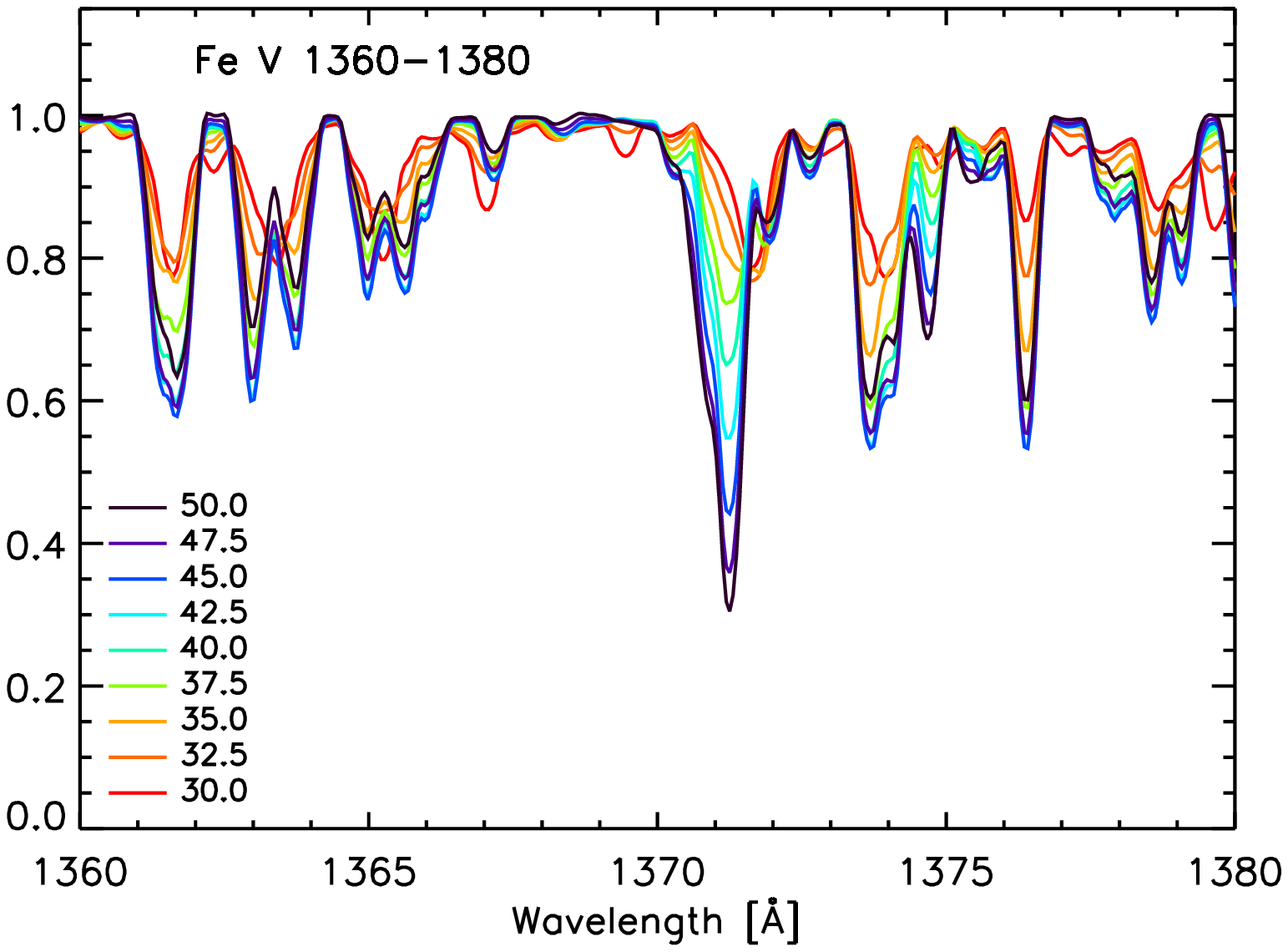} \figurenum{10q}
\caption{ONLY IN ELECTRONIC EDITION}
\end{figure}

\clearpage

\begin{figure}[tbp]
\epsscale{.699} \plotone{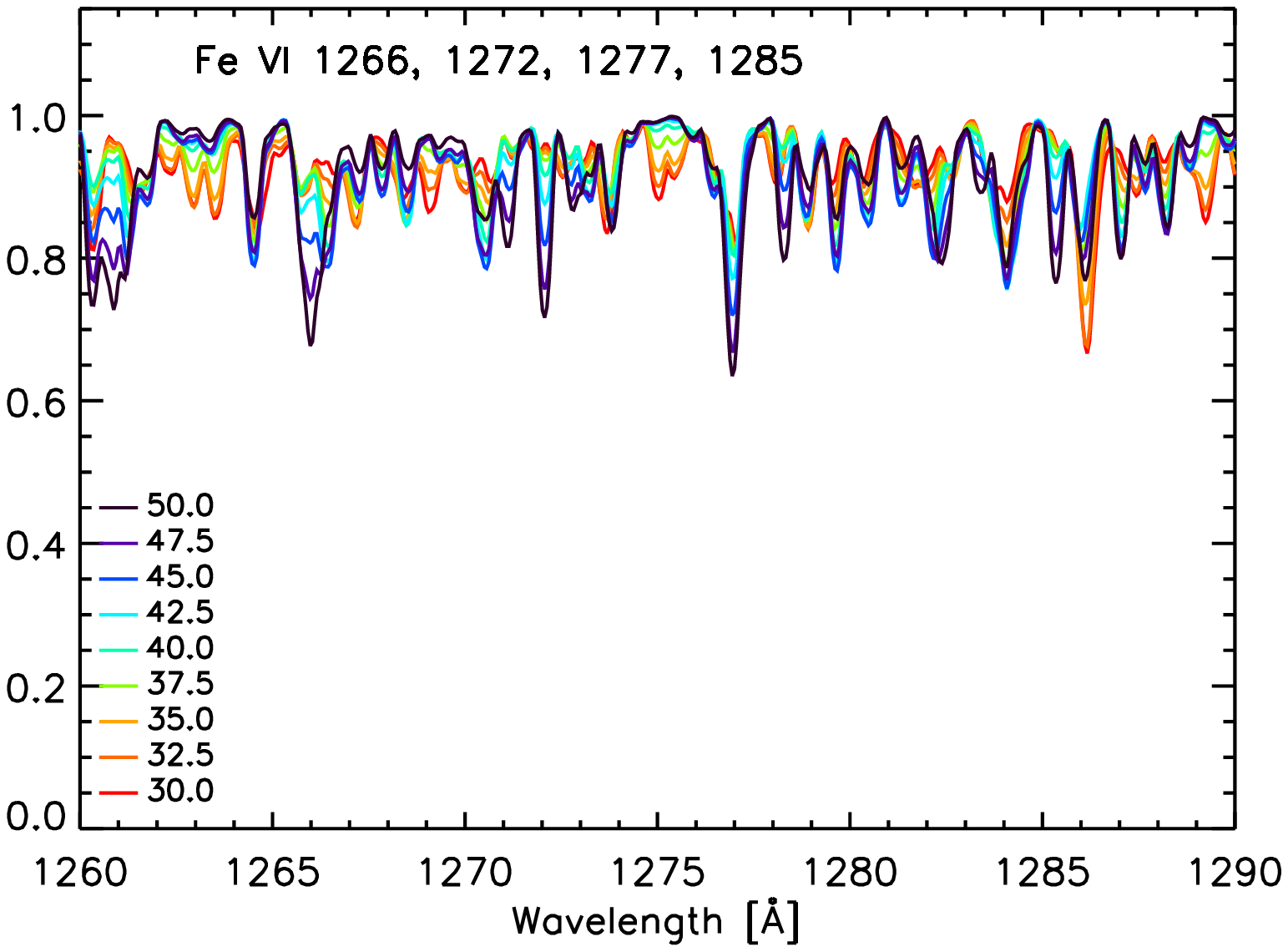} \figurenum{10r}
\caption{ONLY IN ELECTRONIC EDITION}
\end{figure}

\clearpage

\begin{figure}[tbp]
\epsscale{.99} \plotone{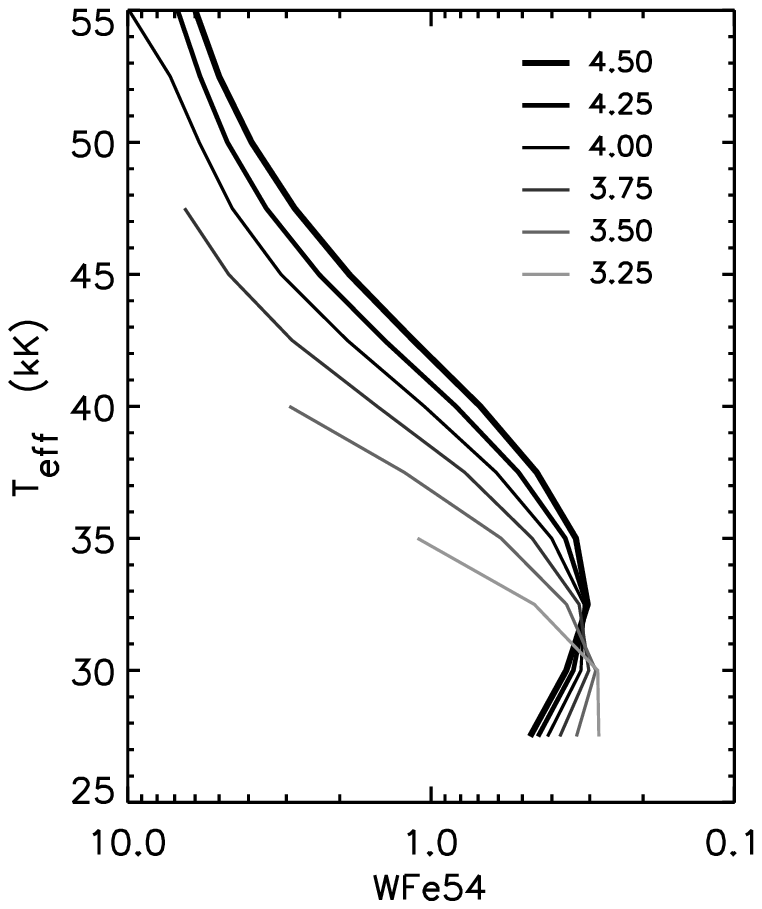} \figurenum{11}
\caption{Iron ionization index, WFe54 (defined in text), as a
function of effective temperature and gravity, for a metallicity $Z =
0.2\,Z_\sun$ and a microturbulent velocity $V_{\mathrm{t}} = 10$\,km\,s$%
^{-1} $. For a given temperature, the index is larger (higher ionization) at lower gravities
as expected from the Saha equation. However, if the gravity is determined from the optical
spectrum (H$\gamma$), then WFe54 is an excellent temperature indicator at 
$T_{\mathrm{eff}} \ge 35,000$~K.  }
\label{WFE45fig}
\end{figure}

\clearpage

\begin{figure}[tbp]
\epsscale{.99} \plotone{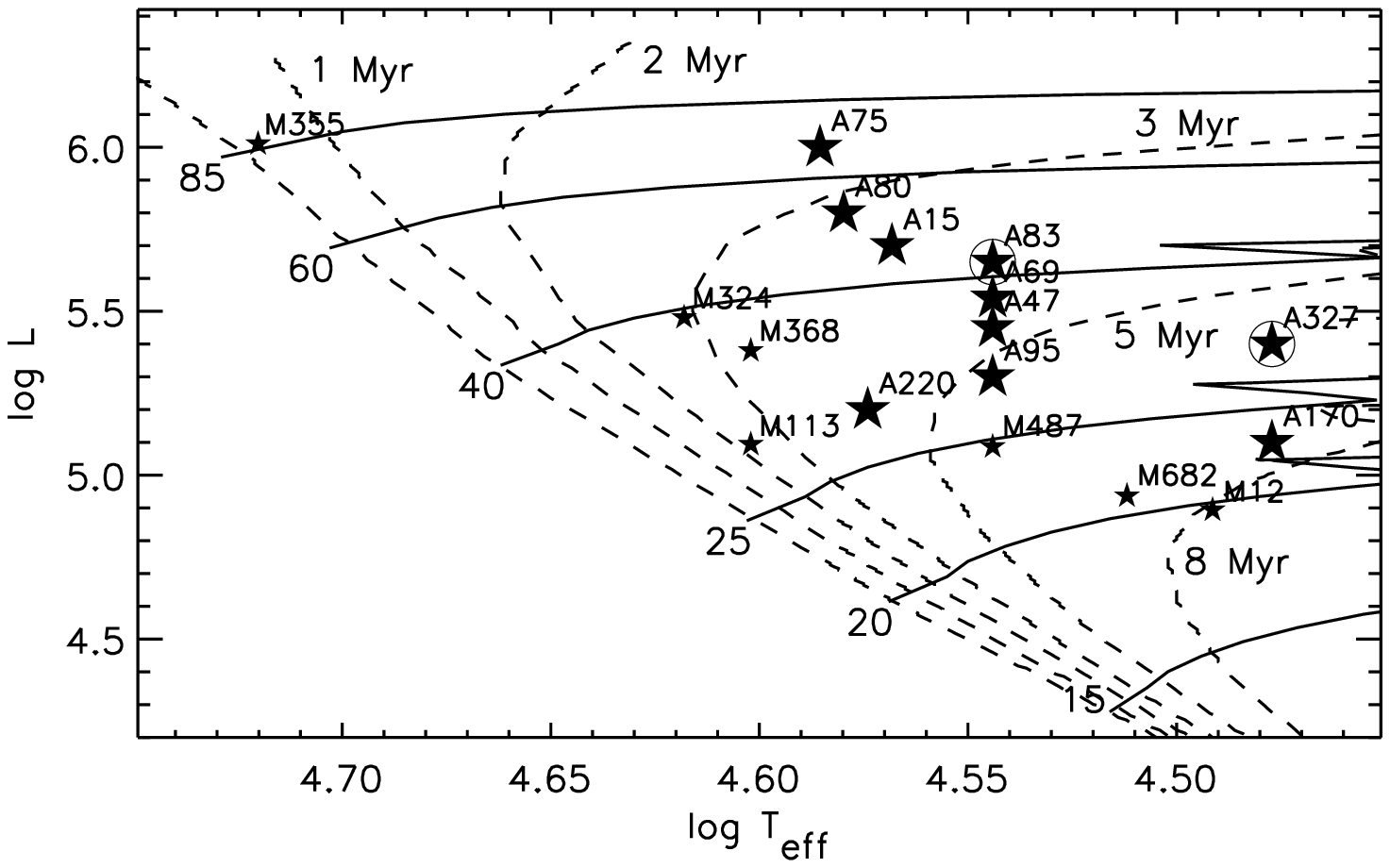} \figurenum{12}
\caption{H-R diagram of the program stars. Each star is identified by its AV (A) number or 
by its MPG (M) number in NGC~346.  Dwarfs are indicated by small star symbols, giants and
supergiants by large stars, and supergiants are circled. Geneva evolutionary tracks  
(thin lines) and  isochrones (dashed lines) for $Z/Z_\sun = 0.2$  are superposed.  
Note the wide spread in ages of stars in the cluster, NGC 346 (see \S 6.7). }
\label{HRfig}
\end{figure}

\clearpage

\begin{figure}[tbp]
\epsscale{.6} \plotone{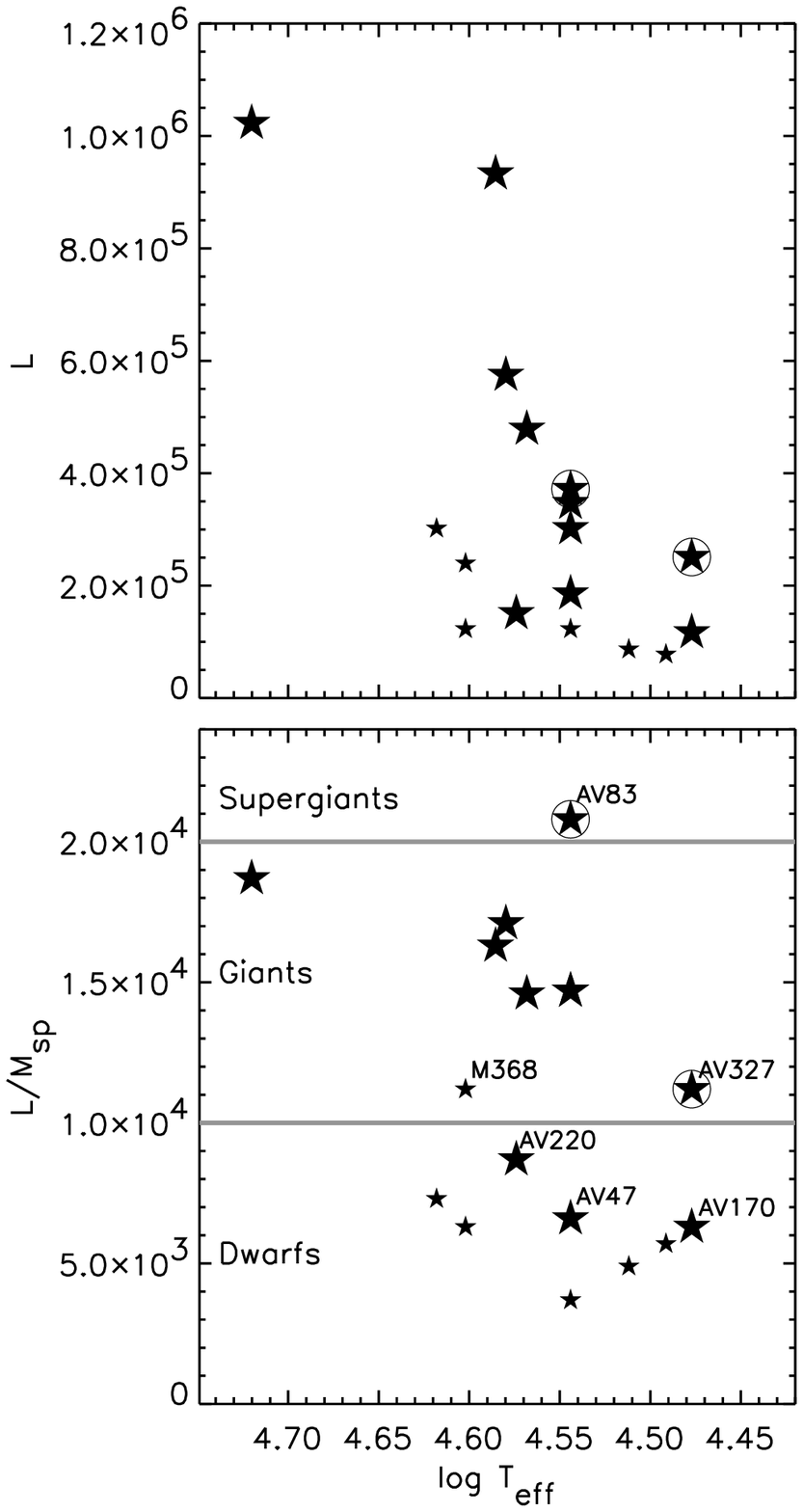} \figurenum{13}
\caption{Effective temperature vs. luminosity (top panel) and ratio of luminosity to spectroscopic mass  
(bottom panel) for stars of different luminosity classes. 
Dwarfs, giants, and supergiants are shown as small stars, large stars, and
circled large stars, respectively.  A few stars are identified and further discussed in the text. 
Although there is little correlation of luminosity class with luminosity, there is a 
somewhat better correspondence with $L/M_{\rm sp}$. }
\label{LMTfig}
\end{figure}

\clearpage

\begin{figure}[tbp]
\epsscale{.99} \plotone{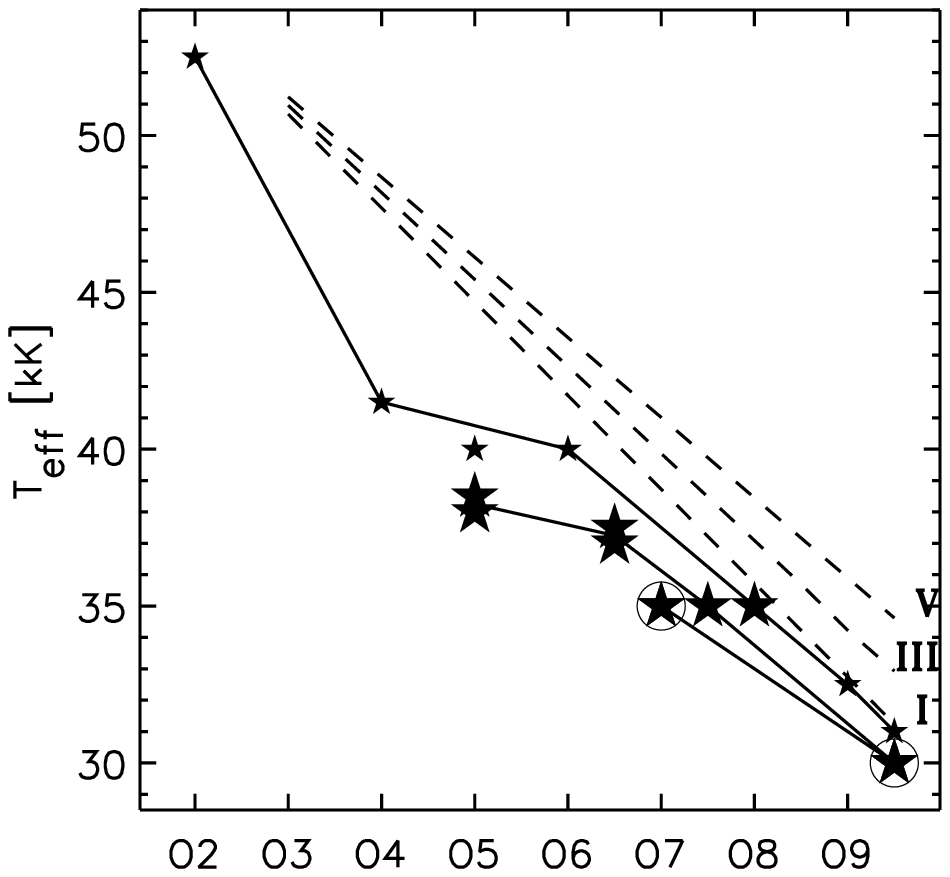} \figurenum{14}
\caption{Temperature calibration of spectral type for SMC
dwarfs (small stars), giants (large stars), and supergiants (circled large stars). 
The dashed lines show the calibration by Vacca et al. (1996), which is considerably
hotter than ours. }
\label{Tefig}
\end{figure}

\clearpage

\begin{figure}[tbp]
\epsscale{.4505} \plotone{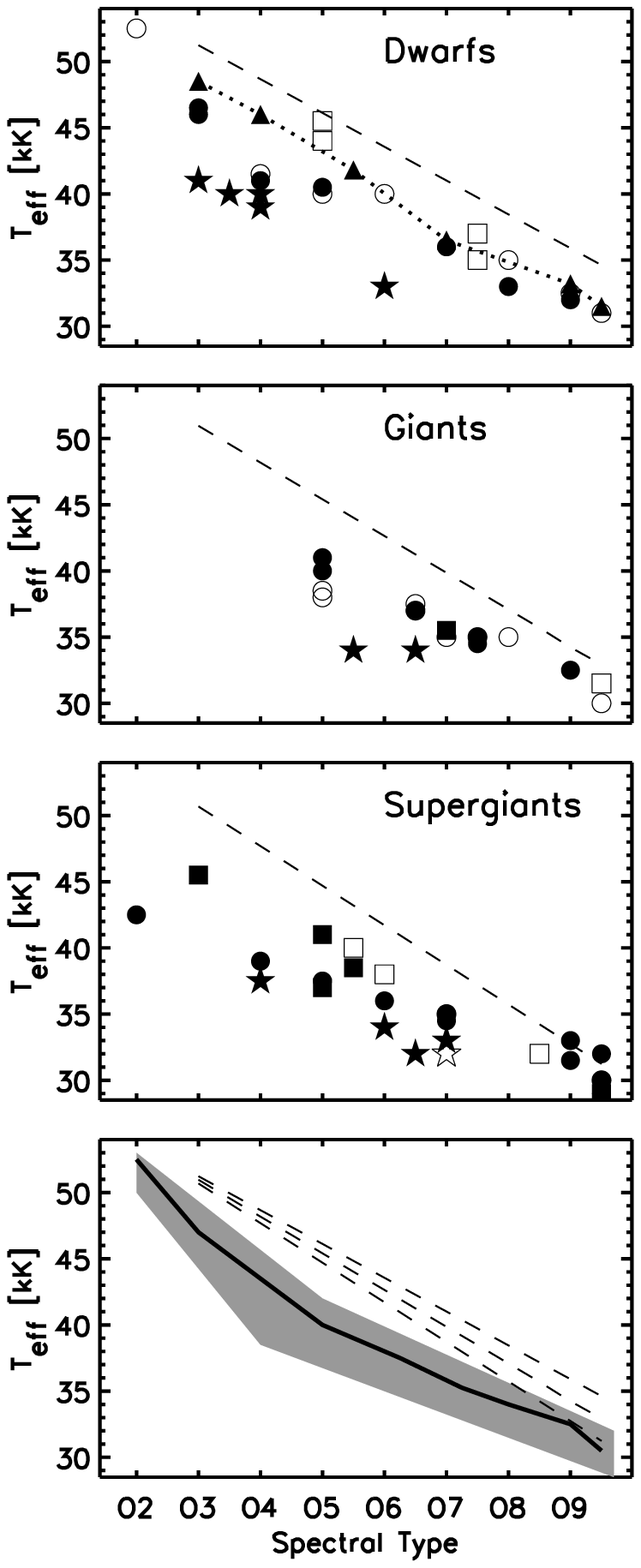} \figurenum{15}
\caption{Recent determinations of effective temperatures of Galactic (full
symbols) and SMC (open symbols) O stars compared to the Vacca et~al. (1996)
relations (dashed lines; Galactic). Studies of Galactic stars include
Martins et al. 2002 (triangles, dotted line), Bianchi~\& Garcia 2002
(stars), Herrero et~al. 2002 (squares), and Repolust et al. 2004 (circles).
Results for SMC stars are extracted from Crowther et~al. 2002 (open star),
Massey et~al. 2004 (open squares), and this paper (open circles). The bottom
panel shows the range of recent $T_{\mathrm{eff}}$\ determinations (shaded
area), the Vacca et~al. (1996) relations, and the typical temperatures that
we assign to each spectral type from a critical comparison (full line). }
\label{Tcompfig}
\end{figure}

\clearpage

\begin{figure}[tbp]
\epsscale{.99}
\plotone{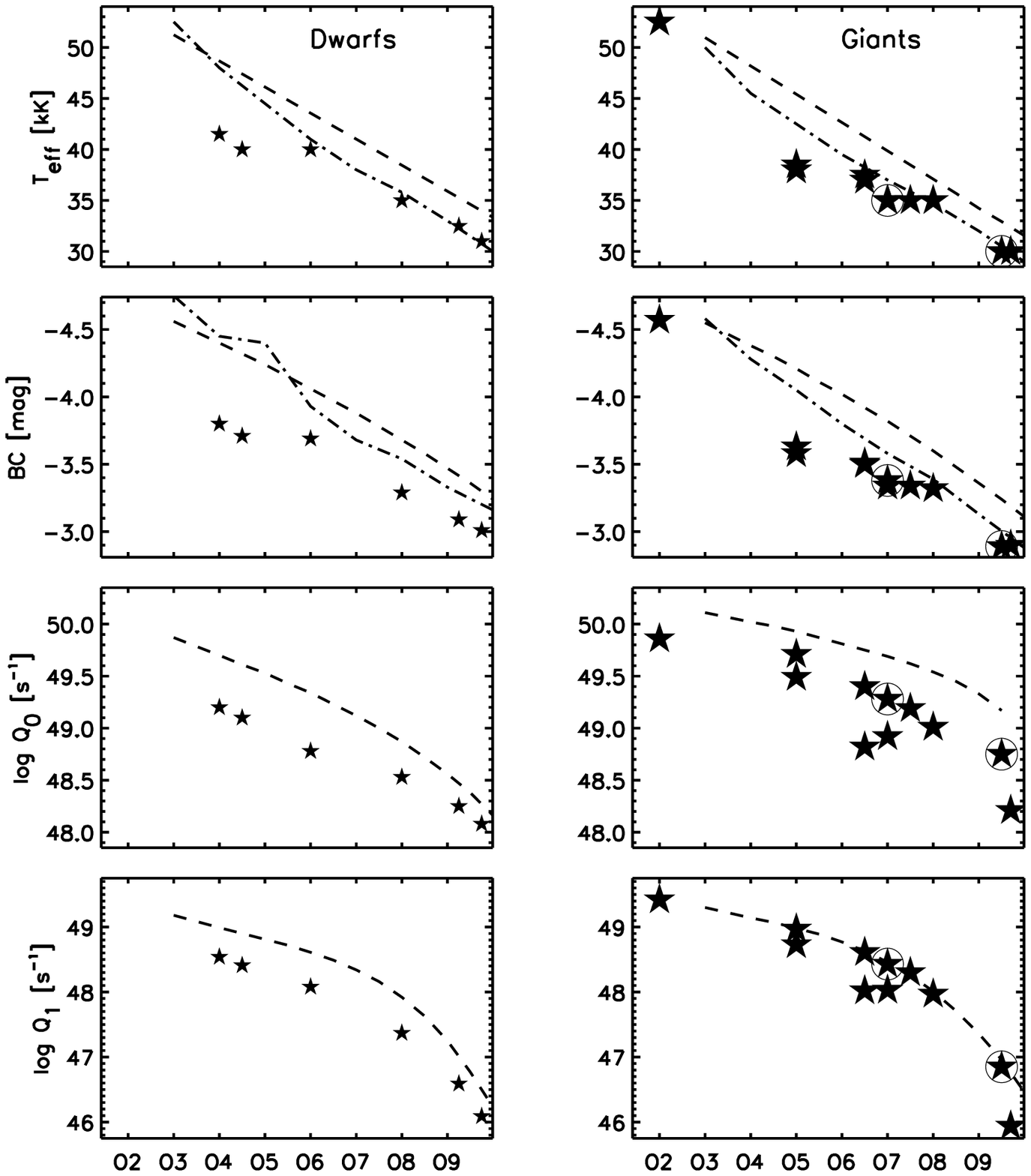}  \figurenum{16}
\caption{Spectral type vs. effective temperature, bolometric correction, \ion{H}{1}- and 
\ion{He}{1}-ionizing luminosities. The dashes and dash-dot lines show the \cite{Vacca96} 
and \cite{SK82} relations, respectively.}
\label{BCQfig}
\end{figure}

\clearpage

\begin{figure}[tbp]
\epsscale{.799} \plotone{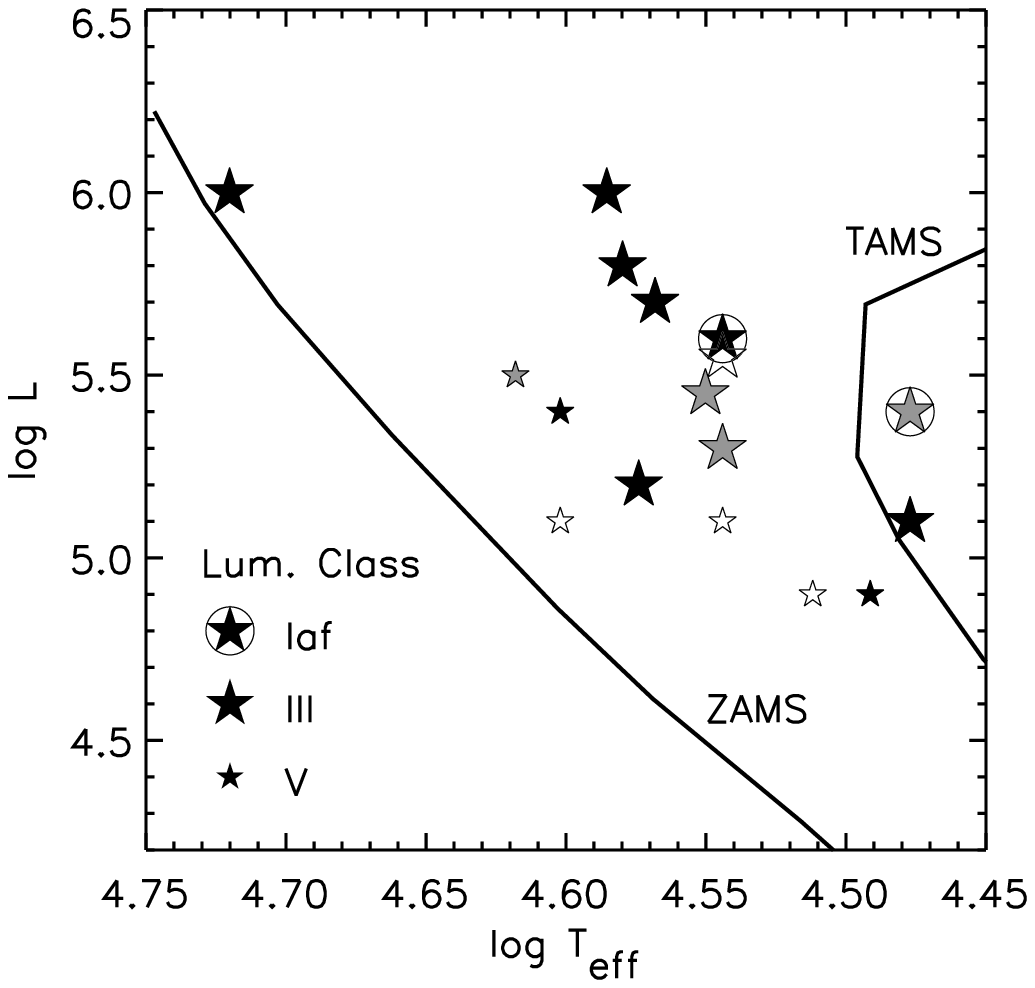} \figurenum{17}
\caption{H-R diagram of SMC stars with symbols shaded according to their N/C surface
enrichment (black: N/C = 25-40 X ; gray: N/C = 6-15 X; white: original (SMC nebular) abundances).
The two lines depict the ZAMS and TAMS (end of core
hydrogen burning) of non-rotating Geneva models. (Rotating models have a cooler TAMS.) 
There is little correlation between nitrogen enrichment and position on
the H-R diagram. }
\label{HRNCfig}
\end{figure}

\clearpage

\begin{figure}[tbp]
\epsscale{.799} \plotone{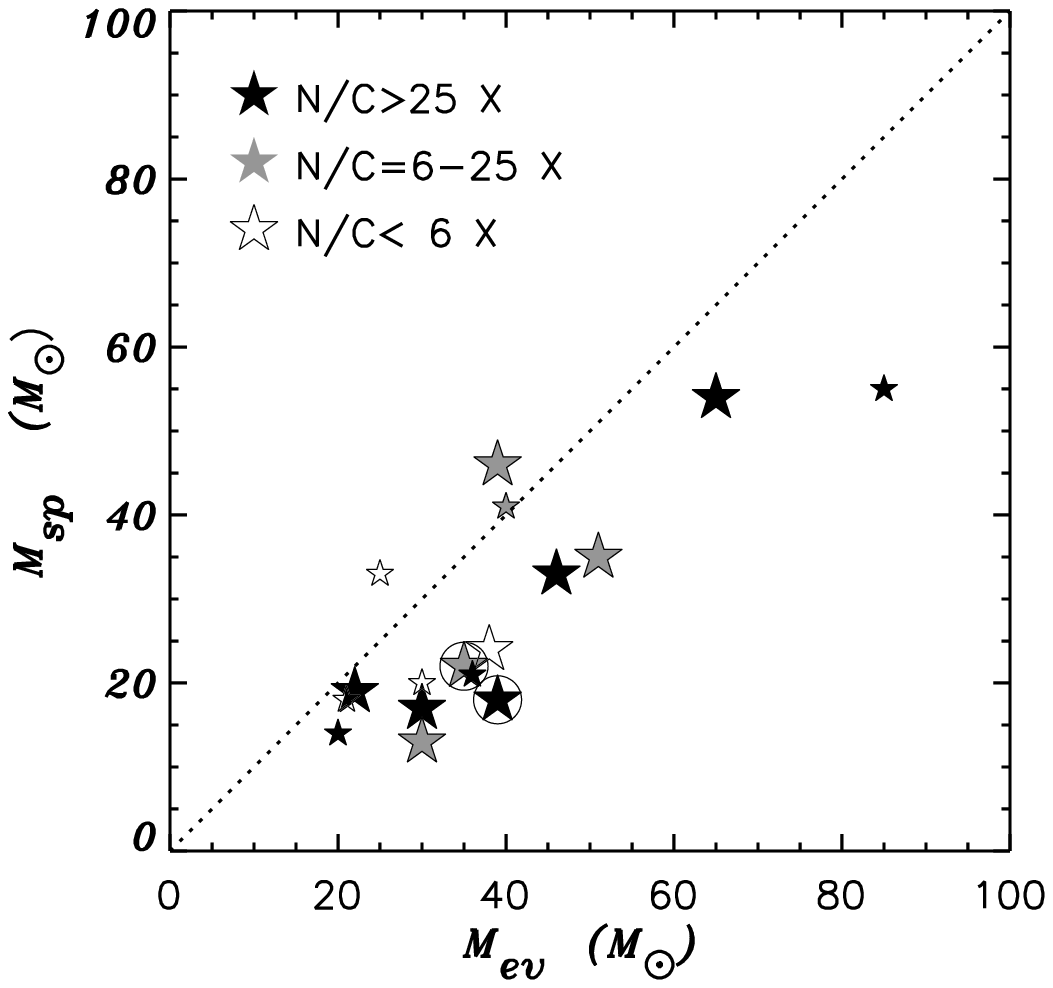} \figurenum{18}
\caption{Spectroscopic vs. evolutionary masses illustrating the
``mass-discrepancy problem'' in which spectroscopically derived masses are systematically
lower than those derived from their positions on the HRD. The N/C surface enrichment is denoted by the shading as in Fig. 17,
and the luminosity class, by the size of the star symbol (with supergiants circled)}.
\label{Massfig}
\end{figure}

\end{document}